\documentclass[aps,prx,twocolumn,superscriptaddress,longbibliography]{revtex4}
\usepackage{graphicx}
\usepackage{dcolumn}
\usepackage{physics}
\usepackage{amsmath}
\usepackage{amssymb}
\usepackage{bm}
\usepackage{array}
\usepackage{soul}
\usepackage{color}
\usepackage{xcolor} 
\usepackage{float}
\usepackage[normalem]{ulem}
\newcolumntype{P}[1]{>{\centering\arraybackslash}p{#1}}
\usepackage[colorlinks=true, breaklinks=true, linkcolor=blue, citecolor=blue, urlcolor=blue]{hyperref}
\usepackage{ytableau}
\renewenvironment{bmatrix}{\ytableausetup{smalltableaux}\begin{ytableau}}{\end{ytableau}}
\usepackage{comment}

\newcommand{\pks}{Max Planck Institute for the Physics of Complex Systems, N\"othnitzer Str.~38, 01187 Dresden, Germany}
\newcommand{\nice}{Université Côte d’Azur, CNRS, Centrale Med, Institut de Physique de Nice, 06200 Nice, France}

\begin{document}

\title{Constructing Large and Structured Decoherence-Free Subspaces in Hybrid Quantum Systems}
\date{\today}

\begin{abstract}
We develop a framework to construct large decoherence-free subspaces with a nontrivial structure.
The construction is based on hybrid quantum systems in which quantum matter is coupled to a dissipative bosonic mode. 
Dissipation imposes a global constraint on the matter by selecting the matter null states in the long-time limit.
The decoherence-free subspaces spanned by matter null states can exhibit exotic quantum properties and emergent symmetries that do not characterize the full Liouvillian dynamics.
As the multiple steady states arising in the system do not rely on Liouvillian symmetries, the associated sequence of conserved quantities contains multi-body and nonlocal contributions.
We show how to construct many-body operators with physically interesting null states by employing representation theory of Lie algebras.
This approach encompasses previously studied models of noiseless quantum codes and allows us to choose models based on their underlying algebraic structure.
We consider explicitly the example of coupling a lossy optical cavity to the directed tunneling operator of quantum particles in one-dimension. 
In this case, the symmetries of the decoherence-free subspace originate in an $\mathfrak{sl}(2,\mathbb{C})$ algebra, leading to an exponentially large number of steady states. 
We characterize the null states of the directed tunneling operator for both fermionic and bosonic particles, employing analytical and numerical methods.
The states exhibit several properties that underline their complex quantum nature: long-range kinetic correlations and volume-law entanglement.
The nonreciprocal nature of the directed tunneling dynamics leads to the emergence in the many-body regime of a symmetry-constrained Liouvillian skin effect. 
Furthermore, the open light-matter system exhibits a strong symmetry stemming from spinless $\eta$-pairing algebras.
We show how the nontrivial properties of the long-time states emerge in the dissipative dynamics from generic initial states, by performing the quantum evolution of the hybrid open system with time-dependent matrix product state.
Our framework opens avenues for dissipatively engineering complex quantum correlations and harnessing nonreciprocity in a controlled manner.
\end{abstract}

\author{Catalin-Mihai Halati}
\affiliation{\pks}
\author{Tony Jin}
\affiliation{\nice}
\author{Roderich Moessner}
\affiliation{\pks}
\maketitle

\section{Introduction}

\begin{figure*}[!hbtp]
\centering
\includegraphics[width=0.8\textwidth]{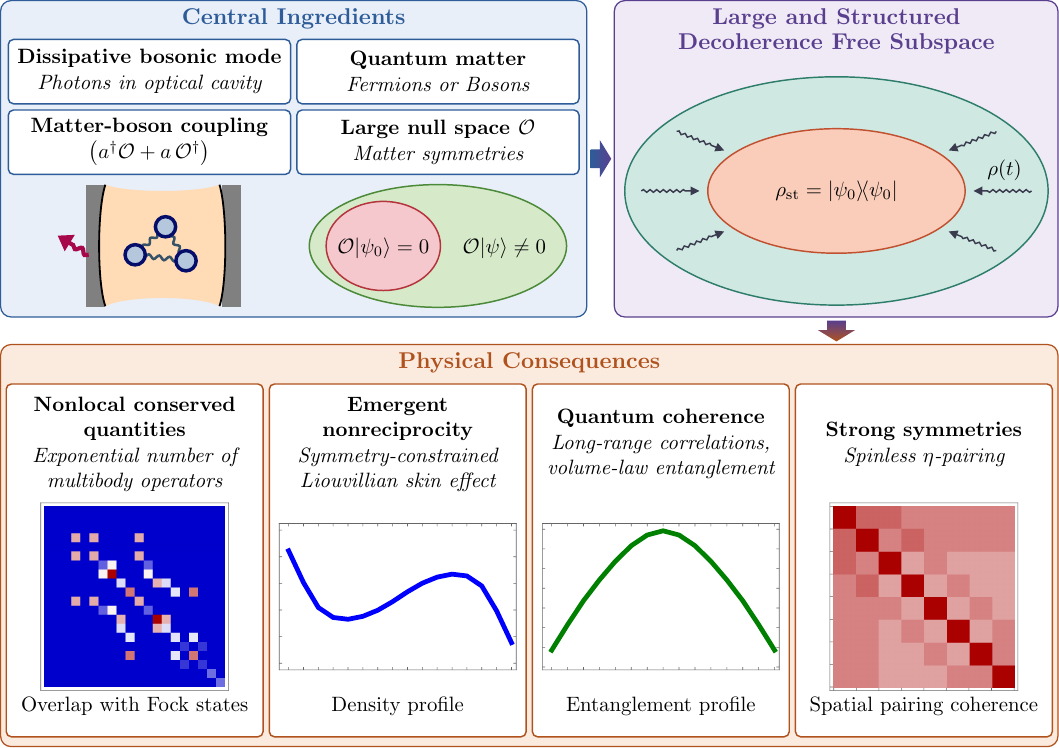}
\caption{
Sketch of the central ingredients we employ in our framework to construct large and structured decoherence-free subspaces, and the main physical consequences.
We depict both the generic ingredients and the results obtained for the coupling between the cavity field to a directed tunneling operator. 
The correspondence between these ingredients and the construction of the decoherence-free subspace is outlined in the extended summary, Sec.~\ref{sec:summary}.
The lower panels sketch the main results: an exponential number of nonlocal conserved quantities, exhibiting a finite overlap with a large number of Fock states; 
nonreciprocal dynamics giving rise to a many-body Liouvillian skin effect in which the density accumulation to an edge is constrained by symmetries; 
long-range complex correlations and entanglement entropy with volume-law scaling; 
and finally, a spinless $\eta$-pairing mechanism emerging as a strong symmetry of the Liouvillian leading to long-range pair coherence.
}
\label{fig:summary}
\end{figure*}

Interacting quantum systems give rise to some of the most striking manifestations of quantum phenomena. The ground states of many-body Hamiltonians can host, for example, long-range entanglement, or degenerate manifolds encoding quantum information \cite{SachdevBook, AmicoVedral2008, Terhal2015}.
However, in practice zero temperature many-body ground states are hard to prepare. In contrast, finite temperature Gibbs states are comparatively easy to prepare, but thermal fluctuations tend to wash out precisely the quantum features of interest \cite{BrownWootton2016, BakshiTang2024}. 
Thus, in equilibrium we encounter a trade-off between how easy it is to prepare a state and what quantum properties it retains.

Driving a quantum system out-of-equilibrium offers an avenue for circumventing the problems encountered in equilibrium.
Drives and engineered environments, by breaking detailed balance, can even stabilize collective behaviors which have no counterpart in equilibrium  \cite{DiehlZoller2008, VerstraeteCirac2009, MuellerZoller2012, MetelmannClerk2015, MivehvarRitsch2021, LandiSchaller2022, SiebererDiehl2025, BelyanskyClerk2025, FruchartVitelli2026}.
One paradigmatic example is passive quantum error correction, where an engineered reservoir continuously removes the entropy injected by noise \cite{KerckhoffMabuchi2010, PastawskiCirac2011, Kapit2016, LihmFischer2018}.
Designing dynamical non-equilibrium protocols is, however, a formidable control problem, as generic drives and baths lead to heating and decoherence.
The development of frameworks which allow to maintain quantum coherence and complex correlations at arbitrarily long times is a central challenge of the field.

One strategy aiming at harnessing quantum properties in out-of-equilibrium settings is dissipative engineering. 
This relies on tailoring the coupling between the quantum system of interest and environments, such that non-trivial target states become attractors of the dissipative dynamics \cite{DiehlZoller2008, VerstraeteCirac2009, MuellerZoller2012, BardynDiehl2012, BardynDiehl2013, BernierKollath2013, LeeClerk2024}.
A generalization of the initial dissipative engineering approaches realizes decoherence-free subspaces, in which a manifold of states is prepared in the long-time limit \cite{ZanardiRasetti1997, LidarWhaley1998}.
The construction of decoherence-free subspaces further offers the opportunity to control metastable and dynamical properties \cite{PolettiKollath2013, ZanardiCamposVenuti2014, SciollaKollath2015, MacieszczakGarrahan2016, KingMorigi2024, HalatiKollath2025}. 
The number and complexity of the states spanning the decoherence-free subspace is directly linked to the nature of the quantum correlations emerging in the dissipative dynamics \cite{Halati2026}.
However, generic recipes in which both the size and the structure of the decoherence-free subspaces can be designed are lacking. 
In our work we propose to make use of a design principle which has shown tremendous success in closed systems: constraints.

In Hamiltonian systems, when an interaction scale dominates, local rules can emerge, obeyed by the configurations contained in the low-energy manifold, e.g.~the ice rule in frustrated magnets \cite{BramwellGingras2001, MoessnerSondhi2001, CastelnovoSondhi2012, GingrasMcClarty2014} and the Rydberg blockade \cite{JakschLukin2000, LukinZoller2001, BrowaeysLahaye2020}.
The constraints lead to extensively degenerate manifolds \cite{MoessnerSondhi2001}, with weak perturbations being able to generate non-trivial emergent phases \cite{Balents2010, SavaryBalents2017}.
Furthermore, the presence of constraints can give rise to new dynamical features as quantum many-body scars and Hilbert-space fragmentation \cite{BernienLukin2017, TurnerPapic2018, SalaPollmann2020, KhemaniNandkishore2020, MoudgalyaRegnault2022}.
The paradigm has been extended recently to global constraints, imposed by long-range interactions \cite{SantosCelardo2016, CelardoBorgonovi2016, WuZhao2026, HosseinabadiMarino2026}.

This raises the question, if one can extend the scope of constraint-based mechanisms to open quantum systems, in order to engineer the structure of decoherence-free subspaces.
Recent works have shown how decoherence-free subspaces are influenced by local kinetic constraints of the Hamiltonian \cite{SrivastavaDutta2025}, or that non-trivial steady states can be obtained from global kinetic constraints \cite{WinterNunnenkamp2026, dosPrazeresMarino2026}.
We show how by dissipatively enforcing a global constraint, we can develop a generic approach for designing decoherence-free subspaces.

Concretely, in this work, we develop a framework for constructing large and structured decoherence-free subspaces in hybrid open quantum system. 
For this, we provide a general recipe, which we exemplify for quantum many-body systems coupled to photonic field of a dissipative cavity, for a schematic overview see Fig.~\ref{fig:summary}. 
Our approach combines notions from quantum many-body physics and quantum optics and incorporates insights and algebraic structures from group representation theory. 
We find particularly rich structure emerging in the decoherence-free subspace, which can be controlled by global light-matter coupling.
The dissipative nature of the cavity imposes a global constraint onto the matter degrees of freedom as the decoherence-free subspace is approached in the long-time limit.
We design the symmetry structure of the constraint by employing representation theory of the underlying Lie algebras.
Our construction yields multiple steady states characterized by symmetry properties which characterize either the Liouvillian dynamics, or only emerge in the decoherence-free subspace.
Salient physical consequences of these structures include, besides the exponentially large number of steady states in the decoherence-free subspace, a robust emergent non-reciprocity, as well as the persistence of quantumness in the form of long-range quantum correlations and volume-law entanglement.

The framework developed in this work allows us to construct novel examples of structured decoherence-free subspaces, by coupling low-dimensional itinerant quantum particles to quantum dissipative light. 
Furthermore, we can shine new light onto previously studied systems in which independent spins-$1/2$ are coupled to the photon field. 
We establish explicitly the emerging nontrivial physical consequences for a system of quantum particles in one dimension globally coupled via directed tunneling processes to a dissipative photonic field. 
We present this setup as a fully worked example of our recipe, for both fermionic and bosonic particles.
Here, the symmetry structure of the decoherence-free subspace originates in two commuting Lie algebras. 
The first $\mathfrak{sl}(2,\mathbb{C})$ algebra determines a many-body Liouvillian skin effect, where the boundary localization competes with multi-particle interference effects and symmetry constraints.
The second algebra leads to a strong symmetry of the open system and manifests itself in an emergent spinless $\eta$-pairing mechanism.
Concretely, the algebraic structure leads to exotic localization and pairing properties of the quantum particles, generating complex quantum correlations.

The paper is organized as follows: Sec.~\ref{sec:summary} represents an extended summary offering a compact guide to key results of our work.
In Sec.~\ref{sec:recipe} we present the recipe to construct structured decoherence-free subspaces spanned by a large number of states. 
A large null space of the non-Hermitian operator describing the matter gives rise to a large number of steady states for the light-matter coupled system, Sec.~\ref{sec:recipe_1}, and non-local conservation laws, Sec.~\ref{sec:recipe_2}.   
We show in Sec.~\ref{sec:construction} how one can construct matter operators with exponentially large null spaces by employing representation theory of Lie algebras.
We exemplify our recipe by coupling a lossy cavity mode to a one-dimensional directed tunneling operator for both bosonic and fermionic quantum particles, Sec~\ref{sec:results_symmetries}. 
The directed tunneling operator satisfies a $\mathfrak{sl}(2,\mathbb{C})$ algebra, which has profound physical consequences.
This leads to an exponentially large decoherence-free subspace and to a symmetry-constrained Liouvillian skin effect.
We show how to construct the null states of the directed tunneling operator analytically in Sec.~\ref{sec:group_theory}.
Our hybrid open system exhibits a strong symmetry realizing a spinless $\eta$-pairing mechanism, Sec.~\ref{sec:pairing_algebra}.
We describe the nontrivial properties of the steady states spanning the decoherence-free subspace obtained by exact diagonalization in Sec.~\ref{sec:results_steady_states}.
We construct the multi-body and nonlocal conserved quantities for a small system in Sec.~\ref{sec:results_conserved}.
In Sec.~\ref{sec:results_tmps} we perform time-dependent matrix product state simulations for the full quantum evolution of the hybrid open system. We show that the nontrivial properties also characterize the mixed steady states.

\section{Extended summary and key results \label{sec:summary}}

In this section we provide an extended summary of our work in  order to outline the main results in a digestible fashion. 
This is motivated by the fact that our work on the one hand constructs a general framework for obtaining large and structured decoherence-free subspaces in light-matter coupled systems and on the other hand it presents a concrete example which we work out in detail.

\emph{Main goal}. We develop a systematic approach to construct decoherence-free subspaces spanned by quantum states characterized by complex quantum correlations.
The decoherence-free subspace of an open quantum system contains all the states that do not dissipatively decay, i.e.~are steady states of the dynamics. It is reached in the long-time limit of the dissipative evolution.
Figure~\ref{fig:summary} sketches the central ingredients: a generic coupling between quantum matter, characterized by either fermionic or bosonic degrees of freedom, and a lossy photonic mode, see Eq.~\eqref{eq:Lindblad} in Sec.~\ref{sec:recipe}. 

Our main requirement is that the matter operator, $\mathcal{O}$, which couples to the cavity mode, (i) is characterized by a large null space, i.e.~the subspace spanned by the states with eigenvalue zero; and (ii) that it does not commute with its Hermitian conjugate. 

The framework engenders dramatic physical consequences. 
Key features include (see Fig.~\ref{fig:summary}): 
(i) a large dissipation free subspace, characterized by the symmetries of $\mathcal{O}$, which can be distinct from the symmetries of the full Liouvillian dynamics; 
(ii) the dissipative dynamics is constrained by a large number of conserved quantities containing nonlocal and multibody operators, deviating from the usual constructions making use only of strong symmetries; 
(iii) additional strong symmetries can arise to further enrich the dissipative dynamics; 
(iv) as the operator $\mathcal{O}$ is non-Hermitian, the quantum dynamics has an emergent non-reciprocal character; 
(v) our construction provides a way to generate long-range complex quantum correlations in dissipative hybrid systems.

\begin{figure}[!hbtp]
\centering
\includegraphics[width=0.42\textwidth]{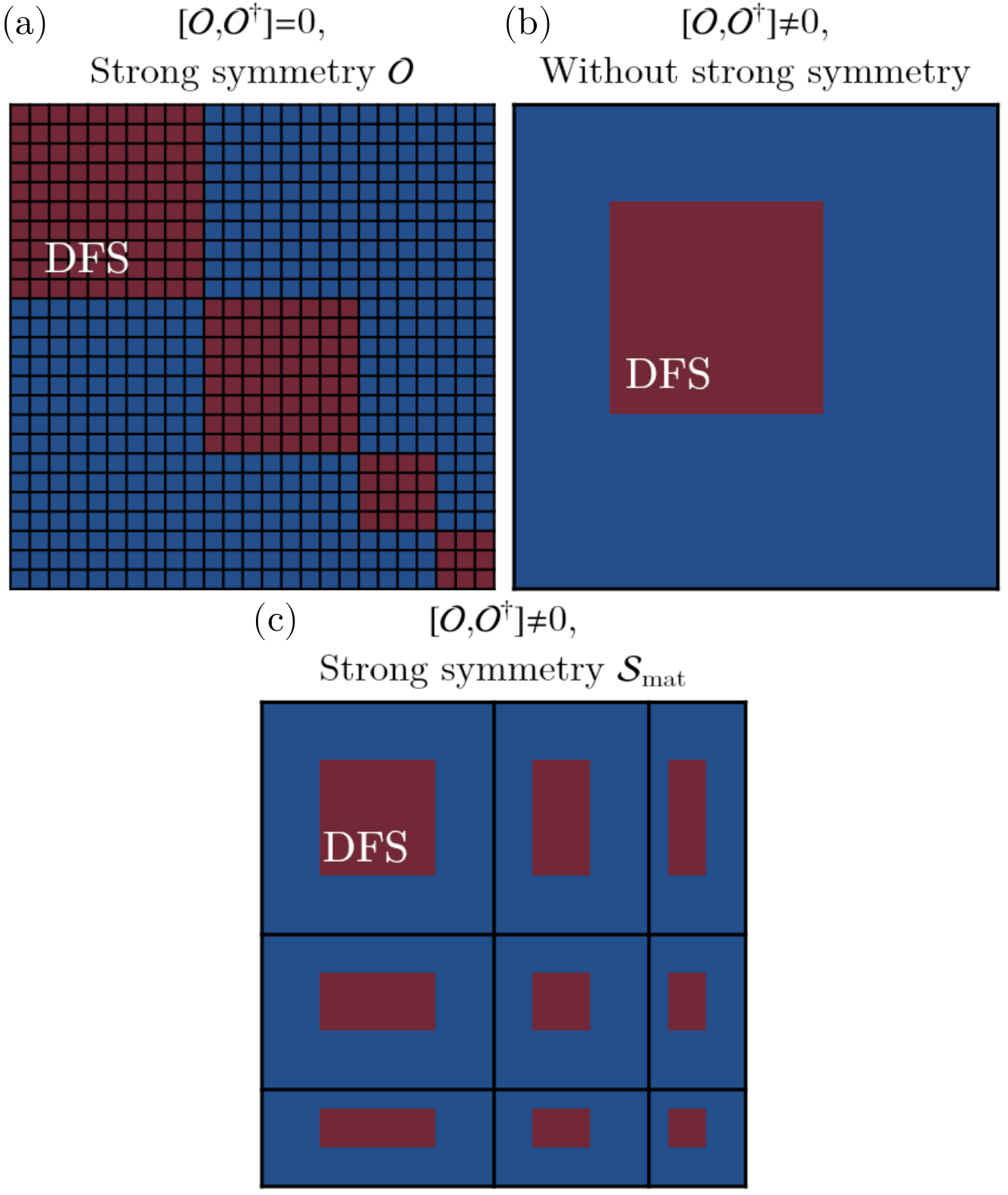}
\caption{
Sketches of the structure of the decoherence-free subspaces (DFS) inside the Hilbert space of the matter density matrices, for the three cases discussed in Sec.~\ref{sec:recipe_1}. The Hilbert space of the matter density matrices is sketched by a two-dimensional square for which the horizontal and vertical directions label the ket and bra wavefunctions.
The black grid represents the subspace structure given by the presence of a strong symmetry. 
Red depicts the subspaces forming the DFS and blue the matter states with a finite lifetime. The photonic states are not considered in this sketch.
}
\label{fig:DFS}
\end{figure}

\emph{Generic construction of steady states}. We begin in Sec.~\ref{sec:recipe} by characterizing the steady states of the dissipative hybrid system, Eq.~\eqref{eq:Lindblad}, depending on the (strong) symmetries of the Liouvillian, $\mathcal{L}$. 
We identify three distinct cases based on two conditions: whether $\mathcal{O}$ commutes with its Hermitian conjugate; and whether the two operators share any symmetries. 
The decoherence-free subspaces corresponding to these three cases are sketched in Fig.~\ref{fig:DFS}.
We focus on the case that $\mathcal{O}$ and $\mathcal{O}^\dagger$ do not commute, but share a symmetry.

Our framework relies on the observation that we do not need a strong symmetry of the Liouvillian in order to construct a large decoherence-free subspace.
A strong symmetry is generated by an operator $\mathcal{S}$ that commutes with the Hamiltonian and with every jump operator and guarantees at least one steady state per symmetry sector~\cite{AlbertJiang2014,BucaProsen2012}.
A large decoherence-free subspace then requires many sectors, which need to be shared by $\mathcal{O}$ and $\mathcal{O}^\dagger$, restricting the possible choices, Sec.~\ref{sec:recipe}.

In contrast, our construction allows us to choose an operator $\mathcal{O}$ whose symmetries are not shared with the Liouvillian $\mathcal{L}$, opening the way to richer decoherence-free structures.
For the case which we focus on, $[\mathcal{O},\mathcal{O}^\dagger]\neq 0$, the steady states are then constructed from the null states of the operator $\mathcal{O}$, Eq.~\eqref{eq:rho}. 
This leads to the existence of multiple steady states even in the absence of a strong symmetry, or within a single strong symmetry sector in its presence.
As a consequence additional symmetry properties emerge at late times in the dissipative dynamics, once the decoherence-free subspace is reached. 
This is the situation we present in our example discussed starting with Sec.~\ref{sec:results_symmetries}.
For strong symmetries the conserved quantities are also the generators of the symmetry \cite{AlbertJiang2014, BucaProsen2012, SeltmannBuca2026}, which in practice means that in most cases one chooses few-body operators or sums over such operators. 
However, in our framework, the emerging conserved quantities need to be constructed from the steady states,  Sec.~\ref{sec:recipe_2}. This results in complex operators containing long-range multibody terms (sketch in Fig.~\ref{fig:summary}). We explicitly construct such conserved quantities in Sec.~\ref{sec:results_conserved}.

\emph{Obtaining non-trivial null states}. The coupling to the lossy photonic mode acts as a global constraint which restricts the steady states to the null states of the matter operator.
One may now ask how to construct the operator $\mathcal{O}$ in practice, in order to ensure a large number of null states with non-trivial properties. 
We provide a systematic recipe to achieve this in Sec.~\ref{sec:construction}, based on representation theory. 
We consider $\mathcal{O}$ to be the sum of single-particle root vectors of the representation of a semisimple Lie algebra. 
One can view the root vector as a generalization of the lowering operator of the spin angular momentum algebra.
This construction ensures the existence of at least one null vector for a single particle, while in the many-body regime the null space is correspondingly larger, generically scaling exponentially with system size.
This gives a practical and systematic way of constructing large null spaces for a variety of representations. 
In particular, we can arrive at models in any dimension, Sec.~\ref{sec:construction}, and recover the known noiseless quantum code construction of independent spins-$1/2$ coupled to a dissipative cavity \cite{ZanardiRasetti1997}.

\emph{Unidirectional directed tunneling model.} 
A particularly interesting application of our framework is a hybrid open quantum system in which the cavity field is coupled to a directed tunneling matter operator defined on a one-dimensional chain with open boundary conditions.
We define the operator $\mathcal{O}$ explicitly in Eq.~\eqref{eq:operator}, Sec.~\ref{sec:results_tmps}, for both bosonic and fermionic degrees of freedom. 
By performing extensive analytical and numerical calculations, we show that all the expected consequences outlined in Fig.~\ref{fig:summary} are realized in this model, Sec.~\ref{sec:results_symmetries} to Sec.~\ref{sec:results_tmps}.

\emph{Underlying spin and pairing symmetries.} 
The central result that allows the generation of large decoherence-free subspace spanned by nontrivial quantum states based on the directed tunneling operator $\mathcal{O}$ is the fact that $\mathcal{O}$ is the lowering operator of a spin-like $\mathfrak{sl}(2,\mathbb{C})$ Lie algebra for both the fermionic and the bosonic case, Sec.~\ref{sec:construction} and Sec.~\ref{sec:symmetry_1}. 
Due to being part of the $\mathfrak{sl}(2,\mathbb{C})$ algebra, $\mathcal{O}$ has an exponentially large number of null states, which we explicitly derive by developing an analytical approach based on representation theory of Lie algebras, Sec.~\ref{sec:group_theory}.
Furthermore, $\mathcal{O}$ commutes with the Casimir operator of a second algebra corresponding to an emergent \emph{spinless $\eta$-pairing} mechanism. 
The spinless $\eta$-pairing algebra consists in the creation and annihilation of delocalized centro-symmetric pairs of particles. 
This is realized as a $\mathfrak{su}(1,1)$ Lie algebra for bosonic particles, and for fermions as a $\mathfrak{su}(2)$ Lie algebra, Sec.~\ref{sec:pairing_algebra}.
However, in contrast to the first spin-like algebra, the second pairing algebra generates a strong symmetry for the open hybrid system, with the number of spinless $\eta$-pairs being a conserved quantity.
Thus, the null states of $\mathcal{O}$ are characterized by two quantum numbers, one stemming from the spin algebra, which can be associated with the center-of-mass, Eqs.~\eqref{eq:sl2_algebra}-\eqref{eq:casimir_sl2}, and the number of $\eta$-pairs originating from the pairing algebras, Eqs.~\eqref{eq:pairing_bosons}-\eqref{eq:casimir_pairing_bosons} and Eqs.~\eqref{eq:pairing_SU2_fermions}-\eqref{eq:casimir_pairing_SU2}. 
However, these two quantum number have additional degeneracies beyond the two quantum numbers, as determined in Sec.~\ref{sec:group_theory_size}.
We discuss the properties of the null states, obtained with exact diagonalization, for different values of the quantum numbers in Sec.~\ref{sec:results_steady_states}.
For the full dissipative evolution, only the number of $\eta$-pairs is a conserved quantity, as the value of the center-of-mass changes throughout the dynamics,  Sec.~\ref{sec:results_tmps}.

\emph{Representation theory analytical approach.}
The fact that the directed tunneling operator $\mathcal{O}$ is part of the triplet $\mathfrak{sl}(2,\mathbb{C})$ Lie algebra implies that we can view each particle as a spin degree of freedom with its size set by the length of the chain and projection by the position of the particle. 
The bosonic, or fermionic, nature of the particles enters as the (anti)-symmetrization of the spin wavefunction. 
The algebraic structure of the problem allows the use of representation theory of Lie algebras for obtaining analytical results, Sec.~\ref{sec:construction} and Sec.~\ref{sec:group_theory}. 
Sec.~\ref{sec:group_theory} provides an analytical solution to the problem of finding the lowest weight states of the representation of the algebra.
The approach is based on the polynomial representation of the bosonic and fermionic states, Sec.~\ref{sec:group_theory_poly}, and leads to an explicit solution for the bosonic states in Eq.~\eqref{eq:q_r_decomposition}. 
Interestingly, there exists a correspondence between each fermionic and bosonic state, allowing to obtain also the fermionic null vectors, Sec.~\ref{sec:bosons_to_fermions}.

\emph{Emerging non-reciprocal dynamics.}
The decoherence-free subspaces that we construct are spanned by the eigenstates of the non-Hermitian operator $\mathcal{O}$. 
Thus, this represents a controlled way to generate and investigate nonreciprocal quantum effects. 
We consider $\mathcal{O}$ to be uni-directional directed tunneling, Eq.~\eqref{eq:operator}, so one might naively expect particle accumulation at the boundary, akin to the non-Hermitian and Liouvillian skin effects.

However, as the null vectors of $\mathcal{O}$ live in symmetry sectors labeled by quantum numbers related to the distribution of particles and pairs of particles, constraints on the accumulation profile arise.
Additionally, we can perform a similarity transformation to map our initial $\mathfrak{sl}(2,\mathbb{C})$ algebra, Eq.~\eqref{eq:sl2_algebra}, to a Hermitian representation of the same algebra, Eq.~\eqref{eq:sl2_Q_algebra}, in which the equivalent directed hopping operator has space-dependent coefficients, Sec.~\ref{sec:symmetry_2}. 
This similarity transformation has the physical effect of inducing a nonlinear potential on the density distribution of the null states, further influencing the profiles (sketch in Fig.~\ref{fig:summary}).
This allows us to introduce the \emph{symmetry-constrained Liouvillian skin effect}, which we discuss at the level of the null states for both bosons and fermions (Fig.~\ref{fig:hardcore_density} and Fig.~\ref{fig:bosons_density} in Sec.~\ref{sec:results_density}), and for the steady states of the dissipative evolution of the hybrid system, Sec.~\ref{sec:results_tmps}.

\emph{Engineering long-range quantum correlations.}
The symmetry structures alluded to in this summary and  described in detail further down, have a deep impact on the quantum properties of the states spanning the decoherence-free subspace: they exhibit nontrivial long-range correlations.
The resulting two main classes of correlations originate in the constraints imposed onto the quantum matter by the Casimirs of the two commuting algebras, Sec.~\ref{sec:results_corr}. 
First are kinetic correlations, where the tunneling events between two neighboring sites are correlated with the tunneling between two other neighboring sites at a distance.
Second, we find $\eta$-pairing correlations, whose nontrivial nature stems from the fact that particle pairs are delocalized and symmetric with respect to the center of the chain, and do not experience the accumulation towards the edge of the chain (sketch in Fig.~\ref{fig:summary}).
Furthermore, the entanglement profile scales with the system size for both  fermionic and bosonic cases (sketch in Fig.~\ref{fig:summary}), Sec.~\ref{sec:results_entanglement}.

All these non-trivial quantum properties persist for mixtures of the null states. 
We characterize the mixed states arising as the long-time state of the dissipative evolution when starting from generic initial states, Sec.~\ref{sec:results_tmps}. 
This we study via time-dependent matrix product states simulations of the dissipative dynamics of the hybrid system. 
These results offer a recipe for choosing simple Fock initial states to prepare correlated quantum states.

\section{Multiple steady states beyond Liouvillian symmetries \label{sec:recipe}}

In this section, we present the recipe for constructing large decoherence-free subspaces and the ingredients permitting control of their structure.
We consider hybrid open quantum systems consisting of matter coupled to a quantum dissipative bosonic mode. 
The matter can have a generic composition, bosonic or fermionic particles, or spin degrees of freedom, while the bosonic mode can model a photon or a phonon mode.
In the following, we refer to the bosonic mode as photon, or cavity, mode, as the example consider later is in particular suitable for realization with ultracold atoms in optical cavities \cite{RitschEsslinger2013, MivehvarRitsch2021}.
For now, we only require a global coupling between the matter and the photon mode and that the photons experience Markovian losses.
In this situation, the dynamics of the density matrix $\rho$ of the hybrid matter-cavity is given by a Gorini-Kossakowski-Sudarshan-Lindblad (GKSL) master equation  \cite{GoriniSudarshan1976, Lindblad_1976, CarmichaelBook, BreuerPetruccione2002}
\begin{align}
\label{eq:Lindblad}
\pdv{t} \rho =\mathcal{L}\rho&= -\frac{i}{\hbar} \left[H, \rho \right]+ \frac{\Gamma}{2}\left(2a\rho a^\dagger-a^\dagger a\rho-\rho a^\dagger a\right),\\
H&=\hbar\delta a^\dagger a -\hbar g \left(a^\dagger \mathcal{O} +a\mathcal{O}^\dagger\right), \nonumber
\end{align}
where $a$ is the annihilation operator of the photon mode of frequency $\delta$, the photons experience losses with a dissipation rate $\Gamma$. 
The matter-cavity coupling has a strength $g$, coupling the cavity operators to a matter operator $\mathcal{O}$ with global support in space.

Open hybrid systems, in which a bosonic field is coupled to quantum matter, are experimentally relevant for several platforms: ultracold atoms in optical cavities \cite{RitschEsslinger2013, MivehvarRitsch2021}; electrons in solid state cavities \cite{FornDiazSolano2019, SchlawinSentef2022}; superconducting circuits \cite{LEHURSchiro2016, ClerkNakamura2020, BlaisWallraff2021, PattonFischer2013}; or trapped ions coupled to phonon modes \cite{TomzaJulienne2019, MonroeYao2021}.
In particular, in the field of ultracold atoms, dissipative quantum light has been experimentally coupled to bosonic atoms \cite{BaumannEsslinger2010, KlinderHemmerich2015b, LandigEsslinger2016, VaidyaLev2018, KesslerHemmerich2021, FerriEsslinger2021, YoungThompson2024}, fermionic atoms \cite{RouxBrantut2020, ZhangWu2021, HelsonBrantut2023,  WuWu2023, ZwettlerBrantut2025, BuhlerBrantut2026} and spin degrees of freedom \cite{MassonParkins2017, MassonBarrett2014, KroezeLev2018, LandiniEsslinger2018, SauerweinBrantut2023}.

\subsection{Steady states \label{sec:recipe_1}}

We aim to obtain multiple steady states, $\mathcal{L}\rho_\text{st}=0$, of the dissipative dynamics, which will span the decoherence-free subspace.
Previously, the main approach to construct multiple steady states in open quantum systems made use of the concept of strong symmetries \cite{AlbertJiang2014, BucaProsen2012}. 
For a strong symmetry the symmetry generator commutes with both the Hamiltonian $H$ and the jump operators characterizing the dissipative processes. 
In this case, each symmetry sector of the Liouvillian $\mathcal{L}$ contains at least one steady state.
However, as there is no upper bound on the number of steady states in a single symmetry sector, the existence of multiple steady states is not necessarily related to the presence of strong symmetries \cite{AlbertJiang2014, BucaProsen2012, SeltmannBuca2026}.

We employ the hybrid matter-cavity system described in Eq.~\eqref{eq:Lindblad} to study the interplay between the size of the decoherence-free subspace and the presence of strong symmetry. 
We can categorize the different situations which arise depending on the value of the commutator $[\mathcal{O},\mathcal{O}^\dagger]$ (we sketch the classification in Fig.~\ref{fig:DFS}):
\begin{enumerate}
 \item $[\mathcal{O},\mathcal{O}^\dagger]=0$, which includes the case in which $\mathcal{O}$ is a Hermitian operator. 
    In this situation, $\mathcal{O}$ commutes with both the Hamiltonian $H$ of the system, Eq.~\eqref{eq:Lindblad}, and with the jump operator $a$, thus, a strong symmetry is present in the system  \cite{AlbertJiang2014, BucaProsen2012}. 
    The different symmetry sectors are spanned by the eigenstates of $\mathcal{O}$, representing a complete basis in the matter Hilbert space. 
    Each symmetry sector contains one matter state, as sketched in Fig.~\ref{fig:DFS}(a). 
    The decoherence-free subspace is spanned by the steady states [red states in Fig.~\ref{fig:DFS}(a)], $\mathcal{L}\rho_\text{st}=0$, given by
    \begin{align} 
    \label{eq:rho_symmetry}
    & \rho^{(i)}_\text{st}=\ket{\alpha; o_i}\bra{\alpha; \tilde{o}_i},
    \end{align}
    where $\ket{o_i}$ and $\ket{\tilde{o}_i}$ are degenerate eigenstates of $\mathcal{O}$ corresponding to the same eigenvalue $o_i$ and the photons are in a coherent state $\ket{\alpha}$, with $\alpha=\frac{g o_i}{\delta-i\Gamma/2}$. 
    The states which have a finite decay rate are the traceless coherences, $\ket{\alpha; o_i}\bra{\alpha; o_j}$, for which $o_i\neq o_j$ \cite{HalatiKollath2025}.
    The implications of the presence of strong symmetry sectors in hybrid systems has been discussed when $\mathcal{O}$ is, e.g.,~the sum of local density terms \cite{HalatiKollath2020b, HalatiKollath2022, TindallMunoz2023, HalatiKollath2025}.
\item $[\mathcal{O},\mathcal{O}^\dagger]\neq 0$ and the identity    operator of the matter Hilbert space, $\mathbb{I}_\text{mat}$, is the only operator which satisfies $[\mathcal{O},\mathbb{I}_\text{mat}]=[\mathcal{O}^\dagger,\mathbb{I}_\text{mat}]= 0$. 
    In this case, no strong symmetries exist and the decoherence-free subspace is spanned by the null states of $\mathcal{O}$:
    \begin{align} 
    \label{eq:rho}
    & \rho^{(i,j)}_\text{st}=\ket{\alpha=0; \mathbf{0}_i}\bra{\alpha=0; \mathbf{0}_j},
    \end{align} 
    where $\ket{\alpha=0}$ is the vacuum of the photon field and $\ket{\mathbf{0}_i}$ is a dark, i.e.~null, state of the matter operator, $\mathcal{O}\ket{\mathbf{0}_i}=0$.
    One can see that $\rho^{(i,j)}_\text{st}$ are steady states from the fact that they commute with the Hamiltonian and are annihilated by the cavity decay processes.
    In this case, sketched in Fig.~\ref{fig:DFS}(b), only one symmetry sector is present, which contains a decoherence-free subspace constructed from the null states of $\mathcal{O}$.
    We note that if $\mathcal{O}$ does not have any null vectors, the hybrid system will have a non-equilibrium steady state with a finite cavity population.
\item $[\mathcal{O},\mathcal{O}^\dagger]\neq 0$ and there exists at least one operator acting on the matter Hilbert space, other than $\mathbb{I}_\text{mat}$, which satisfies $[\mathcal{O},\mathcal{S}_\text{mat}]=[\mathcal{O}^\dagger,\mathcal{S}_\text{mat}] =0$. In this case, sketched in Fig.~\ref{fig:DFS}(c), the strong symmetry generated by $\mathcal{S}_\text{mat}$ gives rise to a structure of symmetry sectors.
    If the quantum numbers of the symmetry do not uniquely identify every null state of $\mathcal{O}$, symmetry sectors containing a decoherence-free subspace spanned by multiple states are present.
    In the case the only shared eigenvectors between $\mathcal{O}$ and  $\mathcal{O}^\dagger$ are null states, then the decoherence-free subspaces only contain the states in Eq.~\eqref{eq:rho}, otherwise also states from Eq.~\eqref{eq:rho_symmetry} are steady states.
\end{enumerate}

For the rest of this work, we consider the third case, $[\mathcal{O},\mathcal{O}^\dagger]\neq 0$.
We show that even if the presence of the strong symmetry $\mathcal{S}_\text{mat}$ only generates a few symmetry sectors, large decoherence-free subspaces can be constructed in each of these sectors.
This situation arises when $\mathcal{O}$ has a high dimensional null space. 
We show how to construct such an operator in Sec.~\ref{sec:construction}. 
This leads to multiple steady states, $\rho^{(i,i)}_\text{st}$, and coherences, $\rho^{(i\neq j)}_\text{st}$, for the hybrid system.
As the steady states are generated by the null states of $\mathcal{O}$, we can think of the approach presented as a global constraint that dynamically emerges due to the coupling to the dissipative cavity mode. 

$\mathcal{O}$ is a non-Hermitian operator, thus, its symmetries are not necessarily symmetries of $\mathcal{O}^\dagger$, and by extension of the Hamiltonian $H$, Eq.~\eqref{eq:Lindblad}. 
This implies that the steady states $\rho^{(i,j)}_\text{st}$ spanning the decoherence-free subspace are not necessarily generated by a strong symmetry.
This construction offers an intriguing opportunity, the decoherence-free subspace generated from the null states of a many-body operator $\mathcal{O}$ is characterized by symmetries that only emerge at late times, having drastic consequences on the quantum dissipative dynamics.
In particular, this offers a controlled approach to impose nonreciprocal dynamics onto the matter degrees of freedom, stemming from the non-Hermitian character of $\mathcal{O}$.

The matter states obtained by tracing the photonic degrees of freedom of the steady states given in Eq.~\eqref{eq:rho}, $\Tr_\text{photon}[\rho^{(i,j)}_\text{st}]=\ket{o_i}\bra{o_j}$, are also steady states of the following matter-only Liouvillian
\begin{align}
\label{eq:Lindblad_matter}
\tilde{\mathcal{L}}_m\rho_m&= -\frac{i}{\hbar} \left[-\frac{\hbar\delta g^2}{\delta^2+\Gamma^2/4}\mathcal{O}^\dagger\mathcal{O}, \rho_m \right] \\
&+ \frac{\Gamma g^2}{2(\delta^2+\Gamma^2/4)}\left(2\mathcal{O}\rho_m \mathcal{O}^\dagger-\mathcal{O}^\dagger \mathcal{O}\rho_m-\rho_m \mathcal{O}^\dagger \mathcal{O}\right). \nonumber
\end{align}
The Liouvillian $\tilde{\mathcal{L}}_m$ can be obtain as an effective matter-only model of the hybrid model, Eq.~\eqref{eq:Lindblad}, \cite{MivehvarRitsch2021}. 
However, for $[\mathcal{O},\mathcal{O}^\dagger]\neq 0$, while the steady states exactly agree, the dynamics is captured only for $g\ll|\delta+i\Gamma/2|$ \cite{JagerBetzholz2022, SchmitJager2026}.
The matter-only effective model allows us to make an interesting connection. The operator $\mathcal{O}^\dagger\mathcal{O}$ is a semi-positive definite operator, thus, the null states of $\mathcal{O}$ are the ground states of the long-range interacting many-body Hermitian operator $\mathcal{O}^\dagger\mathcal{O}$.
Our construction leads to a decoherence-free subspace spanned by the zero-mode ground states of globally interacting Hamiltonians which can be factorized as $\mathcal{O}^\dagger\mathcal{O}$.

\subsection{Conserved quantities \label{sec:recipe_2}}

For every steady state the Liouvillian also posses a conserved quantity, defined as the steady solution of the adjoint Liouvillian \cite{AlbertJiang2014}
\begin{align} 
\label{eq:conserved_charges}
& \mathcal{L}^\dagger \mathcal{C}^{(i,j)}=0,
\end{align}
corresponding to the steady state $\rho^{(i,j)}_\text{st}$ defined in Eq.~\eqref{eq:rho}. 
The correspondence is defined such that for a generic initial state, the overlap with $\mathcal{C}^{(i,j)}$ determines the weight of $\rho^{(i,j)}_\text{st}$ in the steady state mixture  \cite{CarmichaelBook, BreuerPetruccione2002}.  
As discussed in the previous section, we are considering a situation in which the multiple steady states are not fully determined by the presence of a strong symmetry \cite{AlbertJiang2014, BucaProsen2012, SeltmannBuca2026}. 
In this case the conserved quantities are not required to simultaneously commute with the Hamiltonian and the jump operator, $[\mathcal{C}^{(i,j)},H]\neq 0,~[\mathcal{C}^{(i,j)},a]\neq 0$.
The adjoint Liouvillian is defined as
\begin{align}
\label{eq:Lindblad_adjoint}
\pdv{t}\mathcal{C}=\mathcal{L}^\dagger\mathcal{C}&= \frac{i}{\hbar} \left[H, \mathcal{C} \right]+ \frac{\Gamma}{2}\left(2a^\dagger\mathcal{C} a-a^\dagger a\mathcal{C}-\mathcal{C} a^\dagger a\right),
\end{align}
the main differences compared to the Liouvillian defined in Eq.~\eqref{eq:Lindblad} are the sign of the coherent contribution and the exchange of $a$ and $a^\dagger$ in the quantum jump term.

We present how to construct a general conserved quantity for the hybrid system, Eq.~\eqref{eq:Lindblad}, in \ref{app:conserved}.
The conserved quantities have the form
\begin{align}
\label{eq:conserved_C_final}
&\mathcal{C}^{(i,j)}=\mathbb{I}_\text{pho}\otimes \ket{\mathbf{0}_i}\bra{\mathbf{0}_j}\\
&\qquad-\sum_{m>0} \sum_{k=1}^m  \frac{i g}{(k-1)!}\left(\mathcal{K}^{-1}\mathcal{J}\right)^{m-k} \mathcal{K}^{-1} \nonumber\\ &\qquad\qquad\left[\left(a^\dagger\right)^{k-1} \mathcal{O}^\dagger  \rho^{(i,j)}_\text{st} \left(a\right)^k -\left(a^\dagger\right)^{k}  \rho^{(i,j)}_\text{st}\mathcal{O} \left(a\right)^{k-1}\right], \nonumber
\end{align}
where $\mathcal{K}\mathcal{C}=-i\left(\delta+i\frac{\Gamma}{2}\right)a^\dagger a\mathcal{C}+i\left(\delta-i\frac{\Gamma}{2}\right)\mathcal{C}a^\dagger a+ig\left[a^\dagger \mathcal{O} +a\mathcal{O}^\dagger,\mathcal{C}\right]$ and $\mathcal{J} \mathcal{C} = \Gamma a^\dagger\mathcal{C} a$.
From this expression we can see that $\mathcal{C}^{(i,j)}$ contains a term resulting from the tensor product of an identity over the photonic Hilbert space and the matter part of the states, given by the null states of $\mathcal{O}$. 
The second term in Eq.~\eqref{eq:conserved_C_final}, exchanges excitations between the photonic and matter degrees of freedom.
In principle, this can connect all the matter states that are accessible with successive applications of $\mathcal{O}^\dagger$ and $\mathcal{O}$ on the steady states (see \ref{app:conserved}).
The expression from Eq.~\eqref{eq:conserved_C_final} has the potential to produce nonlocal and multibody operators as the conserved quantities of the dissipative dynamics. 
This is in stark contrast with the conserved quantities that are usually considered in the case of strong symmetries which are few-body operators, e.g.~in Refs.~\cite{AlbertJiang2014, MunozPorras2019, LieuGorshkov2020, DuttaCooper2020, HalatiKollath2022, TindallMunoz2023, HalatiKollath2025}.
In Sec.~\ref{sec:results_conserved} we compute the conserved quantities for the chosen example system.

Some simple expressions for Eq.~\eqref{eq:conserved_C_final} can be found by looking for solutions of the form $\mathcal{C}=\mathbb{I}_\text{pho}\otimes C_\text{mat}$ for Eq.~\eqref{eq:Lindblad_adjoint}. 
The Ansatz represents a conserved quantity only if $[C_\text{mat}, \mathcal{O}]=[C_\text{mat}, \mathcal{O}^\dagger]=0$. 
This implies that $C_\text{mat}$ would also commute with the Hamiltonian, Eq.~\eqref{eq:Lindblad}, and generates a strong symmetry.
The Ansatz contains also the trivial solution when $ C_\text{mat}=\mathbb{I}_\text{mat}$, as the identity is always a conserved quantity of the GKSL master equation assuring trace conservation.
Another possibility arises when $\mathcal{O}$ and $\mathcal{O}^\dagger$ share a subset of eigenvectors, for example if $\ket{\tilde{\mathbf{0}}_i}$ are shared null states, then we can have $C_\text{mat}=\ket{\tilde{\mathbf{0}}_i}\bra{ \tilde{\mathbf{0}}_j}$. 
In this situation, $\mathbb{I}_\text{pho}\otimes \ket{\tilde{\mathbf{0}}_i}\bra{ \tilde{\mathbf{0}}_j}$ are conserved quantities stemming from a strong symmetry, see related discussion in Sec.~\ref{sec:recipe_1}.

\section{Constructing operators with large null spaces \label{sec:construction}}

In this section, we explain in detail how to generically construct operators characterized by a large null space. 
For the hybrid model of Sec.~\ref{sec:recipe_1}, Eq.~\eqref{eq:Lindblad} with non-Hermitian $\mathcal{O}$, the photon losses effectively give rise to a dissipative channel that removes matter ``excitations'' until the null states are reached, Eq.~\eqref{eq:rho}.
One way to describe physical processes that decrease the number of matter excitations is with a ``lowering operator''.
Representation theory of Lie algebras is a natural tool to systematically construct such an operator. 
We require that $\mathcal{O}$ has at least one null state for a single particle and that the size of the null space increases with the particle number.
The most general formulation of a lowering operator is the sum of single-particle root vectors of a representation of a Lie algebra, Eq.~\eqref{eq:operator_root} below.
For the reader unfamiliar with representation theory, $\mathcal{O}$ can be thought of as, for all practical purposes, the generalization of the collective decay operator of spin angular momentum, $\sum\sigma^-_n$.

We outline the argument underlying the recipe in Sec.~\ref{sec:recipe_general} and briefly discuss the general case. We then work out a first example in which the physical system is a collection of spins-$1/2$ undergoing collective decay in Sec.~\ref{sec:ex_spin12}. This allows us to introduce the representation theory concepts that we employ in Sec.~\ref{sec:results_symmetries} and Sec.~\ref{sec:group_theory}, and to show how our approach recovers previous results from the literature.
We conclude with Sec.~\ref{sec:itinerant}, showing how to construct models of itinerant particles in any dimension, including the directed tunneling model we discuss in detail starting with Sec.~\ref{sec:results_symmetries}.
While the analytical arguments from the following subsections underline the generality of our approach they are not necessary to understand physical properties of the null states of the directed tunneling model (Sec.~\ref{sec:results_symmetries} and Sec.~\ref{sec:results_steady_states}).

\begin{figure}
\centering
\includegraphics[width=0.47\textwidth]{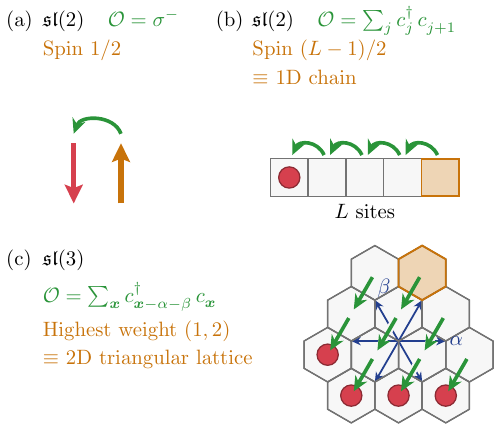}
\caption{
Different physical system one may consider depending on the representation of the Lie algebra  ${\cal O}$ belongs to. We represent here only the single-particle sector which has a simple null space structure. A complex structure with entanglement and quantum correlations emerges in the many-body case, see e.g.~Fig.~\ref{fig:up_and_down_spin_half} and Fig.~\ref{fig:example_nullstates}.  
(a) The spin $1/2$ irreducible representation of $\mathfrak{sl}(2)$, discussed Sec.~\ref{sec:ex_spin12}. 
Although the null eigenspace is trivial for a single-particle, the $\ket{-}$ state, the many-body null space for $N$ spins grows exponentially and has a rich structure, Fig.~\ref{fig:up_and_down_spin_half}. 
(b) Going to higher weight, i.e.~higher spin representation of $\mathfrak{sl}(2)$, amounts to increasing the system size. 
We go from a two-state system to a lattice of size $L$ where the spin is given by $J=\frac{L-1}{2}$.
The lowering operator is $\sum_{j}c^{\dagger}_{j}c_{j+1}$ and the single-particle null eigenvector corresponds to the particle at the left-most site. 
(c) For $\mathfrak{sl}(3)$ the physical space becomes two-dimensional. We represent in blue the root system of $\mathfrak{sl}(3)$ generating the triangular lattice where $\alpha$ and $\beta$ denote the simple roots. The root vector defining $\mathcal{O}$ can be obtained as the translation operator in a given direction. The null states are then located in hyperplanes orthogonal to this root vector and are represented in red. 
\label{fig:Different-cases}}
\end{figure}

\subsection{General case\label{sec:recipe_general}}

In order to define the equivalent of a lowering operator for a generic algebra, we consider the single particle Hilbert space to be associated with a complex representation of a semisimple Lie algebra.
For $N$ particles, we then take $\mathcal{O}$ to be the sum of single-particle root vectors of this representation,
\begin{align}
\label{eq:operator_root}
\mathcal{O}:=\sum_{n=1}^NE^{n}_{\alpha},
\end{align}
where $n$ indexes different replicas of this representation and $\alpha$ is the corresponding root 
\footnote{More generally, one could consider a weight vector instead of a root vector, but for simplicity we restrict to the latter case here.}. 
Our goal is to find the null eigenspace of the many-body operator $\mathcal{O}$ of Eq.~\eqref{eq:operator_root}.
The construction applies to any semisimple Lie algebra~\cite{Humphreys1972, Georgi1999}.
For concreteness we phrase it for $\mathfrak{sl}(m,\mathbb{C})$, the Lie algebra of traceless complex $m\times m$ matrices, with $m\geq 2$ an integer.
This covers all the examples of this work: $m=2$ for the spin-$\frac{1}{2}$ example of Sec.~\ref{sec:ex_spin12} and the directed tunneling model, and $m=3$ for the triangular lattice of Fig.~\ref{fig:Different-cases}(c).
Each of the $N$ particles carries a copy of an irreducible representation $D_n$, so that the Hilbert space is $D^{\otimes N}_{n}$, and $\mathcal{O}$ is the sum over the $N$ replicas of a single root vector $E_\alpha$, Eq.~\eqref{eq:operator_root}.

Consider first a single particle. The triplet $\{H_\alpha, E_\alpha, E_{-\alpha}\}$ forms an $\mathfrak{sl}(2)$ subalgebra, with $H_\alpha:=[E_\alpha,E_{-\alpha}]$ the corresponding element of the Cartan subalgebra. The three operators play the role of $\{J_z, J_-,J_+\}$ in the theory of angular momentum.
We label the basis states of $D_n$ by their weights, i.e.\ their eigenvalues under the commuting generators of the Cartan subalgebra.
These weights form the weight diagram of the representation: the values of the magnetic quantum number on a line for $\mathfrak{sl}(2)$, or the points of a triangular lattice for $\mathfrak{sl}(3)$, see Fig.~\ref{fig:Different-cases}.
The root vector $E_\alpha$ maps a state of weight $\mu$ onto a state of weight $\mu+\alpha$, i.e.\ it translates the weight diagram by the vector $\alpha$.
Under the $\mathfrak{sl}(2)$ subalgebra, $D_n$ thus splits into ordinary spin multiplets, each formed by a string of weights $\mu, \mu+\alpha, \mu+2\alpha,\dots$ along a line parallel to $\alpha$.
Just as $J_-$ annihilates only the bottom rung $\ket{j,-j}$ of a spin-$j$ ladder, $E_\alpha$ annihilates exactly one state per string, the one at its end in the direction $\alpha$. 
The dimension of the null space is the number of strings.
For the representations of Fig.~\ref{fig:Different-cases}, where each weight occurs once, these are the sites with no neighbor in the direction $\alpha$ (red in Fig.~\ref{fig:Different-cases}) \footnote{When a weight occurs with multiplicity larger than one, several strings pass through it and some strings may end in the interior of the weight diagram; the counting ``one null vector per string'' remains valid.}.
The single-particle space therefore need not be a single ladder, its geometry is set by the weight diagram of $D_n$, on which $\mathcal{O}$ acts as a directed hopping operator.

We now turn to many particles. The sum over replicas in Eq.~\eqref{eq:operator_root} is precisely how a generator acts on a tensor product, as for the total spin $J_- = \sum_n \sigma_n^-$.
Hence $\mathcal{O}$ represents $E_\alpha$ on $D_n^{\otimes N}$, and the single-particle argument applies to every irreducible component of $D_n^{\otimes N}$.
Since $D_n^{\otimes N}$ is in general reducible, we decompose it into irreducible representations. 
In the adapted basis $\mathcal{O}$ is block diagonal, and its null space is the direct sum of the null spaces of the blocks.
This change of basis is in general complicated, as already illustrated by Clebsch--Gordan coefficients for spins-$1/2$ in Sec.~\ref{sec:ex_spin12}.

The decomposition proceeds in two stages.
First, permutations of the particles commute with $\mathcal{O}$, and by Schur--Weyl duality, $D^{\otimes N}_{n}$ splits into sectors labeled by the irreducible representations of the symmetric group $S_{N}$, i.e.\ by Young diagrams with $N$ boxes~\cite{FultonHarris, Sagan2001}.
For instance, for $N=4$,
\begin{equation}
D^{\otimes4}_{n}=\begin{bmatrix}\, & \, & \, & \,\end{bmatrix}\oplus3\,\begin{bmatrix}\, & \, & \,\\
\,
\end{bmatrix}\oplus2\,\begin{bmatrix}\, & \,\\
\, & \,
\end{bmatrix}\oplus3\,\begin{bmatrix}\, & \,\\
\,\\
\,
\end{bmatrix}\oplus\begin{bmatrix}\,\\
\,\\
\,\\
\,
\end{bmatrix}~,
\label{eq:Young_tableaux}
\end{equation}
where diagrams with more rows than the dimension of $D_n$ are absent.
Each prefactor is the multiplicity of the corresponding representation, equal to the number of standard Young tableaux of that shape (ways to fill the diagram with $1,\dots,N$ increasing along rows and columns), given by the hook length formula, \cite{zuber_group_theory, Wu_Ki_tung}.
The Young diagram encodes the exchange symmetry: the single row corresponds to bosonic and the single column to fermionic states, the two cases relevant for the directed tunneling model of Sec.~\ref{sec:results_symmetries}.
Second, when $D_{n}$ is the fundamental representation, each Young diagram is an irreducible representation of $\mathfrak{sl}(m,\mathbb{C})$. Otherwise, it is in general reducible and must be decomposed further.
For example, the totally symmetric states of $N$ spins $j>\frac{1}{2}$ contain several values of the total spin, see Sec.~\ref{sec:results_symmetries}.

The dimension of the null space of $\mathcal{O}$ is then the number of bottom rungs summed over all irreducible blocks, weighted by their multiplicities.
Since only the triplet $\{H_\alpha, E_\alpha, E_{-\alpha}\}$ enters, it suffices in practice to decompose $D^{\otimes N}_{n}$ into multiplets of the collective $\mathfrak{sl}(2)$ subalgebra, the route followed in Sec.~\ref{sec:ex_spin12} and~Sec.~\ref{sec:results_symmetries}.
As the multiplicities grow rapidly with $N$, the null space is generically exponentially large in the number of particles.

\subsection{Example: Spin-$1/2$ \label{sec:ex_spin12}} 

In order to exemplify the type of operators captured by Eq.~\eqref{eq:operator_root}, we solve first for the simplest case. This is realized when $D_{n}$ is the fundamental representation of $\mathfrak{sl}(2,\mathbb{C})$, i.e.~a single spin-1/2. 
The single-particle Hilbert space has basis $\{|+\rangle,|-\rangle\}$. 
We choose $\mathcal{O}=\sum_{n}\sigma_{n}^{-}=:J_{-}$, the collective spin-lowering operator. 
Such types of Dicke collective operators were originally considered in the context of superradiance \cite{Dicke1954, KirtonDallaTorre2019}. 
Its null space is spanning the decoherence-free subspace for a collection of $N$ two-level systems undergoing collective decay, extensively studied in the quantum-information literature~\cite{ZanardiRasetti1997, DuanGuo1997, LidarWhaley1998, KempeBaconLidarWhaley2001} and in dissipative engineering approaches using subradiant states \cite{ShankarHolland2021, ReillyHolland2022, VaecairnHolland2025, ChuClerk2026}.
We derive the structure of the decoherence-free subspace from the Lie-algebraic construction above.

For $N=1$, the single-particle sector, the null eigenspace is simply spanned by $|-\rangle$, sketched in Fig.~\ref{fig:Different-cases}.
For $N=2$, the space decomposes as 
\[
D^{\otimes2}_{n}=\begin{bmatrix}\, & \,\end{bmatrix}\oplus\begin{bmatrix}\,\\
\,
\end{bmatrix}~,
\]
the triplet and singlet representations, respectively. 
The triplet
representation has spin~$1$ and is spanned by the basis $\big\{|+,+\rangle,\;|\mathcal{S}\rangle:=\tfrac{1}{\sqrt{2}}(|+,-\rangle+|-,+\rangle),\;|-,-\rangle\big\}$.
Its lowest-weight vector is $|-,-\rangle$, which is a null eigenvector of $\mathcal{O}$. 
The singlet representation has spin~$0$ and consists of the single vector $|\mathcal{A}\rangle:=\tfrac{1}{\sqrt{2}}(|+,-\rangle-|-,+\rangle)$, for which $\mathcal{O}|\mathcal{A}\rangle=0$. 
Interestingly, in this case the annihilation by $\mathcal{O}$ arises from a many-body interference effect -- a phenomenon we will also encounter in more complicated examples presented in Sec.~\ref{sec:itinerant} and Sec.~\ref{sec:results_symmetries}.

\emph{Which Young diagrams survive}.
Since for this example the single-particle representation is two-dimensional, no Young diagram with more than two rows can appear in the decomposition of $D^{\otimes N}_{n}$. 
Any diagram with three or more rows would require antisymmetrizing over at least two identical single-particle states, forcing the result to vanish, as a result of the Pauli exclusion principle.
The surviving shapes are therefore exactly the two-row partitions
of $N$: $\,\begin{bmatrix}\, & \, & \,\\
\, & \,
\end{bmatrix}\cdots$ (a one-row shape being the special case of an empty second row).
The decomposition takes the closed form 
\[
D^{\otimes N}_{n}=\bigoplus^{\lfloor N/2\rfloor}_{k=0}d_{N,J}\,\big[(N-k,k)\big],\qquad J=\frac{N}{2}-k,
\]
where $[(N-k,k)]$ denotes the $\mathfrak{sl}(2,\mathbb{C})$ spin-$J$ multiplet built from the two-row Young diagram with $N-k$ boxes in the first row and $k$ boxes in the second. 
From the Hook-length formula \cite{zuber_group_theory}, we get that each such diagram appears with degeneracy 
\[
d_{N,J}=\binom{N}{k}-\binom{N}{k-1}.
\]
Each of the $d_{N,J}$ copies of the spin-$J$ multiplet contributes exactly one null vector of $\mathcal{O}$,  the lowest-weight state $|J,-J\rangle$.
The dimension of the null subspace is 
\[
\dim\ker\mathcal{O}=\sum^{\lfloor N/2\rfloor}_{k=0}d_{N,J}=\binom{N}{\lfloor N/2\rfloor}.
\]
The central binomial coefficient grows as $2^{N}/\sqrt{N}$ by Stirling's approximation, i.e.\ super-extensively in $N$.
Thus, an exponentially large decoherence-free subspace is obtained from the null states of the collective spin operator. 

\begin{figure}
\begin{centering}
\includegraphics[width=0.95\columnwidth]{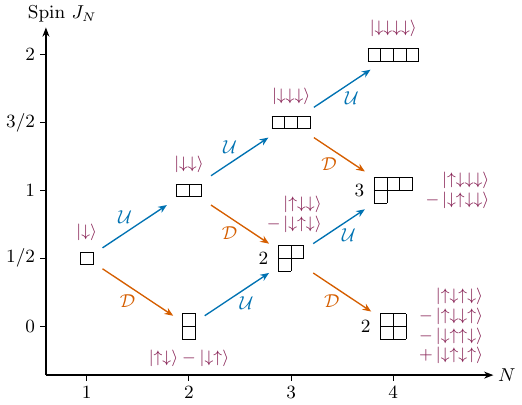}\caption{
\label{fig:up_and_down_spin_half}
Null eigenvectors for the case of the spin $1/2$. Under the condition that the total spin can not be negative, the $\mathcal{U}$, Eq.~\eqref{eq:Up_move}, and $\mathcal{D}$, Eq.~\eqref{eq:Down_move}, operations respectively increase or decrease the spin of the representation by $\frac{1}{2}$ and 
promote a null eigenvector from the $N$ to the $N+1$ particles sector. 
In terms of Young diagrams index, the $\mathcal{U}$ adds a box in the row while $\mathcal{D}$ adds a box in the column. 
The fact that the spin can not be smaller than 0 is reflected by the fact that the diagrams can not have more than 3 rows. 
The degeneracy of each representation, labeled by a number in front of the diagrams, corresponds to the different paths one can take to reach the configuration. 
We also show for each representation a sample of a null eigenvector.
For $N=4$, we see the appearance of states with extensive entanglement and long-range correlations. 
As explained in the main text, the dimension of the null eigenspace grows exponentially with $N$.
}
\par\end{centering}
\end{figure}

\emph{General-$N$ construction via sequential coupling}.
In the following, we describe a systematic recursive construction that produces, for any number of spins $N$, an explicit and automatically orthonormal basis for the null space of $\mathcal{O}$.
The idea is to build up the states of a certain total spin by adding successively one particle at a time, while tracking only the lowest-weight state at each step, sketched in Fig.~\ref{fig:up_and_down_spin_half}. 
The total spin reached after $N$ particles is $J_{N}$, with $J_{1}=\tfrac{1}{2}$. 
At each step, $J_{N+1}=J_{N}\pm\tfrac{1}{2}$, and the corresponding lowest-weight state is generated by one of two elementary recursions. 
The \emph{up-move} $\mathcal{U}$, for which $J_{N+1}=J_{N}+\tfrac{1}{2}$, (see Fig.~\ref{fig:up_and_down_spin_half}) is realized by trivially adding a spin down, 
\begin{equation}
|J_{N+1},-J_{N+1}\rangle=|J_{N},-J_{N}\rangle\otimes|-\rangle_{N+1},   
\label{eq:Up_move}
\end{equation}
since the extremal weight of the larger multiplet is reached in only one way. 
The \emph{down-move} $\mathcal{D}$ (see Fig.~\ref{fig:up_and_down_spin_half}), for which $J_{N+1}=J_{N}-\tfrac{1}{2}$, requires a genuine recoupling, 
\begin{align}
|J_{N+1}&,-J_{N+1}\rangle=-\sqrt{\frac{2J_{N+1}+1}{2J_{N+1}+2}}\,|J_{N},-J_{N}\rangle\otimes|+\rangle_{N+1} \nonumber\\
&+\sqrt{\frac{1}{2J_{N+1}+2}}\,|J_{N},-J_{N}+1\rangle\otimes|-\rangle_{N+1},
\label{eq:Down_move}
\end{align}
where $|J_{N},-J_{N}+1\rangle$, the next-to-lowest state of the previous multiplet, is obtained from $|J_{N},-J_{N}\rangle$ by one application of the collective raising operator $J_{+}=\sum_{N}\sigma^{N}_{+}$.
The normalization of this procedure is given by $J_{\pm}|J,M\rangle=\sqrt{J(J+1)-M(M\pm1)}\,|J,M\pm1\rangle$ 
\footnote{Both formulas are the ordinary Clebsch--Gordan coefficients for coupling spin $J_{N}$ with a single spin-$\tfrac{1}{2}$; they follow directly from diagonalizing $\vec{S}_{1}\cdot\vec{S}_{2}$ in the two-dimensional degenerate-$M$ subspace spanned by $|J_{N},M\mp\tfrac{1}{2}\rangle\otimes|\pm\rangle_{N+1}$, see e.g.~\cite{Sakurai1994,Varshalovich1988}.}. 
Because each step is an ordinary unitary recoupling, the resulting
states are automatically mutually orthonormal across the different
admissible sequences of moves \cite{FultonHarris, Sagan2001}.

Not every sequence of $\mathcal{U}$ or $\mathcal{D}$ moves is admissible. A down-move requires $J_{N}>0$, so only sequences whose partial sums never go negative are allowed. 
The number of admissible sequences ending at spin $J=N/2-k$ after $N$ steps is exactly $d_{N,J}=\binom{N}{k}-\binom{N}{k-1}$,
by the classical ballot-problem/reflection-principle count. This reproduces multiplicities above by an independent combinatorial argument, and confirming $\dim\ker\mathcal{O}=\binom{N}{\lfloor N/2\rfloor}$.
This construction is shown in Fig.~\ref{fig:up_and_down_spin_half}, up to $N=4$ spins.

\subsection{Obtaining models of itinerant particles\label{sec:itinerant}}

In the following, we describe how by considering different representations for the single-particle sector in Eq.~\eqref{eq:operator_root} one can arrive at different geometries. In particular we show how to arrive at the model whose properties we describe in detail starting with Sec.~\ref{sec:results_symmetries}.

\emph{One-dimensional lattice}.
We can consider a spin-$j$ representation of $\mathfrak{sl}(2,\mathbb{C})$ instead of spin-$1/2$. In this case the single-particle states live on a one-dimensional lattice whose sites are labeled by the projection (magnetic) quantum number, with lattice size $2j+1$, see Fig.~\ref{fig:Different-cases}(b).
Thus, the spin lowering operator can be written as a directed tunneling operator defined in the tight-binding model of the one-dimensional chain, $\mathcal{O}=\sum d^\dagger_j d_{j+1}$.
As sketched in Fig.~\ref{fig:Different-cases}(b) the single particle null state is given by the particle being localized on the first site of the chain.
This is the case we study in detail in the following sections,
since it is the natural generalization of the spin-$1/2$ case and exhibits a great deal of nontrivial structure leading dramatic physical consequences.

\emph{Two-dimensional lattices}.
Changing representation also allows one to change the physical dimension of the matter space. 
For instance, one can consider $\mathfrak{sl}(3,\mathbb{C})$, sketch in Fig.~\ref{fig:Different-cases}(c).
The weights of any representation populate a two-dimensional triangular lattice, generated by the two simple roots $\alpha$ and
$\beta$, Fig.~\ref{fig:Different-cases}(c).
This allows for the extension of the one-dimensional chain to a two-dimensional model.
In general, the lattice geometry for any Lie algebra is classified by Cartan's classification of semisimple Lie algebras, which identifies four infinite families and five exceptional algebras~\cite{FultonHarris,Humphreys1972}.

\emph{Generalization}. 
In principle, the construction above is not constrained to the case where the operator $\mathcal{O}$ is only a single root vector. 
For instance, one can take a combination of roots, $\mathcal{O}=\sum_{n,i}a_{i}E^{n}_{\alpha_{i}}$. 
For $\sum_{i}a_{i}E^{n}_{\alpha_{i}}$ to be guaranteed at least one null eigenvector for arbitrary coefficients $a_{i}$, the roots $\alpha_{i}$ must all be negative roots. This is equivalent to all $E_{\alpha_{i}}$ must being lowering operators. 
The lowest-weight vector of the representation is then automatically a null eigenvector of $\mathcal{O}$, since it is annihilated by every negative root vector individually. 
This allows one further control in designing the structure of the null eigenspace. 
For instance, in the two-dimensional triangular lattice, Fig.~\ref{fig:Different-cases}(c), we would no longer have a set of independent one-dimensional chains of various lengths, but something that genuinely reflects the two-dimensional structure of the problem, Fig.~\ref{fig:Different-cases}(c).
We expect that such a setup would give rise to exotic two-dimensional null states populating the large decoherence-free subspace.
We leave the study of these more general and higher dimensional structures to future works. 

\section{Directed tunneling model  \label{sec:results_symmetries}} 

Starting with this section we present a fully worked example of the general framework introduced in Sec.~\ref{sec:recipe} to generate a large decoherence-free subspace. For this we employ the recipe detailed in Sec.~\ref{sec:construction} to obtain a particular form for the matter operator $\mathcal{O}$. 
We consider $\mathcal{O}$ to be a one-dimensional directional tunneling operator. We discuss first the experimental motivation of the model for ultracold atoms in optical cavities. 
We afterwards show how the $\mathfrak{sl}(2,\mathbb{C})$ algebra appears explicitly and briefly discuss the physical consequences of the algebraic structure.
We build explicitly the null eigenvectors for the many-body case in Sec.~\ref{sec:group_theory} and characterize their nontrivial properties in Sec.~\ref{sec:results_steady_states}.

\subsection{The model \label{sec:model}}

\begin{figure}
\begin{centering}
\includegraphics[width=0.9\columnwidth]{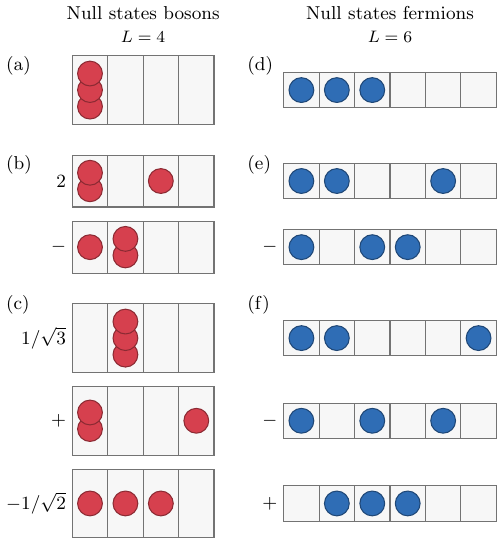}\caption{Sketch of the null eigenvectors $\mathcal{O}$, Eq.~\eqref{eq:operator}, for three quantum particles in $L$ sites, of either bosonic, or fermionic, statistics.
Panels (a) and (d) sketch the fully localized states for bosons and fermions, while the other examples depict superpositions between different Fock states.\label{fig:example_nullstates}}
\par\end{centering}
\end{figure}

Following the recipe in Sec.~\ref{sec:itinerant}, we define our matter operator, Eq.~\eqref{eq:Lindblad}, as a one-dimensional directional tunneling operator on a finite chain, Fig.~\ref{fig:Different-cases}(b),
\begin{align} 
\label{eq:operator}
\mathcal{O}&=\sum_{j=1}^{L-1} d^\dagger_j d_{j+1}, 
\end{align}
where the operators $d_{j}$ can be either bosonic operators, $d_{j}\equiv b_j$, or fermionic operators, $d_{j}\equiv c_j$.
For the results that are independent of the statistics of the particles we refer to $\mathcal{O}$ in its generic form in terms of the $d_{j}$ operator. 
For the results in which the nature of the particles explicitly matters, we use a superscript notation and the expression in terms of the bosonic and fermionic operators, i.e.~$\mathcal{O}^\text{(b)}=\sum_{j=1}^{L-1} b^\dagger_j b_{j+1}$ for the bosonic case and $\mathcal{O}^\text{(f)}=\sum_{j=1}^{L-1} c^\dagger_j c_{j+1}$ for the fermionic case.
In all cases, we consider a chain of length $L$ with $N$ particles, for open boundary conditions (OBC).

The coupling of the directed tunneling operator to a dissipative bosonic mode, has been proposed for ultracold atoms in optical cavities \cite{ZhengCooper2016, LaflammeZoller2017}. 
Proposals to realize this model, Eq.~\eqref{eq:Lindblad} and Eq.~\eqref{eq:operator}, consider atoms confined to a one-dimensional optical lattice in the presence of a linear potential and two-photon Raman transitions employing a cavity photon and a pump photon via an intermediate excited state \cite{ZhengCooper2016, LaflammeZoller2017}. 
In this case the usual tunneling terms are suppressed, and cavity assisted tunneling terms arise, leading to the coupling of the cavity field to the directed tunneling operator $\mathcal{O}$, Eq.~\eqref{eq:operator}.
The cavity frequency $\delta$ in Eq.~\eqref{eq:Lindblad} becomes the detuning of the cavity with respect to the pump frequency.

Previous works have considered the model, Eq.~\eqref{eq:Lindblad} and Eq.~\eqref{eq:operator}, in the context of the generation of atomic currents in the presence of additional atomic kinetic terms \cite{ZhengCooper2016, LaflammeZoller2017, Halati2026}.
The nature of the decoherence-free subspace has been discussed only for systems of up to three fermions, and in the presence of atomic kinetic terms \cite{LiYi2023}.
We further note that in Ref.~\cite{ZhengChen2024} the Bose-Hubbard model with directional tunneling has been shown to be integrable, however, the free and hardcore limits with OBC, realized by $\mathcal{O}$, have not been discussed. 
The coupling of the cavity mode to tunneling terms can arise also in other geometries and usually leads to non-trivial steady states of the coupled system \cite{MaschlerRitsch2008, HabibianMorigi2013, Caballero-BenitezMekhov2015, KollathBrennecke2016,  HalatiKollath2017, ChandaMorigi2022}.

As anticipated, the directional tunneling operator $\mathcal{O}$, Eq.~\eqref{eq:operator}, corresponds in the position Fock base to a non-diagonalizable matrix with a very large null space, see Sec.~\ref{sec:itinerant}. 
We sketch in Fig.~\ref{fig:example_nullstates} the null states of $\mathcal{O}$ for $N=3$ particles, both in the bosonic and fermionic cases. For a system size of $L=4$ sites for bosons and $L=6$ sites for fermions $\mathcal{O}$ has three null states.
In particular, the null space contains more states than the trivial states for which all particles are at the left end of the chain: $\ket{N,0,\dots,0}$, for bosons, Fig.~\ref{fig:example_nullstates}(a); and $\ket{1,\dots,1,0,\dots,0}$, for fermions, Fig.~\ref{fig:example_nullstates}(d).
The other null states contain superpositions of Fock states in which the particles are spread out in the entire chain [see Fig.~\ref{fig:example_nullstates}(b),(c) for bosons and Fig.~\ref{fig:example_nullstates}(e),(f) for fermions].
As argued in Sec.~\ref{sec:recipe}, this leads to a large decoherence-free subspace of the hybrid matter-cavity system, Eq.~\eqref{eq:Lindblad}.
In order to understand the physical properties of the null states of $\mathcal{O}$ we present explicitly its symmetry structure.

\subsection{Symmetries of $\mathcal{O}$ -- $\mathfrak{sl}(2,\mathbb{C})$ algebra structure \label{sec:symmetry_1}}

The central finding ensuring the construction of a large and structured decoherence-free subspace based on the coupling between the cavity mode and the directed tunneling operator is that the operator $\mathcal{O}$ is part of a triplet generating a $\mathfrak{sl}(2,\mathbb{C})$ algebra \cite{KitahamaKatsura2024} (following the framework detailed in Sec.~\ref{sec:recipe} and Sec.~\ref{sec:construction})
\begin{align} 
\label{eq:sl2_algebra}
&[\mathcal{O}_z,\mathcal{O}]=-2\mathcal{O},~[\mathcal{O}_z,\mathcal{O}_+]=2\mathcal{O}_+,~[\mathcal{O}_+,\mathcal{O}]=\mathcal{O}_z,\\
&\mathcal{O}_z=2X-N(L+1),~\text{with}~X=\sum_{j=1}^{L} j  d^\dagger_{j} d_{j},\nonumber \\
&\mathcal{O}_+=\sum_{j=1}^{L-1} j(L-j) d^\dagger_{j+1} d_{j},\nonumber
\end{align}
where $X$ is the center-of-mass operator.
The Casimir operator $\Omega_\mathcal{O}$ of the $\mathfrak{sl}(2,\mathbb{C})$ algebra, which commutes with all the operators of the algebra, is given by
\begin{align} 
\label{eq:casimir_sl2}
\Omega_\mathcal{O}&=2\mathcal{O}_+\mathcal{O}+\frac{1}{2}\mathcal{O}_z^2-\mathcal{O}_z. 
\end{align}
We note that for the $\mathfrak{sl}(2,\mathbb{C})$ algebra, $\Omega_\mathcal{O}$, as defined in Eq.~\eqref{eq:casimir_sl2}, is just a rescaling of the total spin squared.
From the explicit algebraic relations, we see that $\mathcal{O}$ is the lowering operator of the $\mathfrak{sl}(2,\mathbb{C})$ algebra, placing us in the framework described in Sec.~\ref{sec:construction}.

A crucial observation is that $\mathcal{O}_+\neq\mathcal{O}^\dagger$ in the $\mathfrak{sl}(2,\mathbb{C})$ algebra, Eq.~\eqref{eq:sl2_algebra}.
This ensures that we are in the case of Fig.~\ref{fig:DFS}(c) in the classification detailed in Sec.~\ref{sec:recipe_1}.
The structure of the algebra $\mathfrak{sl}(2,\mathbb{C})$ implies that $\Omega_\mathcal{O}$ commutes with $\mathcal{O}$ and $\mathcal{O}_+$, but not with $\mathcal{O}^\dagger$.
Thus, the Casimir $\Omega_\mathcal{O}$ does not commute with the Hamiltonian, Eq.~\eqref{eq:Lindblad}, and it is not a strong symmetry of the Liouvillian.
The symmetry of generated by $\Omega_\mathcal{O}$ only characterizes the symmetry properties of the null states spanning the decoherence-free subspace and not of the entire dissipative dynamics, as discussed in Sec.~\ref{sec:recipe_1}.

We employ the center-of-mass $X$ as a quantum number that labels the null states of $\mathcal{O}$.
Since the null eigenvectors are annihilated by $\mathcal{O}$, they are the lowest weight vectors of their spin representation, Sec.~\ref{sec:construction}, such that we can index them with the eigenvalue of $\mathcal{O}_z$. 
As seen in Eq.~\eqref{eq:sl2_algebra}, $\mathcal{O}_z$ is the same as the center-of-mass operator $X$ up to a constant.

\subsection{Physical picture of the $\mathfrak{sl}(2,\mathbb{C})$ algebra \label{sec:symmetry_2}}

The fact that the directed tunneling operator $\mathcal{O}$ is part of the triplet of a $\mathfrak{sl}(2,\mathbb{C})$ algebra, Eq.~\eqref{eq:sl2_algebra}, gives us an alternative picture to particles moving in the one-dimensional lattice (see Sec.~\ref{sec:itinerant}).
We can view the system of $N$ particles on $L$ sites as $N$ spins of magnitude $S=\frac{1}{2}(L-1)$, with the position of the particle being the $z$ quantum number of the spin, Fig.~\ref{fig:Different-cases}(b).
The fermionic or bosonic nature of the particles enters as an (anti)symmetrization condition on the spin states that are considered (see Sec.~\ref{sec:group_theory}).

The form of the $\mathfrak{sl}(2,\mathbb{C})$ algebra defined in Eq.~\eqref{eq:operator} and Eq.~\eqref{eq:sl2_algebra} is not the one that is usually employed for such an algebra, as $\mathcal{O}_+\neq\mathcal{O}^\dagger$.
In the rest of this section, we show how one can transform between different $\mathfrak{sl}(2,\mathbb{C})$ representations corresponding to the same physical system and analyze the physical implications.
In particular, one can perform a similarity transformation which maps the $\mathfrak{sl}(2,\mathbb{C})$ triplet, Eq.~\eqref{eq:sl2_algebra}, to the following operators
\begin{align} 
\label{eq:sl2_Q_algebra}
Q&=\sum_{j=1}^{L-1} \sqrt{j(L-j)} d^\dagger_j d_{j+1}, \\
Q^{\dagger}&=\sum_{j=1}^{L-1} \sqrt{j(L-j)}  d^\dagger_{j+1} d_{j},~Q_z=2X-N(L+1). \nonumber
\end{align}
These operators are a Hermitian representation of the $\mathfrak{sl}(2,\mathbb{C})$ algebra, satisfying same commutation relations as in Eq.~\eqref{eq:sl2_algebra}, and also corresponding to $N$ spins of magnitude $S=\frac{1}{2}(L-1)$. 
The transformation presented does not enter the analytical results of Sec.~\ref{sec:group_theory}, but enters the discussion of the numerical results in Sec.~\ref{sec:results_steady_states}.

Such a transformation is desirable, because it maps the commonly used representation of the $\mathfrak{sl}(2,\mathbb{C})$ algebra, for which the lowering and raising operators are Hermitian conjugates, Eq.~\eqref{eq:sl2_Q_algebra}, to a representation which can be easily implemented as a tight-binding model.
While the lowering and raising operators $Q$ and $Q^\dagger$ can also be viewed as the kinetic operators on a chain, their space dependent factors reduces their experimental feasibility. 
The transformation that we detail below puts all the space dependent weights in $\mathcal{O}^\dagger$,  leaving $\mathcal{O}$ isotropic.

The fact that one can perform a similarity transformation to map $\mathcal{O}$ to $Q$ relies on the fact that these two operators connect the same states in the many-body Fock basis, with only their matrix elements being different.
Thus, we can construct a diagonal matrix in the Fock basis, $G$, defining the similarity transformation
\begin{align} 
\label{eq:transformation}
&GQG^{-1}=\mathcal{O},~\text{with the entries:}\\
&G^\text{(b)}_{\lambda}=\prod_{j=1}^L  \left[\frac{(j-1)!(L-1)!}{(L-j)!}\right]^{\frac{n_j}{2}},\nonumber \\
&G^\text{(f)}_{\lambda}=\prod_{i=1}^N \prod_{j=i}^{x_i-1} \sqrt{j(L-j)}\nonumber,
\end{align}
where for expression corresponding to bosons $n_j$ are the local occupations of each site of state $\ket{\lambda}$, and for the expression corresponding to fermions $\{x_1,\dots,x_N\}$ are the positions of the particles of state $\ket{\lambda}$. 
In particular, $G_{\lambda_0,\lambda_0} = 1$ for the states in which all particles are localized at the boundary of the chain $\ket{\lambda_0}=\ket{N,0,\dots,0}$ for bosons and $\ket{\lambda_0}=\ket{1,\dots,1,0,\dots,0}$ for fermions.
We explicitly check Eq.~\eqref{eq:transformation} in \ref{app:similarity}.
The similarity transformation performed is an example of imaginary gauge transformation, as for example used in Ref.~\cite{HatanoNelson1996} to map an imbalanced tunneling non-Hermitian Hamiltonian to a Hermitian one.
However, the transformation used in Ref.~\cite{HatanoNelson1996} would not be useful to be applied on $\mathcal{O}$ due to its unidirectional character.

The transformation $G$, Eq.\eqref{eq:transformation}, tells us that the null states of the operators $\mathcal{O}$ and $Q$ are related
\begin{align} 
\label{eq:transformation3}
&\ket{\mathbf{0}_i}=G\ket{\mathbf{q}_i},
\end{align}
where $\ket{\mathbf{0}_i}$ is a null state of $\mathcal{O}$ and $\ket{\mathbf{q}_i}$ is a null state of $Q$. The transformation only changes the coefficients of the many-body Fock states entering in the superposition described by $\ket{\mathbf{q}_i}$. 
Interestingly, this can give us an intuition regarding the physical effects induced by the presence of the $\mathfrak{sl}(2)$ symmetry.
As the transformation $G$ is diagonal in Fock space, we can view it as a potential which favors certain density configurations, enhancing their weight in the null state $\ket{\mathbf{0}_i}$ compared to $\ket{\mathbf{q}_i}$.
In \ref{app:potential} we derive this potential and obtain a nonlinear profile as a function of distance from the boundary, leading to a spreading out of the particles in the bulk of the chain.
We discuss in detail in Sec.~\ref{sec:results_steady_states} the impact on the density profiles of the null states of $\mathcal{O}$.

\section{Explicit derivation of the null eigenvectors \label{sec:group_theory}}

In this section, we develop an analytical solution for the null states of the directed tunneling operator $\mathcal{O}$, defined in Eq.~\eqref{eq:operator}. 
This derivation of a general solution for the null states is based on the representation theory of Lie algebras and provides an analytical way to build the decoherence-free subspace. 
In the following, we show that: (i) an explicit solution for the null states can be derived; (ii) a correspondence exists between the bosonic and fermionic states; (iii) the number of null states grows exponentially with the system size.

Readers primarily interested in the physical properties of the null states spanning the decoherence-free subspace may skip this section on a first reading and proceed directly to Sec.~\ref{sec:results_steady_states}, where these results are presented.

\subsection{Structure of the problem \label{sec:group_theory_setting}}

In Sec.~\ref{sec:symmetry_1} we have identified that our problem had an $\mathfrak{sl}\left(2,\mathbb{C}\right)$
structure.
Notably, the directed tunneling operator $\mathcal{O}$ is identified with the lowering operator of the algebra, $\mathcal{O}\equiv J_{-}$. 
As explained in Sec.~\ref{sec:construction}, finding null eigenvectors of $J_{-}$ thus amounts to finding the lowest weight vectors of the representation.

In the single particle sector, the $L$ sites translate into an irreducible representation of $\mathfrak{sl}\left(2,\mathbb{C}\right)$ with spin $J=\frac{L-1}{2}$. We call this representation $D_{J}$. 
The single null eigenvector is then the lowest weight eigenvector. Indexing the lattice from $0$ to $L-1$, this is simply the vector $\left|0\right\rangle $.

Increasing the number of particles $N$, we now have to deal with tensor products of representations of $D_{J}$ that we denote $D^{\otimes N}_{J}$.
As explained in Sec.~\ref{sec:construction}, the tensor representation is, in general, not irreducible anymore and is decomposed as a direct sum of representations that are indexed by Young diagrams, as in Eq.~\eqref{eq:Young_tableaux}. 

Among these representations, only two are of interest for us: (i) the totally symmetric one (trivial representation of the symmetric group), corresponding to bosons and denoted with the fully horizontal Young diagram; (ii) the totally antisymmetric one (sign representation of the symmetric group) corresponding to fermions and denoted by the fully vertical Young diagram. 
When $D_{J}$ is different from the fundamental spin-$1/2$ representation, these representations themselves are not irreducible. 
They are made of direct sums of irreducible representations of spin $\mathcal{J}$, each forming a Jordan chain of length $2\mathcal{J}+1$. 
For $N=2$, an explicit solution is known from the Clebsch-Gordan coefficients, but the problem becomes quickly intractable for higher values of $N$ \cite{angular_momentum}. 
However, for our purpose, it is enough to focus on a simpler problem, that is to find the null eigenvectors of the bosonic and fermionic representations.
To this end we first introduce the polynomial representation of the ladder algebra. 

\subsection{Polynomial representation \label{sec:group_theory_poly}}

We begin by introducing useful notations and conventions for the polynomial representation of bosonic and fermionic states. In this representation, a particle located at site $k$ is represented by a monomial 
\begin{equation}
\left|k\right\rangle \to u_{k}\left(x\right):=\frac{x^{k}}{k!}.
\label{eq:poly_representation}
\end{equation}
Note that we index the first site of the lattice at $k=0$ and the last at $k=L-1$. Multiparticle bosonic or fermionic states are respectively represented by a (pseudo)permanent or a determinant.

For a state of $N$ bosons, let $\boldsymbol{\nu}:=\left\{ \nu_{j}\right\} ^{N}_{j=1}$ be a decreasing partition, $L-1\geq\nu_{1}\geq\nu_{2}\cdots\geq\nu_{N}\geq0$.
Let also $\boldsymbol{n}:=\left\{ n_{j}\right\} ^{L-1}_{j=0}$ index
the number of times $j$ appears in the partition \textbf{$\boldsymbol{\nu}$},
$n_{j}=\sum^{N}_{m=1}\delta_{\nu_{m},j}$. $n_j$ counts the bosonic occupation number of site $j$. The (pseudo)permanents
$m_{\boldsymbol{\nu}}$ are defined as the polynomials%
\begin{equation}
m_{\boldsymbol{\nu}}=\frac{1}{\prod^{L-1}_{j=0}n_{j}!}\sum_{\sigma\in S_{N}}x^{\sigma.\boldsymbol{\nu}}.\label{eq:pseudopermanent}
\end{equation}
where $x^{\boldsymbol{\nu}}:=x^{\nu_{1}}_{1}x^{\nu_{2}}_{2}\cdots x^{\nu_{n}}_{n}$.
The partition $\boldsymbol{\nu}$ can then be mapped, up to a rescaling,
to a bosonic state in second quantization $\left|\boldsymbol{N}\right\rangle ^{{\rm B}}=\left|n_{0},\cdots,n_{L-1}\right\rangle ^{{\rm B}}$ 
\begin{equation}
m_{\boldsymbol{\nu}}\to\frac{\prod_{a}\nu_{a}!}{\sqrt{\prod_{k}n_{k}!}}\left|\boldsymbol{N}\right\rangle ^{{\rm B}}.
\label{eq:rescaling_bosons}
\end{equation}
As an example, for $N=3$ and $\boldsymbol{\nu}=\left(2,1,1\right)$
we have 
\begin{equation}
m_{\left(2,1,1\right)}=x^{2}_{1}x_{2}x_{3}+x^{2}_{2}x_{1}x_{3}+x^{2}_{3}x_{1}x_{2}
\end{equation}
corresponding to the second quantized state $\frac{2!}{\sqrt{2}}\left|0,2,1,0,\cdots\right\rangle ^{{\rm B}}.$

For the fermionic case, the equivalent of the (pseudo)permanent $m_{\boldsymbol{\nu}}$
are the alternants $a_{\boldsymbol{\mu}}$. For $N$ fermions, let
$\boldsymbol{\mu}$ be a strictly decreasing exponent list $\mu_{1}>\mu_{2}>\cdots>\mu_{N}\geq0$,
the alternant is defined as 
\begin{equation}
a_{\boldsymbol{\mu}}:=\det\left[x^{\mu_{a}}_{b}\right]^{N}_{a,b=1}=\sum_{\sigma\in S_{N}}{\rm sgn}\left(\sigma\right)x^{\sigma.\boldsymbol{\mu}}.\label{eq:alternant}
\end{equation}
and can be mapped, up to a rescaling, to the second quantized fermionic
basis element: 
\begin{equation}
a_{\boldsymbol{\mu}}\to\left(-1\right)^{\frac{N\left(N-1\right)}{2}}\prod_{a}\mu_{a}!\left|\boldsymbol{k}\right\rangle ^{{\rm F}}.
\label{eq:rescaling_fermions}
\end{equation}
where the $N$ fermions occupy the sites $k_{1},\cdots,k_{N}$ and
$\left|\boldsymbol{k}\right\rangle ^{{\rm F}}:=\left|k_{1},\cdots k_{N}\right\rangle ^{{\rm F}}$.
The $\left(-1\right)^{\frac{N\left(N-1\right)}{2}}$ prefactor comes
from the fact that one has to reverse the order of the fermions from the polynomial to the second quantized basis. 

As an example, consider for $N=3$, $\boldsymbol{\nu}=\left(2,1,0\right)$.
The alternant $a_{\left(2,1,0\right)}$ is
\begin{align*}
x^{2}_{1}x_{2}x^{0}_{3}-x^{2}_{2}x_{1}x^{0}_{3}-x^{2}_{3}x_{2}x^{0}_{1}-x^{2}_{1}x_{3}x^{0}_{2}+x^{2}_{3}x_{1}x^{0}_{2}+x^{2}_{2}x_{3}x^{0}_{1}
\end{align*}
which corresponds to the state $-2!\left|1,1,1,0,\cdots\right\rangle.$

In this polynomial representation, Eq.~\eqref{eq:poly_representation}, the generators $J_{-}\equiv\mathcal{O}$, $J_{+}\equiv\mathcal{O}_+$ and $J_{z}\equiv\mathcal{O}_z$ of the $\mathfrak{sl}(2,\mathbb{C})$ algebra, Eq.~\eqref{eq:sl2_algebra}, become respectively 
\begin{align}
J_{-}^{\left(L\right)} & =\sum_{j}\partial_{x_{j}},\quad J^{\left(L\right)}_{+}=\sum_{j}\left(L-1\right)x_{j}-x^{2}_{j}\partial_{x_{j}},\\
J^{\left(N\right)}_{z} & =\sum_{j}x_{j}\partial_{x_{j}}-\frac{N\left(L-1\right)}{2}. \nonumber
\end{align}
In this language, eigenvectors of $J_{-}$ for bosons (fermions) are symmetric (antisymmetric) translationally invariant polynomials. 
We keep the index $\left(L\right)$ to keep track of the lattice size, as it will turn out to be useful for the boson-fermion correspondence discussed in Sec.~\ref{sec:bosons_to_fermions}.

We begin by discussing the case of bosons and show in Sec.~\ref{sec:bosons_to_fermions} that the bosonic null states can be mapped to the fermionic null states.
Translation-invariant symmetric polynomial (TISP) are stable by multiplication and addition and hence form an algebra. 
One can show that a set of generators for the TISPs are given by the centered power sums $q_{r}$ defined as (see \ref{app:TISP_basis})
\begin{equation}
q_{r}:=\sum^{N}_{i=1}\left(x_{i}-\bar{x}\right)^{r},
\end{equation}
with $\bar{x}=\sum_{i}\frac{x_{i}}{N}$. 
A generic eigenvector will then be written as a product, see~\ref{app:TISP_basis} for a proof:
\begin{equation}
\mathcal{S}:=\prod^{N}_{r=2}q^{a_{r}}_{r}\label{eq:S_polynomial}
\end{equation}
and we call $d$ the degree of this polynomial, $d:=\sum^{N}_{r=2}ra_{r}$. For $\mathcal{S}$ to represent a state contained on a lattice of size $L$, we must only consider cases where $d\leq L-1$. 
Remark that the degree of the individual terms in the product does not exceed $N$ as any $q_{r}$ for $r>N$ can be decomposed in terms of $q_{r'}$ with $r'\leq N$. 

\subsection{Spin representation and degeneracies \label{sec:group_theory_size}}

An important point is that the \emph{degree of $\mathcal{S}$ fixes the total spin $\mathcal{J}$ of the representation} this vector belongs to. 
Indeed, acting with $J_{z}$ on $\mathcal{S}$ gives
\begin{equation}
J_{z}\mathcal{S}=\left(d-\frac{d^{{\rm B}}_{\max}}{2}\right)\mathcal{S}
\end{equation}
with $d^{{\rm B}}_{\max}:=N\left(L-1\right)$. 
Hence, the spin of the representation is $\mathcal{J}\left(d\right)=\frac{d^{{\rm B}}_{\max}}{2}-d$ and its dimension is $d^{{\rm B}}_{\max}-2d+1$. 
For the ladder to exist at all one needs that the rung at the bottom is below the rung at the top, hence we must have $d\leq\frac{d^{{\rm B}}_{\max}}{2}$.

\begin{figure}
\centering{}\includegraphics[width=0.8\columnwidth]{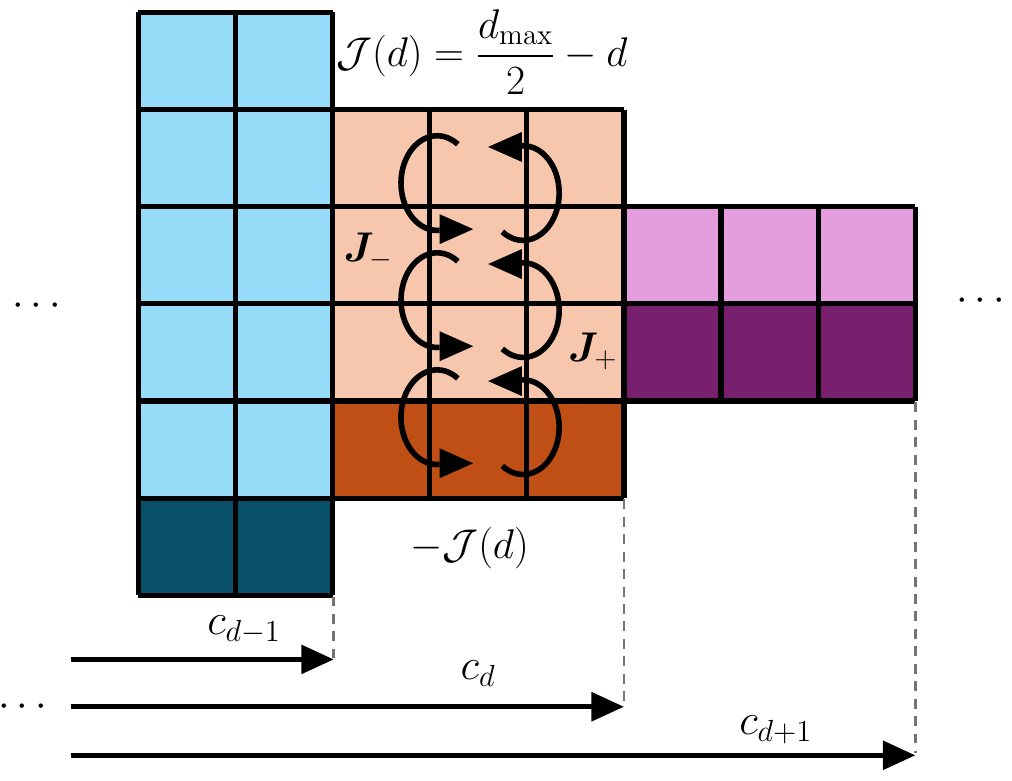}\caption{Jordan chain structure of the null eigenvectors. Each column represents
a Jordan chain corresponding to a representation of spin $\mathcal{J}\left(d\right)=\frac{d_{\max}}{2}-d$.
The $J_{+}$, $J_{-}$ operators allow to respectively move upwards
or downwards on a given ladder. The bottom rung in darker color are the
null eigenvectors. The degeneracy can be counted by subtracting the
number of states with degree $d-1$ from the number of states with
degree $d$.}
\label{fig:Jordan_chains}
\end{figure}

To know the degeneracy $\Omega_{j}$ of a representation of spin $j\left(d\right)$,
we proceed in two steps. First, one can count all the polynomials
with degree $d$. Let us call this number $c_{d}$. Among these polynomials
some are lowest weight eigenvectors of a spin representation and some
of them are just in the middle of a Jordan chain of a higher spin
representation. In order to single-out the elements of grade $d$
that are the lowest weight of a representation of spin $j\left(d\right)$,
we can simply take the difference $c_{d}-c_{d-1}$, see Fig.~\ref{fig:Jordan_chains} for
a visual explanation,
\begin{equation}
\Omega_{j\left(d\right)}=c_{d}-c_{d-1}.
\end{equation}
$c_{d}$ counts the number of ways one can take products of $N$ monomials
$x^{k_{j}}_{j}$ with $k_{j}\leq L-1$ in order to get a polynomial
of grade $d$. This is equivalent to finding the number of ways one
can partition $d$ particles in a box of size $N\times\left(L-1\right)$.
This is precisely what is counted by the \emph{Gaussian binomial coefficient} \cite{Weisstein_qBinomialCoefficient, KitahamaKatsura2024}:
\begin{equation}
\binom{L+N-1}{N}_{q}:=\prod^{N}_{k=1}\frac{\left(1-q^{L-1+k}\right)}{1-q^{k}}=\sum_{d}c_{d}q^{d}.
\end{equation}
The weighting of the coefficient $q^{d}$ in the expression is exactly $c_{d}$. Depending on the scalings of $N$, $L$ and $d$, the partitioning problem admits different asymptotic scaling but outside of pathological case, for large $N$, $\Omega_j$ scale exponentially with $N$ (up to power-law corrections), leading to large decoherence free subspace.
The degeneracies of the null states have been previously computed in Ref.~\cite{KitahamaKatsura2024}, proving a conjecture regarding the Jordan decomposition of a quadratic fermionic Liouvillians \cite{Prosen2010b}.

\subsection{Explicit decomposition}

In order to arrive at our goal to determine the null bosonic and fermionic states, we need to understand how to decompose the polynomial $\mathcal{S}$ in the bosonic basis. 
We consider a product of $q_{r}$ centered polynomials, $\mathcal{S}=q_{\mu_{1}}q_{\mu_{2}}\cdots$, and introduce the non-centered polynomials $p_{r}:=\sum^{N}_{i=1}x^{r}_{i}$.
Each $q_{r}$ polynomial can be decomposed in terms of $p_{r}$ polynomials through the binomial expansion
\begin{equation}
q_{r}=\sum^{r}_{s=0}\left(-1\right)^{r-s}\binom{r}{s}N^{s-r}p^{r-s}_{1}p_{s},
\end{equation}
where we used the fact that $p_1=N\bar{x}$.

For a product of multiple $p_{r}$ polynomials, $p_{\boldsymbol{\rho}}:=p_{\rho_{1}}p_{\rho_{2}}\cdots$, we can write the decomposition of this product in terms of the monomials $m_{\boldsymbol{\nu}}$,
\begin{equation}
p_{\boldsymbol{\rho}}=\sum_{\boldsymbol{\nu}}R_{\boldsymbol{\rho\nu}}m_{\boldsymbol{\nu}}.
\end{equation}
The matrix element $R_{\boldsymbol{\rho\nu}}$ counts all the ways
one can rearrange the partition $\boldsymbol{\rho}$, with only groupings
and permutations to obtain the partition $\boldsymbol{\nu}$. For
instance 
\begin{equation}
R_{\left(1,1,1\right)\left(1,1,1\right)}=6,\quad R_{\left(2,1,1\right)\left(3,1\right)}=2.
\end{equation}
Computing arbitrary elements of $R_{\boldsymbol{\rho\lambda}}$ is amenable to a subset sum problem which is known to be NP-complete \cite{subset_problem}, see the footnote \footnote{To see this, choose $\nu$ to be just two numbers $\boldsymbol{\nu}=\left(A,B\right)$.
Then finding the $\boldsymbol{\rho}$s such that $R_{\boldsymbol{\rho\nu}}\neq0$
and $A+B=\left|\boldsymbol{\rho}\right|$amounts to find sub-multisets
of $\rho$ that sums to $A$ which is precisely the subset sum problem.} for an explanation. Hence, we do not expect in general to find an explicit formula for the matrix $R$.
However, if $\boldsymbol{\rho}$ is given, computing all non-zero $R_{\boldsymbol{\rho\lambda}}$ is an easy task that can be achieved in polynomial time. 

For the case where $\mathcal{S}$ is made of a single element, we can have an analytical expression as we only need to decompose the products $p^{r-s}_{1}p_{s}$. We show in the ~\ref{app:q_r_proof} that
\begin{equation}
q_{r}=\frac{\left(-1\right)^{r}}{N^{r}}\sum_{\boldsymbol{\nu}\vdash r}\frac{r!}{\prod\nu_{a}!}\left(\sum^{\ell\left(\boldsymbol{\nu}\right)}_{a=1}\left(1-N\right)^{\nu_{a}}+N-\ell\left(\boldsymbol{\nu}\right)\right)m_{\boldsymbol{\nu}}\label{eq:q_r_decomposition}
\end{equation}
where $\boldsymbol{\nu}\vdash r$ means all the partition of $r$
and $\ell\left(\nu\right)$ is the number of non-zero elements of
$\boldsymbol{\nu}$. 

The derivation presented above gives a systematic procedure to decompose a TISP, Eq.~\eqref{eq:S_polynomial}, into the original
bosonic basis, which will allow us to recover the bosonic states in second quantization. 
We provide some examples of the full procedure further below, in Sec.~\ref{sec:example_analytic_formula}.
In the next subsection, we show how to obtain the fermionic states from the bosonic ones.

\subsection{Fermion-boson correspondence \label{sec:bosons_to_fermions}}

\begin{figure}
\centering{}\includegraphics[width=0.9\columnwidth]{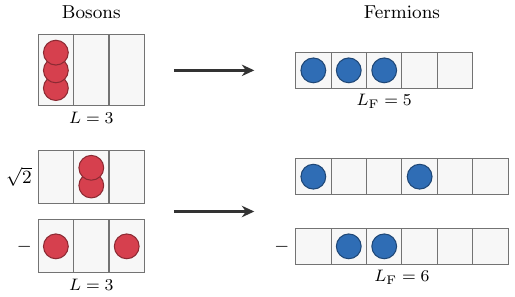}\caption{Illustration of the ``fermionization procedure'' introduced in Sec.~\ref{sec:bosons_to_fermions}. Top: The correspondence of the fully localized sates, a condensate bosonic state gets mapped to a filled Fermi sea. 
Bottom: Fermionization of the states considered in the $N=2$ example. In both cases the bosonic and fermionic state are lowest weight vectors of a representation with the same spin. Remark that the system size gets \emph{increased} when going from bosons to fermions. We employed the same procedure to get the states shown in Fig.~\ref{fig:example_nullstates}. 
}\label{fig:fermionization}
\end{figure}

The null eigenvectors for the fermionic case correspond to translation-invariant antisymmetric polynomials (TIAP). 
In order to get a TIAP from a TISP $\mathcal{S}$, it suffices to multiply the polynomial $q_{r}$ by the Vandermonde determinant $\Delta_{N}:=\prod_{1\leq i<j\leq N}\left(x_{i}-x_{j}\right)$. 
The correspondence between bosonic null eigenvectors and fermionic ones can actually be done one-to-one, thanks to the following commutation properties.
By viewing $\Delta_{N}$ as an operator itself, one can show that
\begin{align}
\left[J_{-},\Delta_{N}\right] & =0,\nonumber \\
J^{\left(L+N-1\right)}_{z}\Delta_{N} & =\Delta_{N}J^{\left(L\right)}_{z},\\
J^{\left(L+N-1\right)}_{+}\Delta_{N} & =\Delta_{N}J^{\left(L\right)}_{+},\nonumber 
\end{align}
The above relations tell us that whenever a polynomial $\mathcal{S}$ of degree $d$ is a \emph{bosonic} null eigenvector of $J_{-}$, $\Delta_{N}S$ is a \emph{fermionic} null eigenvector of $J_{-}$. 
From the commutation relation with $J_{z}$, we see that $\Delta_{N}S$ belongs to the same spin representation as $\mathcal{S}$ with $\mathcal{J}\left(d,N\right)=\frac{d_{\max}}{2}-d$,
$d_{\max}=n\left(N-1\right)$ but on an \emph{enlarged space} of lattice size $L_{{\rm F}}:=L+N-1$ \footnote{An alternative interpretation could be that the representation is of a smaller spin but on the same lattice size. We will not adopt this convention in the paper.}. 
The degree of $\Delta_{N}\mathcal{S}$ is $d_{{\rm F}}:=d+\binom{N}{2}$ and
$d_{{\rm F},\max}$ corresponds to the state with all the fermions packed at the left-end of the lattice, i.e.~$d_{{\rm F,\max}}=\sum^{L_{{\rm F}}-1}_{L_{{\rm F}}-N}=N\left(L_{{\rm F}}-N\right)+\binom{N}{2}$.
In terms of these notations, the value of the spin becomes 
\begin{equation}
\mathcal{J}\left(d_{{\rm F}},L_{{\rm F}}\right)=\frac{d_{{\rm F,\max}}}{2}-d_{{\rm F}}+\frac{1}{2}\binom{N}{2}.
\end{equation}
As an example of this mapping, consider the simplest bosonic null
eigenvector for $N$ particles where all the particles condensate
on the left most site. In polynomial language, it is just a constant
$x^{0}_{1}\cdots x^{0}_{N}$. Multiplying by the Vandermonde we obtain
the Fermi sea state on $L+N-1$ sites, see Fig.~\ref{fig:fermionization}.

Let us now see how to do this mapping in practice. Start from a bosonic
null eigenvector $\mathcal{S}$. From the previous construction, we can decompose
it in terms of the pseudopermanents
\begin{equation}
\mathcal{S}=\sum_{\boldsymbol{\nu}}\alpha_{\boldsymbol{\nu}}m_{\boldsymbol{\nu}}.
\end{equation}
Let's suppose $\alpha_{\boldsymbol{\nu}}$ is known. Now consider the corresponding fermionic null eigenvector $\Delta_{N}\mathcal{S}$. We are looking for the decomposition of the latter in terms of the alternants
\begin{equation}
\Delta_{N}\mathcal{S}=\sum_{\boldsymbol{\mu}}\beta_{\boldsymbol{\mu}}a_{\boldsymbol{\mu}}.
\end{equation}
A convenient parametrization of the Vandermonde is in terms of the
staircase partition $\boldsymbol{\delta}:=\left(N-1,N-2,\cdots,1\right)$,
$\Delta_{N}=\sum_{\sigma\in S_{N}}{\rm sgn}\left(\sigma\right)x^{\sigma.\boldsymbol{\delta}}$.
Combining this expression with the definitions of the pseudopermanent
and alternant, Eq.~(\ref{eq:pseudopermanent}) and Eq.~(\ref{eq:alternant}),
we obtain after a few elementary manipulations: %
\begin{align}
\label{eq:result_fermions}
\Delta_{N}\mathcal{S} & =\sum_{\boldsymbol{\nu}}\sum_{\sigma'\in S_{N}}\frac{\alpha_{\boldsymbol{\nu}}}{\boldsymbol{n}!\left(\nu\right)}a_{\boldsymbol{\delta}+\sigma'.\boldsymbol{\nu}}
\end{align}
where we defined for convenience $\boldsymbol{n}!\left(\nu\right):=\prod^{L-1}_{j=0}n_{j}!$
with $n_{j}$s the occupation numbers of the bosonic state. So the
recipe to obtain a fermionic null eigenvector from a pseudopermanent
$m_{\boldsymbol{\nu}}$ is to add the staircase partition to all (distinguishable)
permutations of a given $\boldsymbol{\nu}$. We show below an example
of this procedure. 

\subsection{Example for $N=2$ particles \label{sec:example_analytic_formula}}

In this subsection, we work out in detail the fermion-bosons correspondence for the simplest example of $N=2$ particles and a bosonic lattice of size $L$ to illustrate the machinery presented above. The corresponding fermionic lattice has one site more $L_{{\rm F}}=L+1$. 

For bosons, the first trivial null eigenvector is the constant polynomial
$x^{0}_{1}x^{0}_{2}$. The degree of this polynomial is $0$ corresponding
to the total spin $\mathcal{J}\left(0\right)=\frac{d_{\max}}{2}=L-1$.
The corresponding second quantized state is $\left|2,0,\dots\right\rangle ^{{\rm B}}$.
Applying the Vandermonde to this polynomial yields the state $x_{1}x^{0}_{2}-x_{2}x^{0}_{1}$
with grade $d=1$ and the corresponding fermionic state is $\left|1,1,0,\cdots\right\rangle ^{{\rm F}}$. 

Now consider the centered polynomial $q_{2}:=\sum^{2}_{i=1}\left(x_{i}-\bar{x}\right)^{2}$. It can be decomposed on the pseudopermanent basis as
\begin{align*}
q_{2} & =\frac{1}{2}\left(m_{(2,0)}-2m_{(1,1)}\right).
\end{align*}
After rescaling the pseudopermanents map to the bosonic states (up
to an overall prefactor)
\begin{equation}
\sqrt{2}\left|1,0,1,\cdots\right\rangle -\left|0,2,0,\cdots\right\rangle ,
\end{equation}
which is a null state as one can verify explicitly, depicted in Fig.~\ref{fig:fermionization}. 

The fermionic null eigenvector is $\Delta_{2}q_{2}$. Equation \eqref{eq:result_fermions} tells us that the (pseudo)permanents are decomposed onto alternants as 
\begin{align}
m_{\left(2,0\right)} & \to a_{\left(3,0\right)}-a_{\left(2,1\right)}, & m_{\left(1,1\right)}\to a_{\left(2,1\right)},
\end{align}
so that 
\begin{equation}
\Delta_{2}q_{2}=\frac{1}{2}\left(a_{\left(3,0\right)}-3a_{\left(2,1\right)}\right).
\end{equation}
After rescaling accordingly to Eq.~\eqref{eq:rescaling_fermions}, the alternants map to the fermionic states (up
to an overall prefactor)
\begin{equation}
\left|1,0,0,1,\cdots\right\rangle -\left|0,1,1,0\right\rangle,
\end{equation}
fermionic state sketched in Fig.~\ref{fig:fermionization}.

The other null eigenvectors for the case $N=2$ are obtained by considering
$q^{m}_{2}$ with $2m\leq L-1$ and the associated $\Delta_{2}q^{m}_{2}$.
In that case, it is possible to obtain an explicit expression for the
pseudopermanent decomposition 
\begin{equation}
\left|\Phi_{m}\right\rangle ^{{\rm B}}=\sum_{j=0}^{2m}(-1)^j b_j^\dagger b_{2m-j}^\dagger \ket{\varnothing},
\end{equation}
with the corresponding fermionic state being 
\begin{equation}
\left|\Phi_{m}\right\rangle ^{{\rm F}}=\sum^{m}_{j=0}\left(-1\right)^{j}c^{\dagger}_{2m+1-j}c^{\dagger}_{j}\left|\varnothing\right\rangle,
\end{equation}
with $\ket{\varnothing}$ the vacuum state.

\section{\boldmath Emergent spinless $\eta$-pairing algebras \label{sec:pairing_algebra}}

We have discussed in Sec.~\ref{sec:results_symmetries} and Sec.~\ref{sec:group_theory} the implication of the spin-like $\mathfrak{sl}(2,\mathbb{C})$ algebra satisfied by the directed tunneling operator, $\mathcal{O}$, Eq.~\eqref{eq:operator}, for both fermionic and bosonic quantum particles.
However, interestingly, there exists a second algebra which commutes with the $\mathfrak{sl}(2,\mathbb{C})$ spin algebra.  
This algebra gives rise to an emergent \emph{spinless $\eta$-pairing} mechanism in the hybrid system.
The triplet of the algebra, $(K_+, K_-, K_z)$, contains raising, $K_+$, and lowering, $K_-$, operators corresponding to the creation and annihilation of pairs of two particles. 
As we show in Sec.~\ref{sec:pairing_algebra_bosonic} and Sec.~\ref{sec:pairing_algebra_fermionic}, the pairs are formed from delocalized particles correlated in their respective positions.
In the following we refer to this structure as \emph{spinless $\eta$-pairing algebra}.
After presenting the bosonic and fermionic spinless $\eta$-pairing algebras, Sec.~\ref{sec:pairing_algebra_bosonic} and Sec.~\ref{sec:pairing_algebra_fermionic}, we discuss the connection with the $\eta$-pairing mechanism occurring in the Hubbard model \cite{Esslerbook, Yang1989}.

The spinless $\eta$-pairing algebra takes different forms depending on the quantum statistics: a $\mathfrak{su}(1,1)$ algebra for bosons and a $\mathfrak{su}(2)$ algebra for fermions.
We discuss the two cases separately. 
These two algebras cannot be mapped from one to the other by the transformation described previously relating the bosonic and fermionic null states of $\mathcal{O}$ (see Sec.~\ref{sec:bosons_to_fermions}). 
In particular, as we see in Sec.~\ref{sec:results_steady_states}, even in two equivalent bosonic and fermionic symmetry sectors (related by the correspondence, Sec.~\ref{sec:bosons_to_fermions}), the number of distinct eigenvalues of the bosonic and fermionic pairing algebras can be different.
They thus represent distinct pairing mechanisms.

The Casimir operators of the pairing algebras not only commute with $\mathcal{O}$, and the corresponding $\mathfrak{sl}(2,\mathbb{C})$ algebra, but also with $\mathcal{O}^\dagger$. 
This implies that the pairing Casimir generates a strong symmetry in the open hybrid system,  Sec.~\ref{sec:pairing_consequences}.
In each symmetry sector we obtain a large decoherence-free subspace, as sketched in Fig.~\ref{fig:DFS}(c) and discussed in Sec.~\ref{sec:recipe_1}.

\subsection{\boldmath Bosonic spinless $\eta$-pairing algebra \label{sec:pairing_algebra_bosonic}}

For bosons we have a $\mathfrak{su}(1,1)$ non-compact pairing algebra, which holds only for $L$ odd system sizes:
\begin{align} 
\label{eq:pairing_bosons}
K_z^\text{(b)}=&\sum_{j=1}^{L} \left(n_{j}+\frac{1}{2}\right)=N+L/2, \\
K_+^\text{(b)}=&\sum_{j=1}^{\frac{L-1}{2}} (-1)^{j-1} b^\dagger_j b^\dagger_{L+1-j} +\frac{1}{2}(-1)^\frac{L-1}{2}b^\dagger_\frac{L+1}{2}b^\dagger_\frac{L+1}{2},\nonumber\\
K_-^\text{(b)}=&\sum_{j=1}^{\frac{L-1}{2}} (-1)^{j-1} b_j b_{L+1-j}+\frac{1}{2}(-1)^\frac{L-1}{2}b_\frac{L+1}{2}b_\frac{L+1}{2}, \nonumber
\end{align}
with the triplet of operators $(K_+^\text{(b)}, K_-^\text{(b)}, K_z^\text{(b)})$ satisfying the commutation relations
\begin{align} 
\label{eq:pairing_bosons_2}
&[K_z^\text{(b)},K_+^\text{(b)}]=2K_+^\text{(b)},~[K_z^\text{(b)},K_+^\text{(b)}]=-2K_+^\text{(b)},\\
&[K_+^\text{(b)},K_-^\text{(b)}]=-K_z^\text{(b)}. \nonumber
\end{align}
The raising operator of the algebra, $K_+^\text{(b)}$, corresponds to creation of a \emph{centro-symmetric $\eta$-pair}, while $K_-^\text{(b)}$ removes such a pair. 
The null states of $K_-^\text{(b)}$ correspond to the states without any pairs.
From the expressions in Eq.~\eqref{eq:pairing_bosons} we can see that bosonic particles forming the pairs are created at symmetric positions with respect to the middle of the chain, $j$ and $L+1-j$, in a superposition throughout the chain. 
Thus, the application of $K_+^\text{(b)}$ to an empty chain results in a superposition with equal weight of all two-particle states with a center-of-mass  $X=L+1$. The sign of each state in the superposition is given by the parity of the first occupied site.

The Casimir operator of the bosonic spinless $\eta$-pairing algebra counts the number of pairs in the system, $N_p$, 
\begin{align} 
\label{eq:casimir_pairing_bosons}
&\Omega_K^\text{(b)}=\frac{1}{2}\left(N+L/2\right)^2-\left(N+L/2\right)-2K_+^\text{(b)}K_-^\text{(b)},
\end{align}
with the eigenvalues $2\left(N_p-\frac{N}{2}-\frac{L}{4}\right)\left(N_p+1-\frac{N}{2}-\frac{L}{4}\right)$ indexed by the number of the pairs $N_p$, which ranges from $0$ to $\left[\frac{N}{2}\right]$.

\subsection{\boldmath Fermionic spinless $\eta$-pairing algebra \label{sec:pairing_algebra_fermionic}}

In the case of fermionic particles, we have a $\mathfrak{su}(2)$ pairing algebra, which holds only for $L$ even system sizes.
In terms of the fermionic operators, the pairing algebra is 
\begin{align} 
\label{eq:pairing_SU2_fermions}
K_+^\text{(f)}&=\sum_{j=1}^{L/2} (-1)^{j-1} c^\dagger_j c^\dagger_{L+1-j}, \\
K_-^\text{(f)}&=\sum_{j=1}^{L/2} (-1)^{j-1} c_{L+1-j} c_j , \nonumber\\
K_z^\text{(f)}&=\sum_{j=1}^{L} \left(n_{j}-\frac{1}{2}\right)=N-L/2, \nonumber
\end{align}
with the commutation relations
\begin{align} 
\label{eq:pairing_SU2_fermions_2}
&[K_z^\text{(f)},K_+^\text{(f)}]=2K_+^\text{(f)},~[K_z^\text{(f)},K_-^\text{(f)}]=-2K_-^\text{(f)},\nonumber\\
&[K_+^\text{(f)},K_-^\text{(f)}]=K_z^\text{(f)}. 
\end{align}
As in the case of the bosonic pairing algebra, the triplet $(K_+^\text{(f)}, K_-^\text{(f)}, K_z^\text{(f)})$ defines a pairing algebra which removes, or adds, \emph{centro-symmetric $\eta$-pairs} of delocalized fermionic particles positioned symmetrically with respect to the center of the chain, at $j$ and $L+1-j$.
While the action of $K_+^\text{(f)}$ onto an empty state is the same as in the bosonic case, in the presence of additional particles the resulting state will be very different due to the Pauli exclusion principle.
$K_z$ measures the deviations of the filling from $1/2$. 
For the results presented in the next sections we focus on half-filling for the fermionic case, corresponding to the sector $K_z=0$.

The Casimir operator of the fermionic spinless $\eta$-pairing algebra is 
\begin{align} 
\label{eq:casimir_pairing_SU2}
&\Omega_K^\text{(f)}=2K_+^\text{(f)}K_-^\text{(f)}+\frac{1}{2}\left(N-L/2\right)^2-\left(N-L/2\right),  \\
&K_+^\text{(f)}K_-^\text{(f)}=\sum_{j,j'=1}^{L/2} (-1)^{j+j'}c^\dagger_{j} c^\dagger_{L+1-j}c_{L+1-j'} c_{j'}, \nonumber 
\end{align}
with its eigenvalues $2\left(N_p-\frac{N}{2}+\frac{L}{4}\right)\left(N_p+1-\frac{N}{2}+\frac{L}{4}\right)$ being related to the number of pairs $N_p$, which ranges from $0$ to $\left[\frac{N}{2}\right]$.

We note that by performing the Jordan-Wigner transformation, one can derive the hardcore bosons equivalent of the algebra defined in Eq.~\eqref{eq:pairing_SU2_fermions_2}. As $K_+^\text{(f)}$ and $K_-^\text{(f)}$ are a non-local operators in the fermionic representation, the Jordan-Wigner string does not vanish.
Thus, in the hardcore bosonic algebra, the sign of each term within $K_+^\text{(f)}$ and $K_-^\text{(f)}$ depends on the parity of the number of particles located between the operators at $j$ and $L+1-j$.

\subsection{Consequences of the pairing algebras \label{sec:pairing_consequences}}

The two spinless $\eta$-paring algebras presented above give rise to a mechanism in which pairs at finite momentum are created or annihilated.
To understand better the physical nature of this pairing mechanism we rewrite the $K_-^\text{(f)}$ and $K_-^\text{(b)}$ in momentum space
\begin{align} 
\label{eq:pairing_momentum}
K_-^\text{(f)}&=\sum_{n=1}^{L/2} (-1)^{n-1}\, c_{k_n}c_{\pi-k_n}, \\
K_-^\text{(b)}&=\sum_{n=1}^{\frac{L-1}{2}} (-1)^{n-1}\, b_{k_n} b_{\pi-k_n}+\frac{1}{2}(-1)^{\frac{L-1}{2}}\, b_{\pi/2}\, b_{\pi/2}, \nonumber
\end{align}
where we have used the sine-Fourier transform defined for open boundary conditions, with $k_n=\frac{\pi n}{L+1}$.
The pairs contain particles with momentum $k$ and $\pi-k$, resulting in a finite momentum of $\pi$.
This represents a spinless analogue to the Hubbard model $\eta$-pairs \cite{Esslerbook, Yang1989}, up to the phase structure given by $(-1)^n$. 
In the spinfull $\eta$-pairs of the Hubbard model, particles with momentum $k$ and $\pi-k$ and opposite spin are paired, i.e.~the annihilation operator has the form $\sum c_{k,\uparrow}c_{\pi-k,\downarrow}$.

The double algebra structure of our system, in which a spin-like algebra is accompanied by a pairing algebra, resemble algebraic structures that play a role in many-body integrable quantum systems. 
As mentioned above, in the one-dimensional Hubbard model one has symmetries stemming from the $\mathfrak{su}(2)$ algebra of spin rotations and from the $\eta$-pairing $\mathfrak{su}(2)$ algebra \cite{Esslerbook, Yang1989}.
Another example are the isospin and quasi-spin  $\mathfrak{su}(2)$ algebras appearing in seniority models in nuclear physics \cite{KruppaZarand2022, Talmibook, Georgi1999}, related to Richardson-Gaudin pairing models \cite{Richardson2002, DukelskySierra2004}.
In these instances, the algebras offer the symmetry structure for the \emph{Hamiltonian} necessary to obtain the integrable nature and its conserved charges.
However, this is very different from our open system described by a \emph{Liouvillian}, Eq.~\eqref{eq:Lindblad}, where the spin-like algebra determines symmetries only for the decoherence-free subspaces reached at long times, while the pairing algebras appear as strong symmetries of the dissipative dynamics at all times.

In the rest of this section, we highlight the main consequences stemming from the spinless $\eta$-pairing algebras on the structure of the decoherence-free subspace: 
(i) one can construct null states of  $\mathcal{O}$ by adding or removing pairs of particles; 
(ii) the impact on the density distribution of the null states due to the delocalization of the particles forming the pairs; 
(iii) the Liouvillian strong symmetry originating from the pairing symmetry.

The $\mathfrak{sl}(2,\mathbb{C})$ algebra, Eq.~\eqref{eq:sl2_algebra}, defined by the triplet $(\mathcal{O}, \mathcal{O}_+,\mathcal{O}_z)$, commutes, in the fermionic case with the $\mathfrak{su}(2)$, Eq.~\eqref{eq:pairing_SU2_fermions}, and in the bosonic case with the $\mathfrak{su}(1,1)$ algebra, Eq.~\eqref{eq:pairing_bosons}. 
The fact that these are commuting algebras has a crucial impact on the construction of the null states of $\mathcal{O}$. 
It implies that if $\ket{\mathbf{0}}$ is a null state of $\mathcal{O}$, $K_-\ket{\mathbf{0}}$ and $K_+\ket{\mathbf{0}}$ are also null states of $\mathcal{O}$.
Thus, one can construct null states with a larger number of particles, by adding a pair of delocalized particles via the operator $K_+$.
Alternatively, if we apply $K_-$ to a null state from a symmetry sector with a finite $N_p$ and particle number $N$, we obtain a null state with $N_p-1$ and particle number $N-2$.

In the null states of the directed tunneling operator, one expects that the particles will accumulate at the boundary.
However, the particles forming the pairs are correlated in their position within a delocalized superposition. 
Thus, the centro-symmetric pairs of particles are not affected by boundary accumulation, but remain fully delocalized. This result has a profound effect on the density distribution of the particles. 
We can also understand this effect from the fact that the pairing algebras commute with the similarity transformation, Eq.~\eqref{eq:transformation}, (see \ref{app:potential}).
We discuss in more detail the role of the pairs in the density distribution in Sec.~\ref{sec:results_density}.

For both the fermionic and the bosonic pairing algebras, the corresponding Casimir operators, $\Omega_K^\text{(f)}$ and $\Omega_K^\text{(b)}$, commute not only with $\mathcal{O}$ and $\mathcal{O}_+$, but also with $\mathcal{O}^\dagger$.
Hence, $\Omega_K$ commutes with the Hamiltonian, $H$, and trivially with the jump operator $a$, and, thus, it determines a strong symmetry of the open hybrid system, Eq.~\eqref{eq:Lindblad}.

We have shown in Sec.~\ref{sec:results_symmetries} and Sec.~\ref{sec:pairing_algebra} that the null states of $\mathcal{O}$ are labeled by two quantum numbers, the center-of-mass $X$ and the number of pairs $N_p$.
However, for the Liouvillian defined in  Eq.~\eqref{eq:Lindblad}, the two quantum numbers play a very different role. 
The center-of-mass $X$ only labels the states within the decoherence-free subspace; in contrast, the number of pairs $N_p$ is a conserved quantity of the entire evolution [realizing the decoherence-free subspace structure sketched in Fig.~\ref{fig:DFS}(c)].
While in Sec.~\ref{sec:results_steady_states} we treat the two quantum numbers on an equal footing when discussing the properties of the null state of $\mathcal{O}$, in Sec.~\ref{sec:results_tmps}, we emphasize their very different role in the dissipative dynamics.

\section{Properties of the steady states \label{sec:results_steady_states}}

In this section, we discuss the physical behavior of the steady states, Eq.~\eqref{eq:rho}, of the open hybrid system, Eq.~\eqref{eq:Lindblad}. 
We focus on how the symmetry properties, induced by the Casimir operators of the algebras presented above, manifest themselves in observables. 
We first analyze the properties of the null states of the directed tunneling operator $\mathcal{O}$, Eq.~\eqref{eq:operator}, which are pure states. 
The mixed states reached in the long-time dissipative dynamics starting from generic initial states are in turn analyzed in Sec.~\ref{sec:results_tmps}.
In the following, we present the most striking properties of the null states spanning the decoherence-free subspace: the density profiles, Sec.~\ref{sec:results_density}, long-range kinetic and pairing correlations, Sec.~\ref{sec:results_corr} and entanglement scaling, Sec.~\ref{sec:results_entanglement}.

Before presenting the results, we first comment on obtaining the null states of $\mathcal{O}$ and how we present the results for the different observables.
The states obtained with the analytical exact formulas, Eq.~\eqref{eq:q_r_decomposition} and Eq.~\eqref{eq:result_fermions}, are not eigenstates of the Casimirs of the spinless $\eta$-pairing algebras, Sec.~\ref{sec:pairing_algebra}.
In order to discuss the results in the symmetry sectors of both the spin algebra and the pairing algebra, we obtain the null states by performing exact diagonalization employing the symmetries generated by the Casimirs of the two algebras.
We use this approach in both the bosonic and fermionic cases. We note that for the fermionic case, the exact diagonalization calculation is done in the Hilbert space of hardcore bosonic particles, making use of the Jordan-Wigner transformation.
We show results for system sizes and symmetry sectors for the bosonic and fermionic cases that are related by the boson-fermion correspondence, Sec.~\ref{sec:bosons_to_fermions}.

The symmetry sectors are labeled by two quantum numbers, the center-of-mass, $X$, Eq.~\eqref{eq:sl2_algebra}, and the number of pairs, $N_p$, determining the eigenvalues of the pairing Casimir $\Omega_K$,  [Eq.~\eqref{eq:casimir_pairing_bosons} for bosons and Eq.~\eqref{eq:casimir_pairing_SU2} for fermions].
Most symmetry sectors have multiple null states, thus, in the numerical exact diagonalization procedure one usually obtains a particular basis within the sector.
In order to ensure that the states we analyze represent a random basis within the sector, we first apply a Haar unitary on the states obtained from the numerical routine.
In order to improve the presentation of the results and to focus on the main physical features of the states, we present averages over the expectation values taken with each basis state. 
Thus, we define the average of an observable $M$ in a symmetry sector $(X,N_p)$
\begin{align} 
\label{eq:average}
\langle \langle M \rangle \rangle =\frac{1}{N_s} \sum_{i=1}^{N_s} \bra{X,N_p; i} M \ket{X,N_p; i},
\end{align}
where $i$ runs over all $N_s$ null states in the sector. 
In the presented results we also show the standard deviation of this averaging procedure as error bars, in order to quantify the spread of the values of the observables.
In \ref{app:numerical_results} we also show results for each state separately for certain observables.

\subsection{Symmetry-constrained Liouvillian skin effect \label{sec:results_density}}

Our framework, Eq.~\eqref{eq:Lindblad}, represents a controlled approach to generate the physics of nonreciprocal quantum systems \cite{MetelmannClerk2015, SiebererDiehl2025, BarzanjehRabl2026, FruchartVitelli2026}. This leads us to a novel type of many-body Liouvillian skin effect. 
While our model, Eq.~\eqref{eq:Lindblad}, contains a fully reciprocal coupling between the photonic and matter degrees of freedom, the dissipative dynamics induced by the photonic loss induces an effective nonreciprocal dynamics.
This is underlined by the fact that steady states are constructed from the null eigenstates of a non-Hermitian operator $\mathcal{O}$.

\begin{figure}[!hbtp]
\centering
\includegraphics[width=0.49\textwidth]{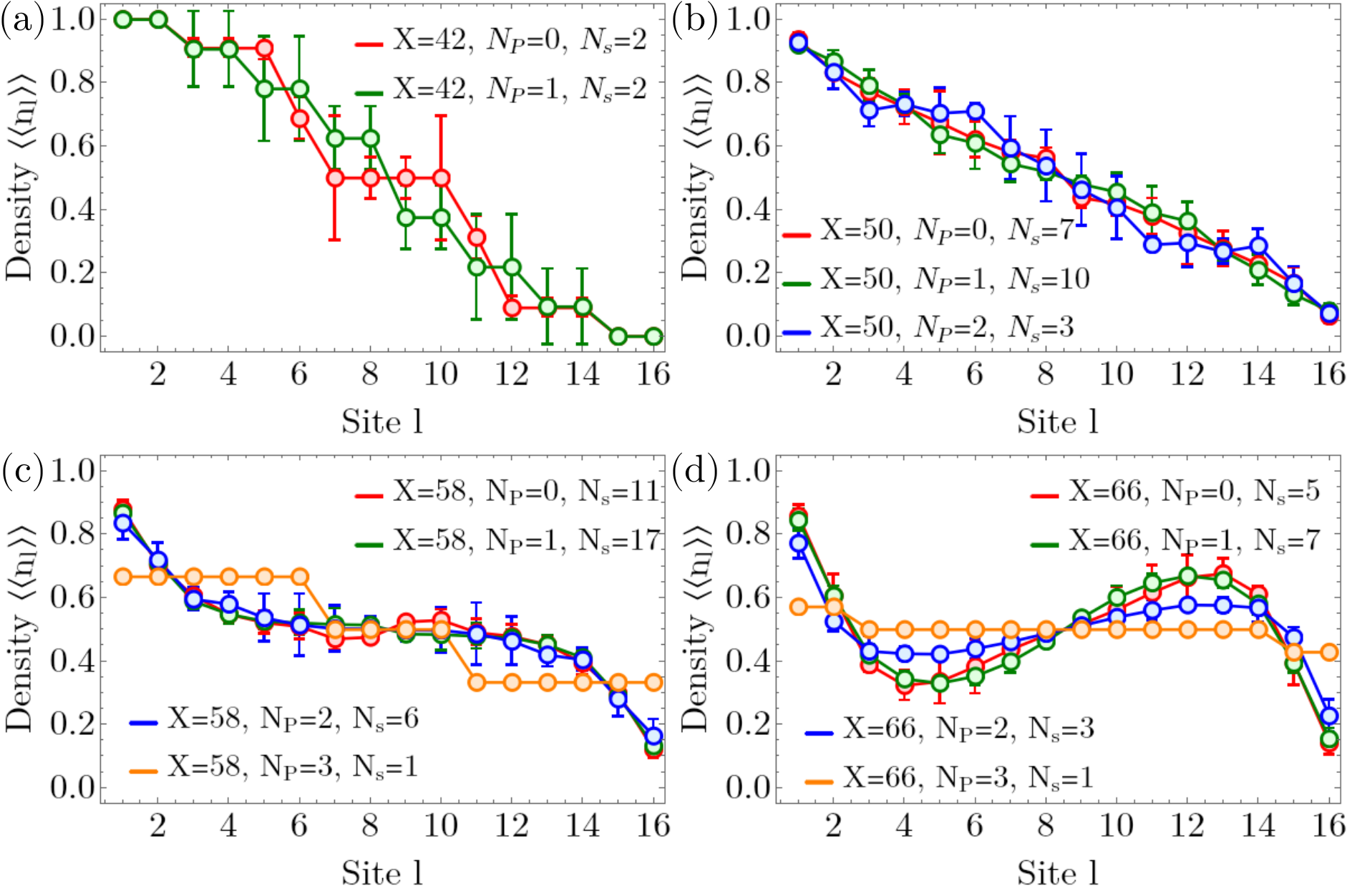}
\caption{
The average density profile, $\langle \langle n_i \rangle \rangle$, of the null states $\mathcal{O}^\text{(f)}$ for fermionic particles for $N=8$ particles in $L=16$ sites for several symmetry sectors: (a) $X=42$, (b) $X=50$, (c) $X=58$, (d) $X=66$. 
Each panel contains several curves corresponding to the possible values of the pairing quantum number $N_p$. 
As discussed around Eq.~\eqref{eq:average}, the density profiles shown are obtained by averaging over the expectation values of a random basis in each $(X, N_p)$ symmetry sector, while the number of states in each sector is $N_s$. The error bars represent the standard deviation stemming from the averaging over the random basis. The density profile of each state from panel (c) is shown in \ref{app:numerical_results}.
}
\label{fig:hardcore_density}
\end{figure}

In the following, we first briefly review previous instances of non-Hermitian and Liouvillian skin effects, which we use to contrast the \emph{symmetry-constrained Liouvillian skin effect} introduced in this work. 
Afterwards we discuss how the fermionic and bosonic particles accumulate at the boundary in the null states of $\mathcal{O}$. We emphasize the role of the constrains imposed by the symmetry structure, which lead to density profiles in stark contrast to exponential boundary localization. 
After focusing on the fermionic case, we highlight the role of the particle statistics by presenting the differences that appear for bosonic particles. 
We conclude this section with a brief comment on the role of the boundary conditions.

\begin{figure}[!hbtp]
\centering
\includegraphics[width=0.48\textwidth]{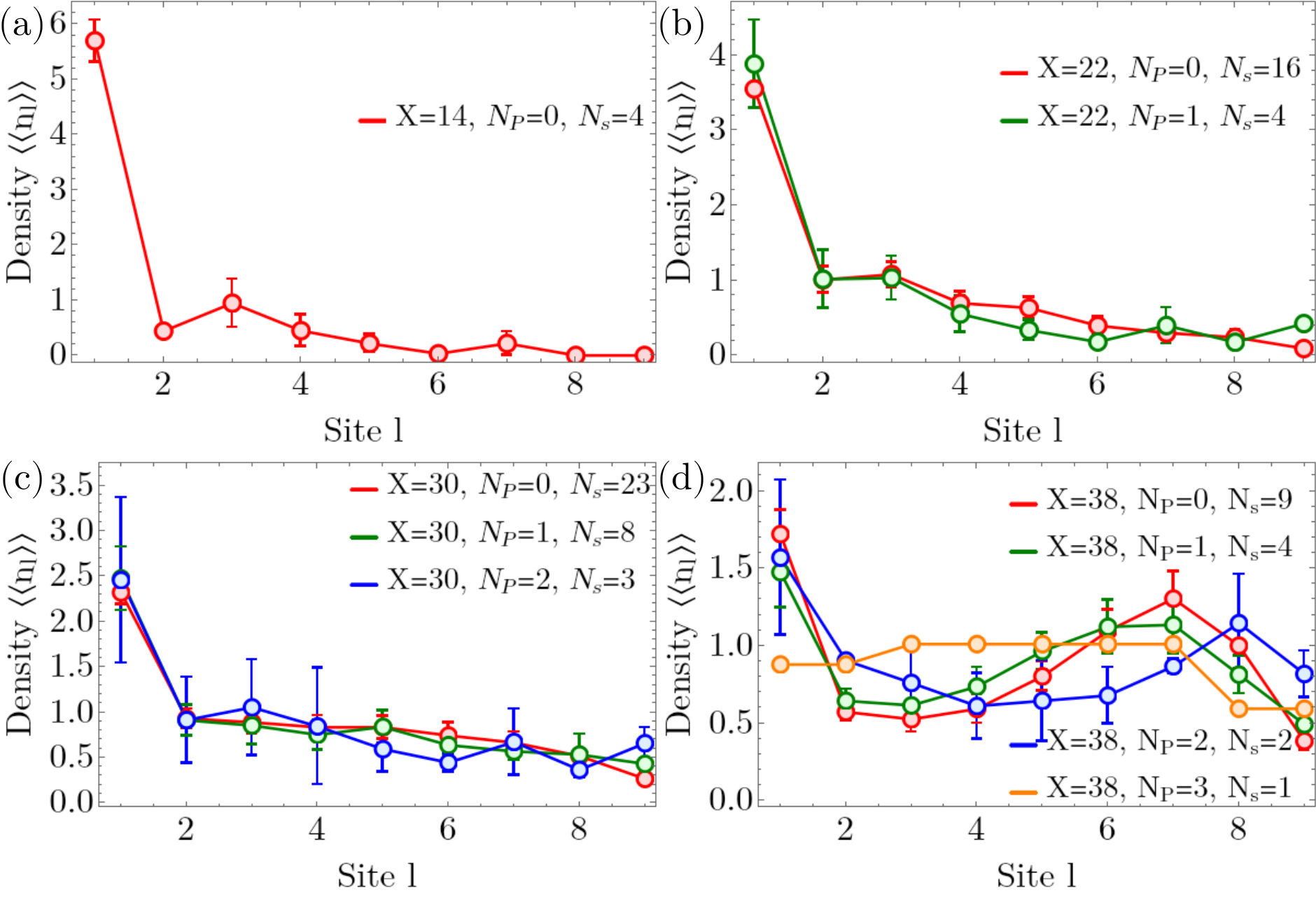}
\caption{
The average density profile, $\langle \langle n_i \rangle \rangle$, of the null states of $\mathcal{O}^\text{(b)}$ for bosonic particles for $N=8$ particles in $L=9$ sites for several symmetry sectors: (a) $X=14$, (b) $X=22$, (c) $X=30$, (d) $X=38$. Each panel contains several curves corresponding to the possible values of the pairing quantum number $N_p$. 
The results are obtained with the same procedure as for Fig.~\ref{fig:hardcore_density}.
}
\label{fig:bosons_density}
\end{figure}

\emph{Non-Hermitian skin effects}. 
We consider unidirectional tunneling processes, Eq.~\eqref{eq:operator}, related to the non-Hermitian Hatano-Nelson model of imbalanced tunneling \cite{HatanoNelson1996}. 
The Hatano-Nelson, and related models, exhibit the non-Hermitian skin effect \cite{AshidaUeda2020, BergholtzKunst2021, GohsrichKunst2025, BrighiNunnenkamp2024}. The skin effect manifests as an exponential accumulation of particles at the boundary and extreme sensitivity between open and periodic boundary conditions.
One can expect similar phenomenology also for our case; however, as we discuss in the following with respect to Fig.~\ref{fig:hardcore_density} and Fig.~\ref{fig:bosons_density}, important differences appear.
One first technical difference is that in the Hatano-Nelson model one performs an imaginary gauge transformation to map the non-Hermitian model to a Hermitian Hamiltonian and compute the spectrum \cite{HatanoNelson1996}. This approach fails in the fully unidirectional situation of Eq.~\eqref{eq:operator}. 
Thus, for us a very different symmetry structure emerges from the two commuting algebras, as discussed in Sec.~\ref{sec:results_symmetries} and Sec.~\ref{sec:pairing_algebra}.

An approach to physically realize non-Hermitian phenomena is through GKSL master equations in the presence of dissipative processes. 
In such settings the equivalent Liouvillian skin effect has been discussed \cite{SongWang2019, LiuChen2020, HagaUeda2021, YangBergholtz2022, HamanakaYoshida2023, EkmanBergholtz2024, CaiDong2025}.
This Liouvillian skin effect has the same main feature as in the case of non-Hermitian Hamiltonians, namely eigenmodes localized at the boundaries for open boundary conditions, which become extended in the periodic case.
In the following, we show how the presence of symmetries in the decoherence-free subspace results in an inherently different type of Liouvillian skin effect.

\emph{Symmetry-constrained Liouvillian skin effect}. 
The null states of the unidirectional tunneling operator $\mathcal{O}$, Eq.~\eqref{eq:operator}, exhibit the expected tendency towards localization at the boundary. However, the constraints resulting from the fixed quantum numbers of the symmetries have a crucial impact on the density profiles.
It is important to note that the symmetries have a non-trivial structure only in the many-body regime. For a single particle the fully localized state is the unique null state of $\mathcal{O}$ (see Sec.~\ref{sec:itinerant}).
To analyze how the particles accumulate at the edge of the chain we depict the density profiles for fermions in Fig.~\ref{fig:hardcore_density} and bosons in Fig.~\ref{fig:bosons_density}. 
Each panel contains results for a fixed value of the center-of-mass $X$ and different values for the number for pairs $N_p$.

For fermions the fully localized state corresponds the first $N$ sites of the chain being occupied (sector $X=36$ for the parameters in Fig.~\ref{fig:hardcore_density}).
By increasing the center-of-mass quantum number the fermions start spreading away from the boundary, e.g.~in Fig.~\ref{fig:hardcore_density}(a) for $X=42$ only the first two sites are fully filled, while in Fig.~\ref{fig:hardcore_density}(b) for $X=50$ the density profile decays almost linearly away from the boundary.
This is in stark contrast with the exponential localization observed in other models exhibiting the skin effect. 
Only in higher dimensions has algebraically decaying accumulation  been discussed \cite{ZhangSun2025}.

\emph{Potential from similarity transformation}.
The similarity transformation, Eq.~\eqref{eq:transformation}, mapping the $\mathcal{O}$ algebra, Eq.~\eqref{eq:sl2_algebra}, to the $Q$ algebra, Eq.~\eqref{eq:sl2_Q_algebra}, effectively creates a non-linear potential for the density of particles (see \ref{app:similarity}). 
The potential has two minima, one at the boundary and one in the second half of the chain.
Our results show that the second minimum can be populated for sectors with higher values of the center-of-mass.
We observe that by increasing the values of the center-of-mass the density distribution becomes flatter in the center of the chain, Fig.~\ref{fig:hardcore_density}(c). 
For large enough $X$, a second peak far away from the accumulation boundary appears, Fig.~\ref{fig:hardcore_density}(d).
In \ref{app:numerical_results} we show the density profiles of the $Q$ algebra. While they exhibit a slow decay away from the boundary, with the slope determined by the value of the center-of mass, there is no prominent second peak present as in Fig.~\ref{fig:hardcore_density}(d).

\emph{$\eta$-pairing symmetry}.
The $\eta$-paired particles, unlike their unpaired counterparts, are delocalized throughout the chain and do not accumulate at the boundary. 
This originates in the symmetry structure of our model, as the spinless $\eta$-pairing algebra, Eq.~\eqref{eq:pairing_SU2_fermions}, commutes with both the spin algebra, Eq.~\eqref{eq:sl2_algebra} and with the similarity transformation, Eq.~\eqref{eq:transformation}.  
Thus, the pairing quantum number has a strong impact on the density profile, especially in the sectors in which a large number of the particles are paired [orange curves in Fig.~\ref{fig:hardcore_density}(c) and Fig.~\ref{fig:hardcore_density}(d)].
In these cases the fermionic density has an almost flat profile without a clear accumulation at the boundary.
Furthermore, the $K_z=0$ value in the fermionic $\eta$-pairing algebra occurring at half-filling determines a particle-hole plus inversion symmetry in the density profiles, Fig.~\ref{fig:hardcore_density}.

\emph{Bosonic accumulation}.
In Fig.~\ref{fig:bosons_density} we show the results obtained for bosonic atoms, which exhibit similar features as discussed above for fermionic atoms. 
The main differences, in contrast to fermions, originate from the different nature of the localized state. 
For bosons in the localized state, all atoms are on the first site of the lattice [corresponding to $X=8$ for the parameters in Fig.~\ref{fig:bosons_density}].
While a density peak at the first site is present even for large center-of-mass sectors, its height is drastically reduced, e.g.~from $\langle\langle n_1\rangle\rangle\approx 6$ for $X=6$ in Fig.~\ref{fig:bosons_density}(a) to $\langle\langle n_1\rangle\rangle\lesssim 2$ for $X=38$ in Fig.~\ref{fig:bosons_density}(d), with the other particles spreading into the chain.
Similarly to the fermionic case, at large $X$ a second density peak appears in the second half of the chain [see Fig.~\ref{fig:bosons_density}(d)] due to the potential induced by the similarity transformation, Eq.~\eqref{eq:transformation}. 
The presence of multiple pairs further flattens the density profile also in the bosonic case [Fig.~\ref{fig:bosons_density}(c) and Fig.~\ref{fig:bosons_density}(d)].

\emph{Periodic boundary conditions}.
We briefly discuss the effect of boundary conditions for our setup. The cavity-matter open model, Eq.~\eqref{eq:Lindblad} and Eq.~\eqref{eq:operator}, exhibits a profound sensitivity to the choice of boundary conditions. 
However, this goes beyond the change of localized to extended nature of the zero eigenmodes of the Liouvillian.
The boundary conditions have a deep impact on the symmetry structure of the problem, altering the phenomenology of the system.
For periodic boundary conditions, the operator $\mathcal{O}=\sum_{j=1}^{L} d^\dagger_j d_{j+1}$ commutes with its Hermitian conjugate, as they can be both diagonalized in the momentum basis. 
This implies the case of Fig.~\ref{fig:DFS}(a), Sec.~\ref{sec:recipe_1}, and the steady states of the hybrid system are given by Eq.~\eqref{eq:rho_symmetry}. 
Each momentum Fock state of the particles is a steady state of the dissipative dynamics, as $\mathcal{O}$ is the generator of a strong symmetry (see case 1.~in the classification made in Sec.~\ref{sec:recipe_1}). 
This implies a drastic change in number and nature of the steady states compared to the open boundary conditions case in which the steady states are constructed from the null states of $\mathcal{O}$. 
As the symmetry structures of the model have such a different nature depending on the boundary conditions, we expect that also the physical behavior of the observables will differ considerably.

To summarize, we establish the \emph{symmetry-constrained Liouvillian skin effect} occurring in dissipative dynamics with emergent non-reciprocal tunneling processes, where in the many-body regime the symmetries of the Liouvillian and of the decoherence-free subspace constrain the edge accumulation profile of the particles.

\subsection{Long-range kinetic and pairing correlations \label{sec:results_corr}}

We analyze the long-range quantum coherence properties of the null states of $\mathcal{O}$ by investigating the behavior of complex four-body quantum correlations.
We focus on two types of correlations, closely related to the underlying symmetry properties of our system: long-range \emph{kinetic} and \emph{pairing} correlations.
We show that the null states exhibit correlations both at short- and long-range. 
We identify the origin of the strong correlations in the symmetry properties stemming from the algebraic structures described in Sec.~\ref{sec:results_symmetries} and Sec.~\ref{sec:pairing_algebra}.

\begin{figure}[!hbtp]
\centering
\includegraphics[width=0.49\textwidth]{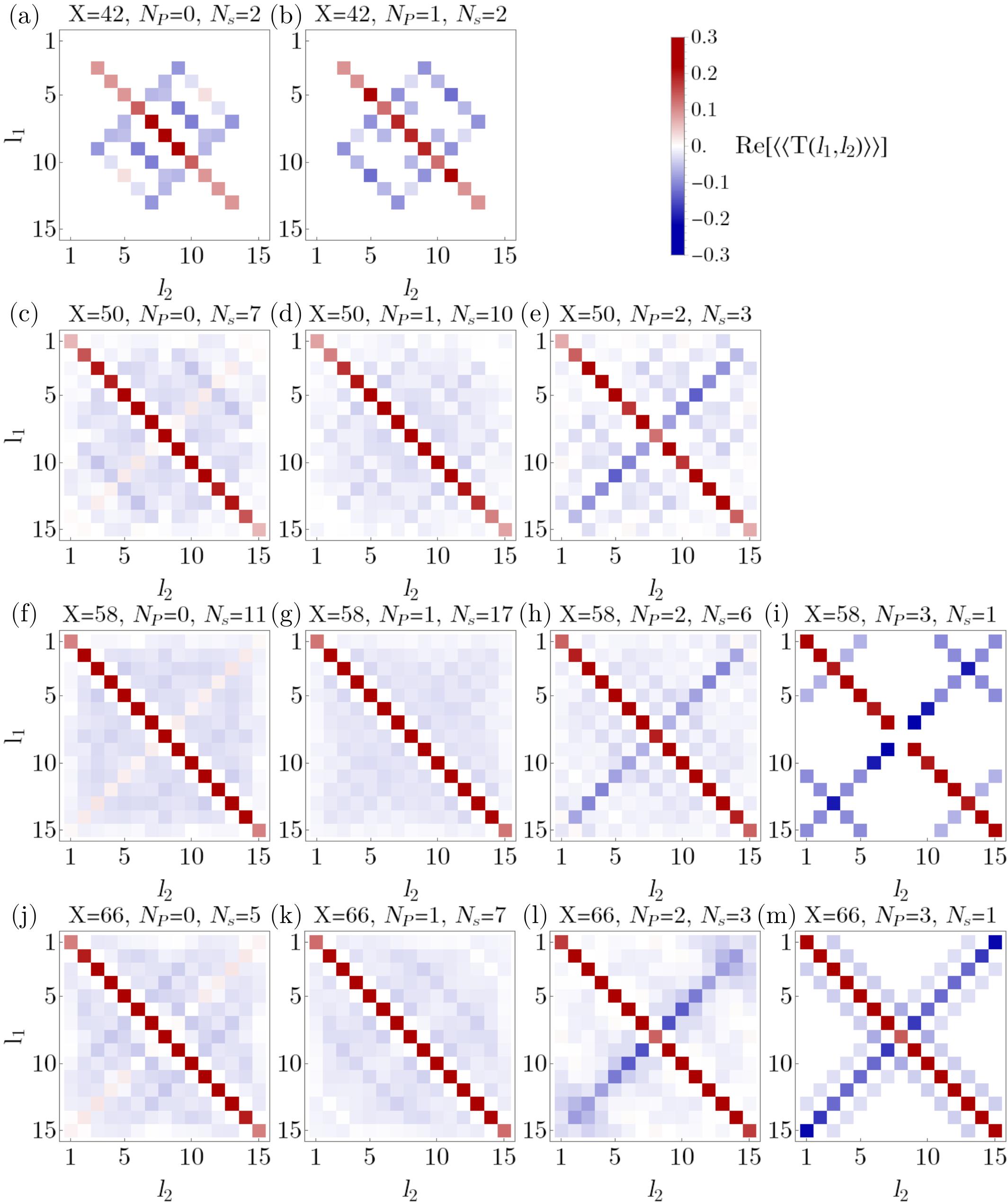}
\caption{
The average kinetic correlations $\Re[\langle\langle T(l_1,l_2)\rangle\rangle]$, defined in Eq.\eqref{eq:kinetic_corr}, of the null states $\mathcal{O}^\text{(f)}$ for fermionic particles for $N=8$ particles in $L=16$ sites for several symmetry sectors, defined by their center-of-mass, $X$, and number of pairs, $N_p$, quantum numbers. The sectors considered are the same as in Fig.~\ref{fig:hardcore_density}. 
As discussed around Eq.~\eqref{eq:average}, the results shown are obtained by averaging over the expectation values of a random basis in each $(X, N_p)$ symmetry sector, the number of states in each sector is $N_s$.
}
\label{fig:hardcore_kinetic_corr}
\end{figure}

\begin{figure}[!hbtp]
\centering
\includegraphics[width=0.48\textwidth]{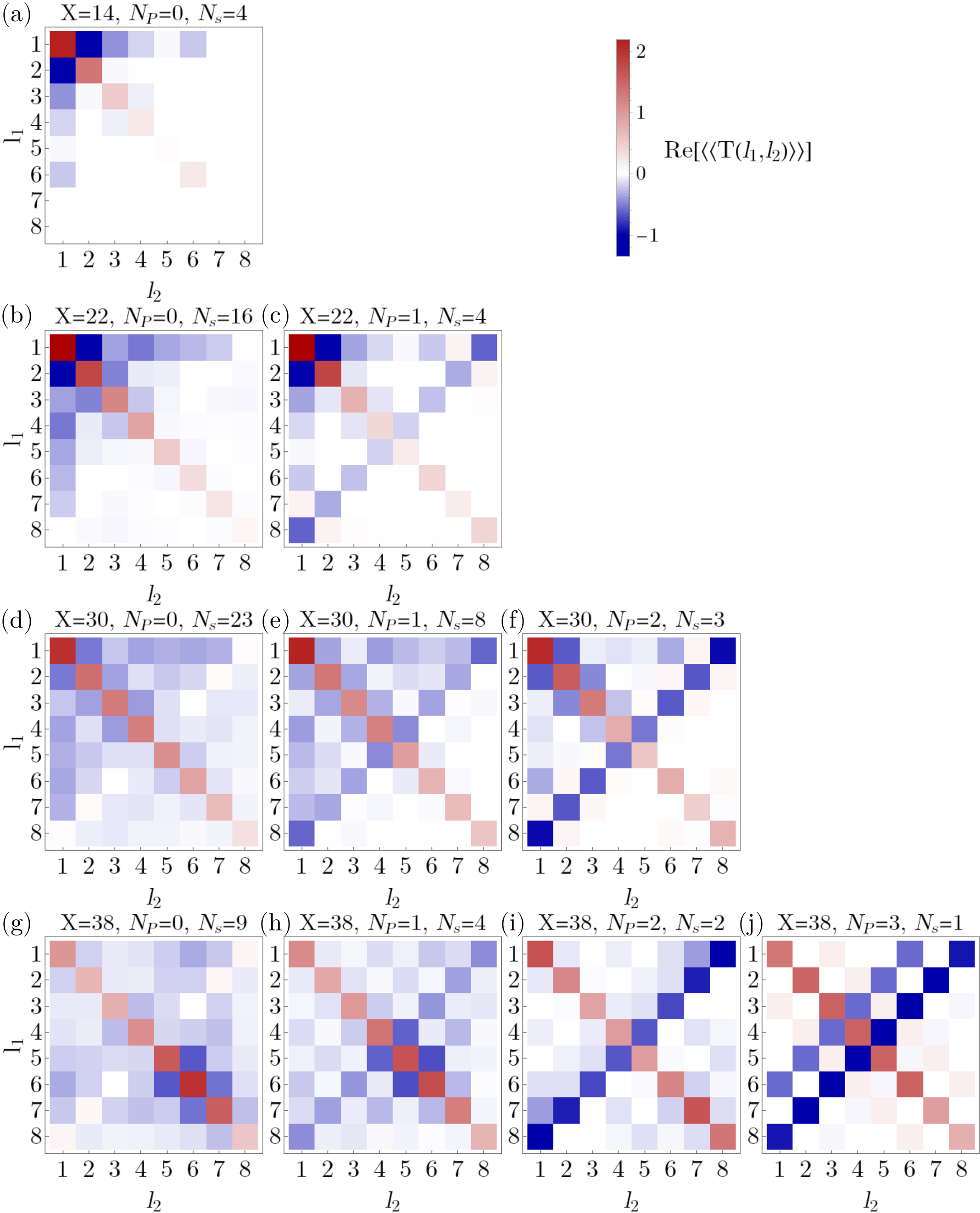}
\caption{
The average kinetic correlations $\Re[\langle\langle T(l_1,l_2)\rangle\rangle]$, defined in Eq.\eqref{eq:kinetic_corr}, of the null states $\mathcal{O}^\text{(b)}$ for bosonic particles for $N=8$ particles in $L=9$ sites for several symmetry sectors, defined by their center-of-mass, $X$, and number of pairs, $N_p$, quantum numbers. The sectors considered are the same as in Fig.~\ref{fig:bosons_density}. 
The results are obtained with the same procedure as for Fig.~\ref{fig:hardcore_kinetic_corr}.
}
\label{fig:bosons_kinetic_corr}
\end{figure}

We define the \emph{kinetic correlations} as 
\begin{align} 
\label{eq:kinetic_corr}
T(l_1,l_2)&=d^\dagger_{l_1+1} d_{l_1}d^\dagger_{l_2} d_{l_2+1},
\end{align}
describing two correlated tunneling events occurring on the bonds $(l_1, l_1+1)$ and $(l_2, l_2+1)$, at a distance $|l_1-l_2|$.
Here, $d_l$ may be a fermionic or a bosonic operator.
Summing $T(l_1,l_2)$ over both $l_1$ and $l_2$ yields the unweighted equivalent to the $\mathcal{O}_+\mathcal{O}$ term of the $\Omega_\mathcal{O}$ Casimir from Eq.~\eqref{eq:casimir_sl2}. 
This term vanishes for the null states of $\mathcal{O}$, thus, we can expect this constraints to have a significative impact of the structure of the expectation value of the correlation matrix $T(l_1,l_2)$.
The diagonal terms, $T(l_1=l,l_2=l)$, reduce to density-density correlations between the sites $l$ and $l+1$. 
For fermions  $T(l_1=l,l_2=l)$ measures if tunneling can occur from site $l+1$ to site $l$. Similarly, the first diagonal $|l_1-l_2|=1$ always vanishes for fermions, but it can be finite for bosons.

We present the results for the kinetic correlation matrix, $\langle\langle T(l_1,l_2)\rangle\rangle$, Eq.~\eqref{eq:kinetic_corr}, first focusing on the fermionic case, Fig.~\ref{fig:hardcore_kinetic_corr}, and following the structure outlined below:
(i) spatial extension of the correlations for small center-of-mass;
(ii)  uniform spreading of the kinetic correlations by increasing $X$;
(iii) role of the paired particles;
(iv) impact of the particle statistics, differences occuring for the bosonic case, Fig.~\ref{fig:bosons_kinetic_corr}.

The kinetic correlations are restricted to the middle of the chain if the center-of-mass has a small value, e.g.~Fig.~\ref{fig:hardcore_kinetic_corr}(a)-(b) for $X=42$. 
This is due to the fact that the first sites are fully filled, while the last sites are empty [density profile in Fig.~\ref{fig:hardcore_density}(a)]. 
The impact of the blocked sites can be already seen from the vanishing values of the density-density correlations on the diagonal of $\langle\langle T(l_1,l_2)\rangle\rangle$.
However, for the bonds on which tunneling is possible coherence is present, quantified by the finite offdiagonal values for $\langle\langle T(l_1,l_2)\rangle\rangle$.

The range of the correlations is extended for symmetry sectors corresponding to larger values of $X$, with the correlations having an almost uniform spatial distribution, e.g.~Fig.~\ref{fig:hardcore_kinetic_corr}(f)-(g) for $X=58$.
The particle-hole together with inversion emergent symmetry determines a symmetric distribution of the fermionic correlations.
This determines the extension of the finite values of the kinetic correlations also towards the boundary of the chain with a smaller occupancy.
Furthermore, as discussed above, the null state nature determines that the positive diagonal entries of $\langle\langle T(l_1,l_2)\rangle\rangle$ are exactly canceled by the mostly negative offdiagonal kinetic correlations.

The presence of pairs of particles has a strong impact on the long-range kinetic correlations. 
We can see this impact in the emergence of strong correlations between bonds symmetric with respect to the center of the chain [antidiagonal correlations in Figs.~\ref{fig:hardcore_kinetic_corr}(e), (h), (l) for $N_p=2$]. 
The centro-symmetric paired particles need to tunnel together in a correlated fashion in order to remain paired, i.e.~$j\!\to\! j\!+\!1$ is correlated with $L\!+\!1\!-\!j\!\to\! L\!-\!j$.
The antidiagonal correlations become dominant when most of the particles are paired [Figs.~\ref{fig:hardcore_kinetic_corr}(i), (m) for $N_p=3$]. 
The finite entries away from the diagonal and antidiagonal for $N_p=3$ are due to the motion of the few unpaired particles.

\begin{figure}[!hbtp]
\centering
\includegraphics[width=0.49\textwidth]{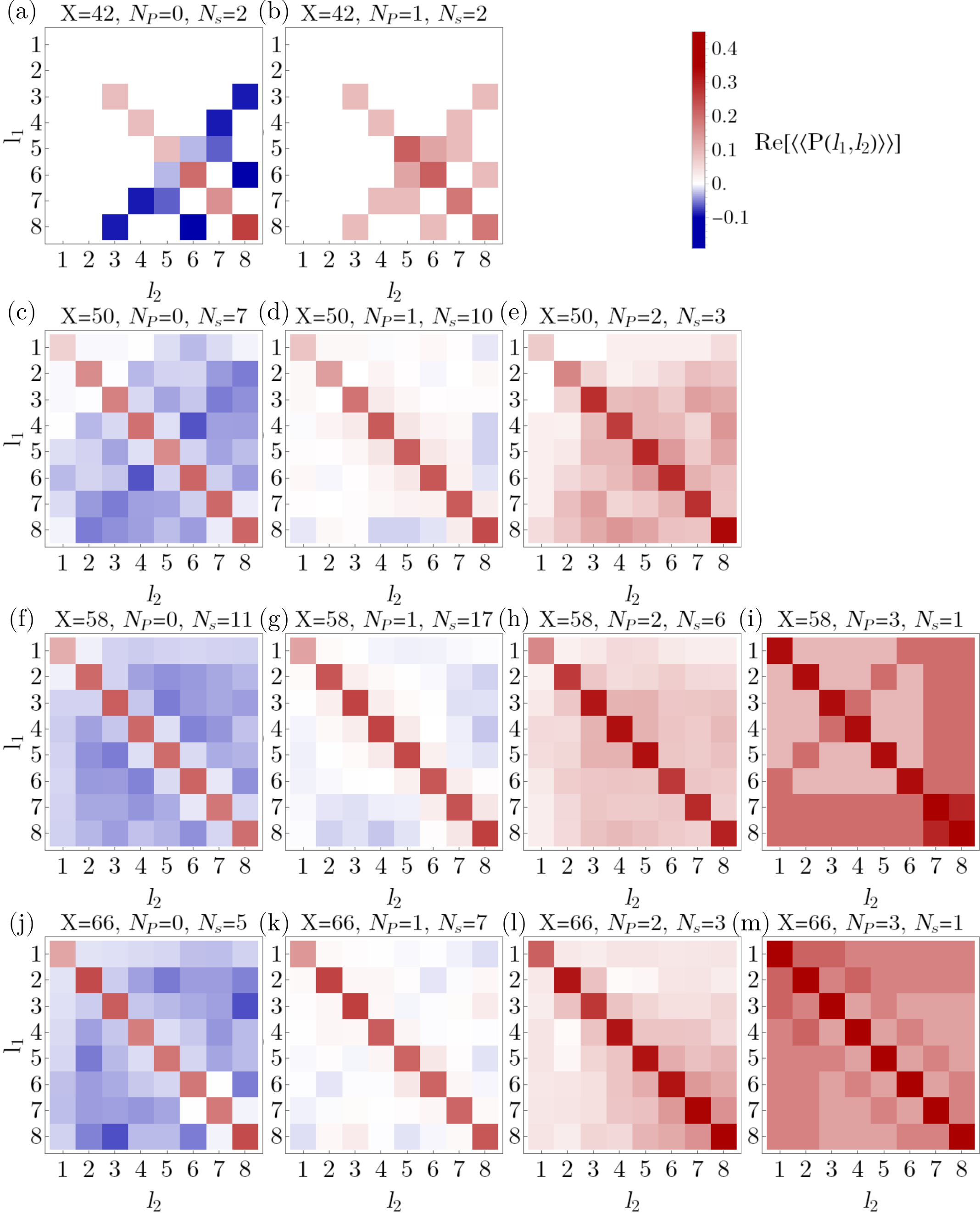}
\caption{
The average fermionic $\eta$-pairing  correlations $\Re[\langle\langle P(l_1,l_2)\rangle\rangle]$, defined in Eq.~\eqref{eq:pairing_corr}, of the null states $\mathcal{O}^\text{(f)}$ for fermionic particles for $N=8$ particles in $L=16$ sites for several symmetry sectors, defined by their center-of-mass, $X$, and number of pairs, $N_p$, quantum numbers. The sectors considered are the same as in Fig.~\ref{fig:hardcore_density}. 
The results are obtained with the same procedure as for Fig.~\ref{fig:hardcore_kinetic_corr}.
}
\label{fig:hardcore_pairing_corr}
\end{figure}

The main difference in the bosonic case, Fig.~\ref{fig:bosons_kinetic_corr}, is the asymmetric spatial distribution of the kinetic correlations $\langle\langle T(l_1,l_2)\rangle\rangle$. This stems from the pronounced accumulation of the bosonic particles at the boundary (see Fig.~\ref{fig:bosons_density}).
Similar to the fermionic case, the extent of the kinetic correlations is controlled by the value of the center-of-mass $X$.
However, due to the large bosonic particle number on the first site, by increasing $X$ the correlations spread first for $l_1=1$ and $l_2>1$ and only afterwards finite correlations between sites in the bulk emerge [see Fig.~\ref{fig:bosons_kinetic_corr}(b) for $X=22$ in contrast to Fig.~\ref{fig:bosons_kinetic_corr}(g) for $X=38$].
The strong antidiagonal correlations due to the presence of pairs also occurs for bosons, Fig.~\ref{fig:bosons_kinetic_corr}(f),(i),(j).

\begin{figure}[!hbtp]
\centering
\includegraphics[width=0.49\textwidth]{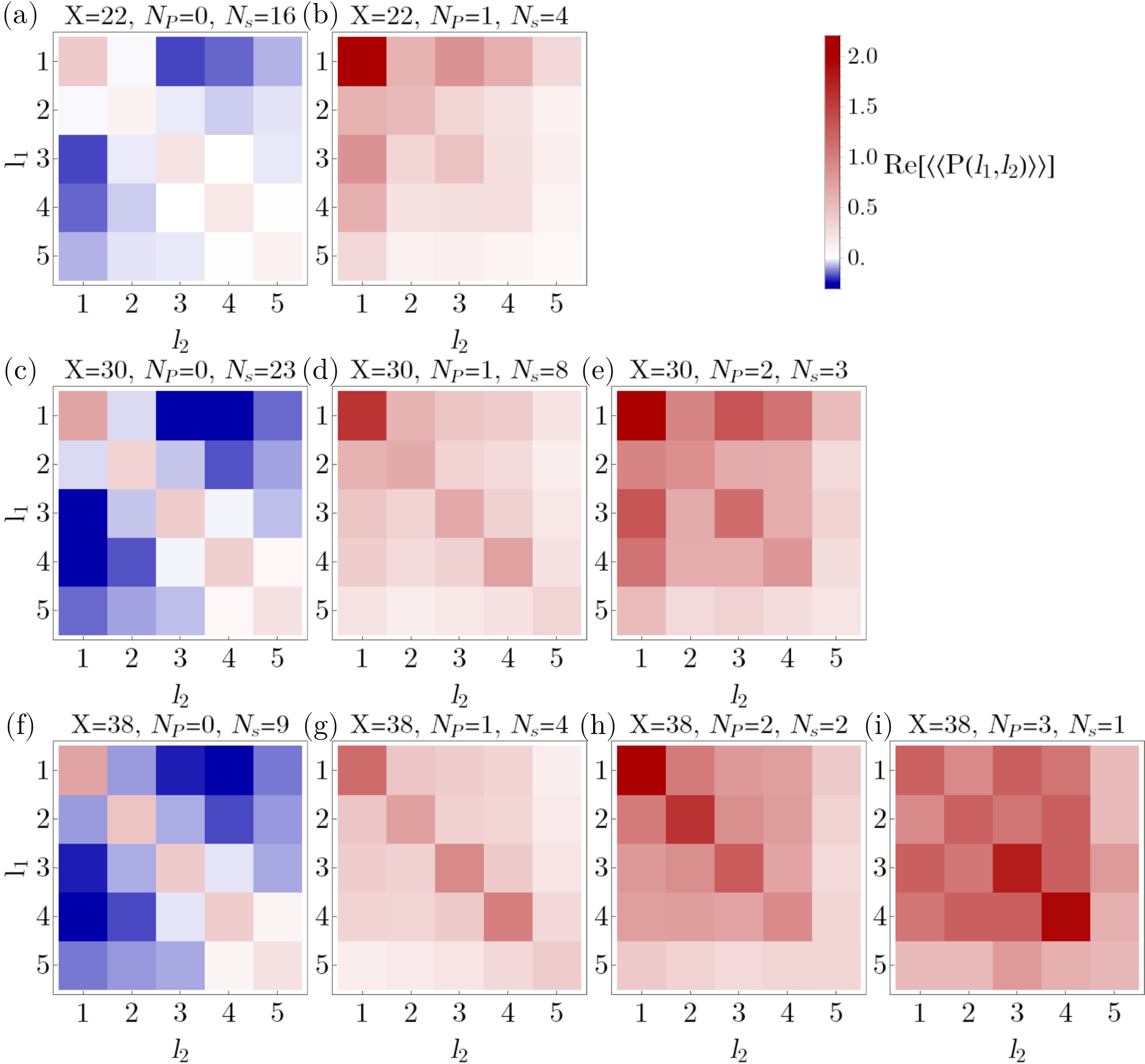}
\caption{
The average bosonic $\eta$-pairing correlations $\Re[\langle\langle P(l_1,l_2)\rangle\rangle]$, defined in Eq.~\eqref{eq:pairing_corr_2}, of the null states $\mathcal{O}^\text{(b)}$ for bosonic particles for $N=8$ particles in $L=9$ sites for several symmetry sectors, defined by their center-of-mass, $X$, and number of pairs, $N_p$, quantum numbers. The sectors considered are the same as in Fig.~\ref{fig:bosons_density}. 
The results are obtained with the same procedure as for Fig.~\ref{fig:hardcore_kinetic_corr}.
}
\label{fig:bosons_pairing_corr}
\end{figure}

The \emph{pairing correlations} are defined in analogy with the spinless $\eta$-pairing algebras described in Sec.~\ref{sec:pairing_algebra}. For the fermionic case we have
\begin{align} 
\label{eq:pairing_corr}
P(l_1,l_2)&=(-1)^{l_1+l_2}c^\dagger_{l_1} c^\dagger_{L+1-l_1}c_{L+1-l_2} c_{l_2},
\end{align}
while for bosons we use
\begin{align} 
\label{eq:pairing_corr_2}
P(l_1,l_2)&=A(l_1,l_2)(-1)^{l_1+l_2}b^\dagger_{l_1} b^\dagger_{L+1-l_1}b_{L+1-l_2} b_{l_2}, \\
\text{with}~&A(l_1,l_2)=1,~\text{for}~l_1\neq\frac{L+1}{2},~l_2\neq\frac{L+1}{2}, \nonumber \\
&A(l_1,l_2)=\frac{1}{2},~\text{either}~l_1=\frac{L+1}{2},~\text{or}~l_2=\frac{L+1}{2}, \nonumber \\
&A(l_1,l_2)=\frac{1}{4},~l_1=l_2=\frac{L+1}{2}. \nonumber
\end{align}
In both cases by summing the pairing correlations matrix, $P(l_1,l_2)$, over both $l_1$ and $l_2$ we obtain the Casimirs of the $\eta$-pairing algebras up to a constant, Sec.~\ref{sec:pairing_algebra}. Thus, the values of the correlations will be constrained by the number of paired particles in the system, $N_p$.
The diagonal entries of $P(l_1,l_2)$ represent the centro-symmetric density-density correlations $n_j n_{L+1-j}$, while the offdiagonal entries represent the coherence between the $\eta$-pairs at different positions.

In the discussion below, we highlight the following key points regarding the behavior of the $\eta$-pairing correlations in the null states of $\mathcal{O}$, $\Re[\langle\langle P(l_1,l_2)\rangle\rangle]$ (see Fig.~\ref{fig:hardcore_pairing_corr} for the fermionic case):
(i) pairing coherence in the absence of pairs;
(ii) spatial distribution of the long-range pairing correlations;
(iv) impact of the particle statistics, differences occuring for the bosonic case (Fig.~\ref{fig:bosons_pairing_corr}).

The diagonal of the pairing matrix $\Re[\langle\langle P(l_1,l_2)\rangle\rangle]$ is given by density-density correlations, Eq.~\eqref{eq:pairing_corr_2}. 
A finite value of the expectation value of $\langle n_j n_{L+1-j} \rangle$ does not imply the existence of paired particles. 
A state is characterized by a finite number of pairs if the value $\langle \Omega_K \rangle$ [defined in Eq.~\eqref{eq:casimir_pairing_SU2} and  Eq.~\eqref{eq:casimir_pairing_bosons}] is larger than the lowest eigenvalue corresponding to $N_p=0$.
This implies that for $N_p=0$ the finite positive values on the diagonal need to be compensated by negative off-diagonal entries in order to obtain a vanishing value of the expectation value of, $\langle K_+ K_- \rangle$.
This can be observed in Figs.~\ref{fig:hardcore_pairing_corr}(a),(c),(f),(j), where even for $N_p=0$ finite correlations can be seen. 
While this is a sign of the non-trivial quantum coherence in the null states, it does not imply the existence of pairs.

\begin{figure}[!hbtp]
\centering
\includegraphics[width=0.48\textwidth]{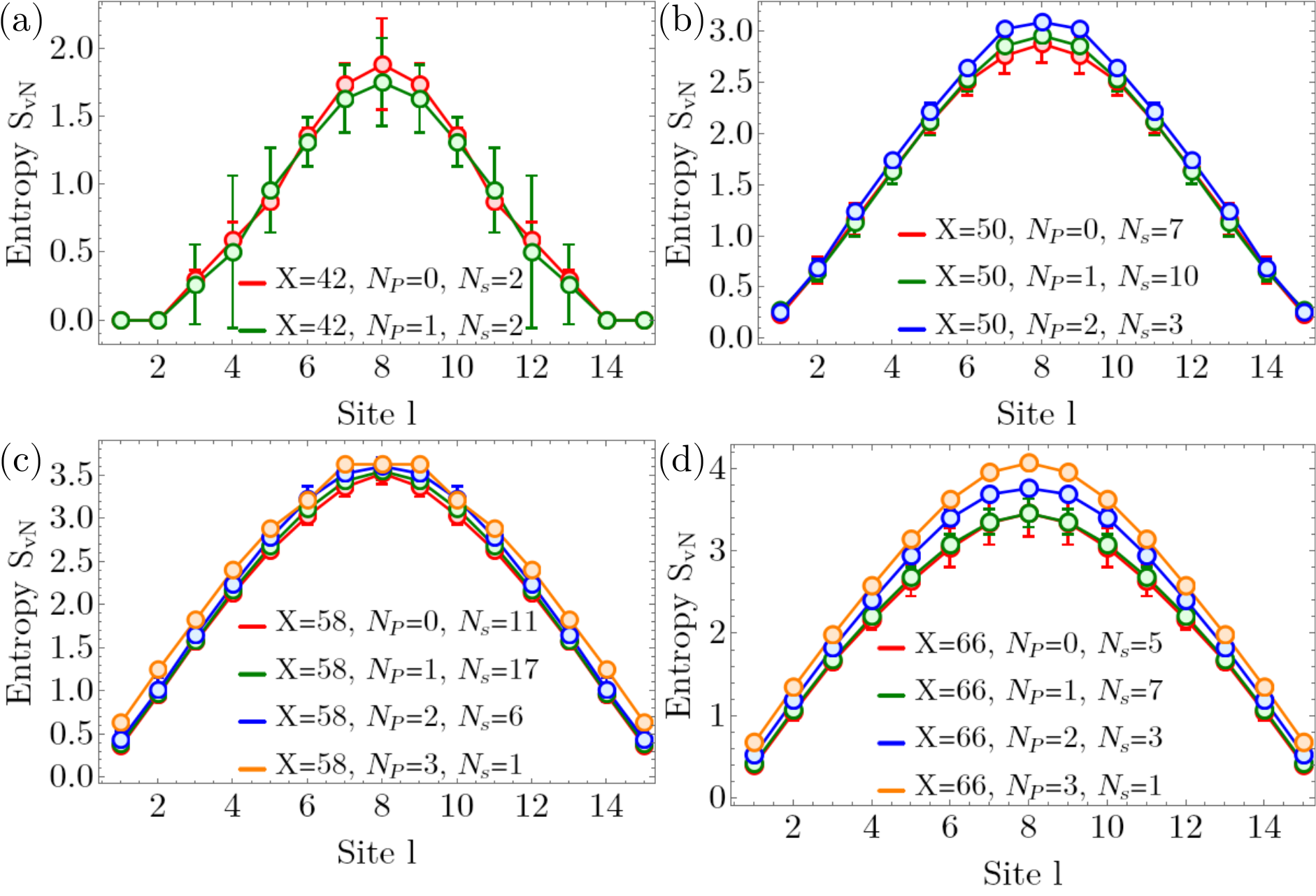}
\caption{
The average entanglement entropy $S_\text{vN}$ of the null states $\mathcal{O}^\text{(f)}$ for fermionic particles for $N=8$ particles in $L=16$ sites for several symmetry sectors: (a) $X=42$, (b) $X=50$, (c) $X=58$, (d) $X=66$. Each panel contains several curves corresponding to the possible values of the pairing quantum number $N_p$. 
The results are obtained with the same procedure as for Fig.~\ref{fig:hardcore_density}.
}
\label{fig:hardcore_entropy}
\end{figure}

An important conclusion of our results is that in the symmetry sector characterized by a finite number of $\eta$-pairs, the pairs are accompanied by long-range coherences, e.g.~Figs.~\ref{fig:hardcore_pairing_corr}(h), (i).
For fermions, the positions of the unpaired particles influence directly the position of the particles forming the $\eta$-pair, through the hardcore constraints.
This interplay of unpaired and paired fermions, can constrain the pairs to have a larger weight in the middle of the chain, e.g.~Fig.~\ref{fig:hardcore_pairing_corr}(h), or a more uniform distribution, e.g.~Fig.~\ref{fig:hardcore_pairing_corr}(l).
However, up to the modulation stemming from the spatial distribution of the pairs, the long-range correlations show weak decay with the distance, e.g.~for $N_p=3$ in Figs.~\ref{fig:hardcore_pairing_corr}(i), (m).

The pairing between bosonic atoms has a similar phenomenology as in the fermionic case, see Fig.~\ref{fig:bosons_pairing_corr}.
The main difference again stems from the larger occupation of the first site in the bosonic case, which determines a larger weight for the pairs which include the first site [Figs.~\ref{fig:bosons_pairing_corr}(b),(d),(g) for $N_p=1$ and Figs.~\ref{fig:bosons_pairing_corr}(e),(h) for $N_p=2$].
For larger values of $X$, the main contribution comes from pairs distributed throughout the chain, with pair coherence over a larger number of sites, Fig.~\ref{fig:bosons_pairing_corr}(i) for $X=38$ and $N_p=3$.

\subsection{Entanglement properties \label{sec:results_entanglement}}

We compute the von Neumann entanglement entropy $S_\text{vN}(l)$ of an embedded subsystem of length $l$ in the chain of length $L$ for the null states of $\mathcal{O}$.
Our results show that besides the fully localized sectors, the null states of $\mathcal{O}$ exhibit finite values of the entanglement entropy scaling with the length of the embedded subsystem.

\begin{figure}[!hbtp]
\centering
\includegraphics[width=0.48\textwidth]{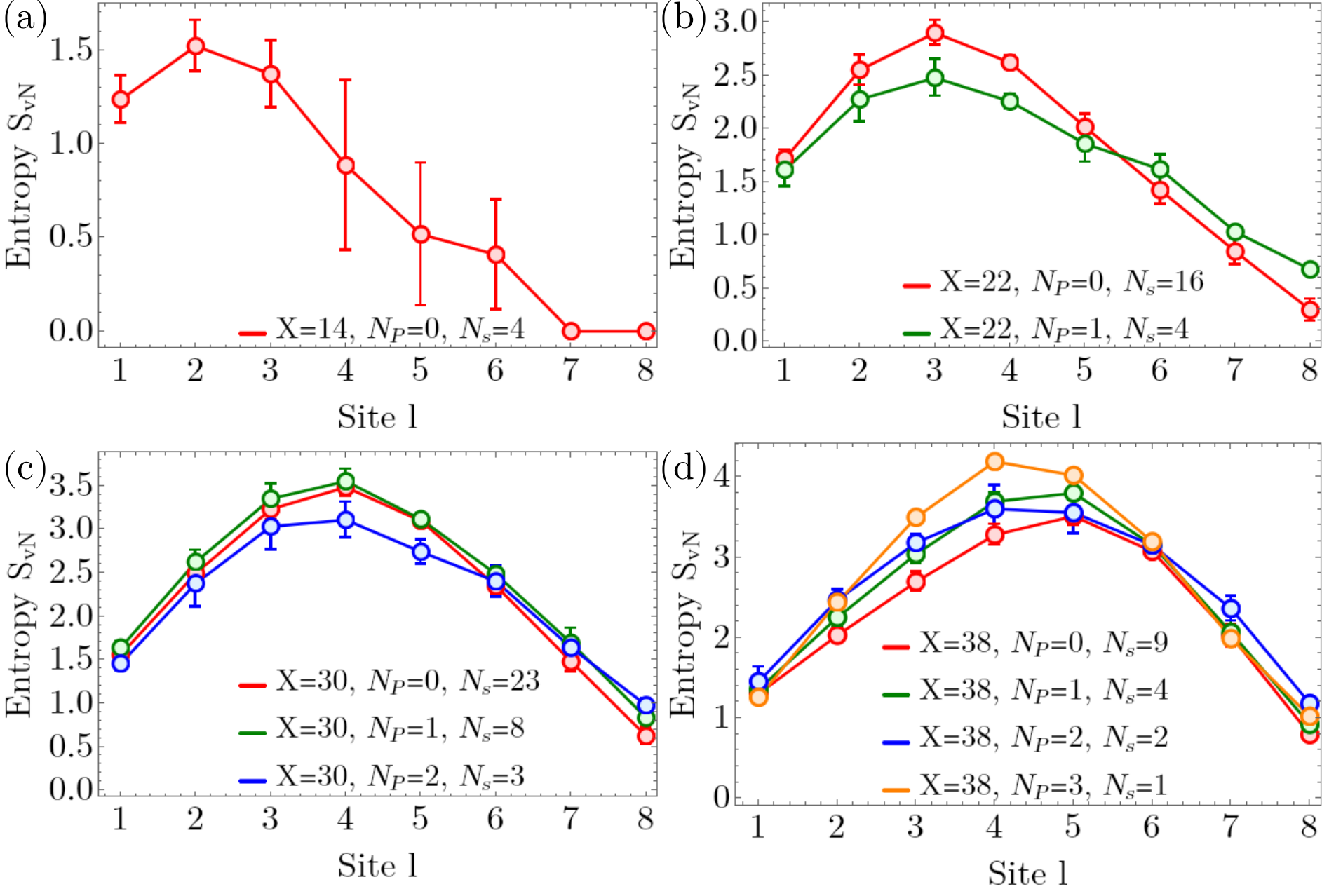}
\caption{
The average entanglement entropy $S_\text{vN}$ of the null states of $\mathcal{O}^\text{(b)}$ for bosonic particles for $N=8$ particles in $L=9$ sites for several symmetry sectors: (a) $X=14$, (b) $X=22$, (c) $X=30$, (d) $X=38$. In each center-of-mass sector we consider also the pairing quantum numbers $N_p$. Each panel contains several curves corresponding to the possible values of the pairing quantum number $N_p$. 
The results are obtained with the same procedure as for Fig.~\ref{fig:hardcore_density}.
}
\label{fig:bosons_entropy}
\end{figure}

The results for fermions are depicted in Fig.~\ref{fig:hardcore_entropy} and for bosons in Fig.~\ref{fig:bosons_entropy}.
Regardless of the nature of the particles, we obtain non-zero values for the entanglement entropy for the null states of $\mathcal{O}$. This shows that generically the superpositions arising to satisfy the null state condition for $\mathcal{O}$ lead to entanglement between the quantum particles.
Moreover, we obtain that the von Neumann entropy $S_\text{vN}(l)$ depends on the size of the subsystem $l$ with a peak at a certain size, which implies a volume-law scaling for the entanglement. 
In contrast, for a one-dimensional chain, an area-law for the entanglement entropy would correspond to a constant value of $S_\text{vN}(l)$.
Interestingly, the height of the entanglement peak increases for sectors with larger center-of-mass, for both fermions, Fig.~\ref{fig:hardcore_entropy}, and bosons, Fig.~\ref{fig:bosons_entropy}.

The detailed structure of the entanglement profile is dependent on the particle statistics.
For fermionic particles, Fig.~\ref{fig:hardcore_entropy}, the peak of $S_\text{vN}(l)$ is in the middle of the chain regardless of the center-of-mass symmetry sector, due to the particle-hole plus inversion symmetry emerging for fermions (discussed in Sec.~\ref{sec:results_density}). 
Thus, even though the fermionic particles are mostly localized in the first half of the chain, their corresponding holes generate entanglement also in the second half of chain.
For bosons, where this symmetry is absent, the entanglement peak follows the value of $X$ and the corresponding distribution, moving towards the center of the chain for larger values of $X$, Fig.~\ref{fig:bosons_entropy}.

\section{Conserved quantities for small fermionic systems\label{sec:results_conserved}}

\begin{figure}[!hbtp]
\centering
\includegraphics[width=0.49\textwidth]{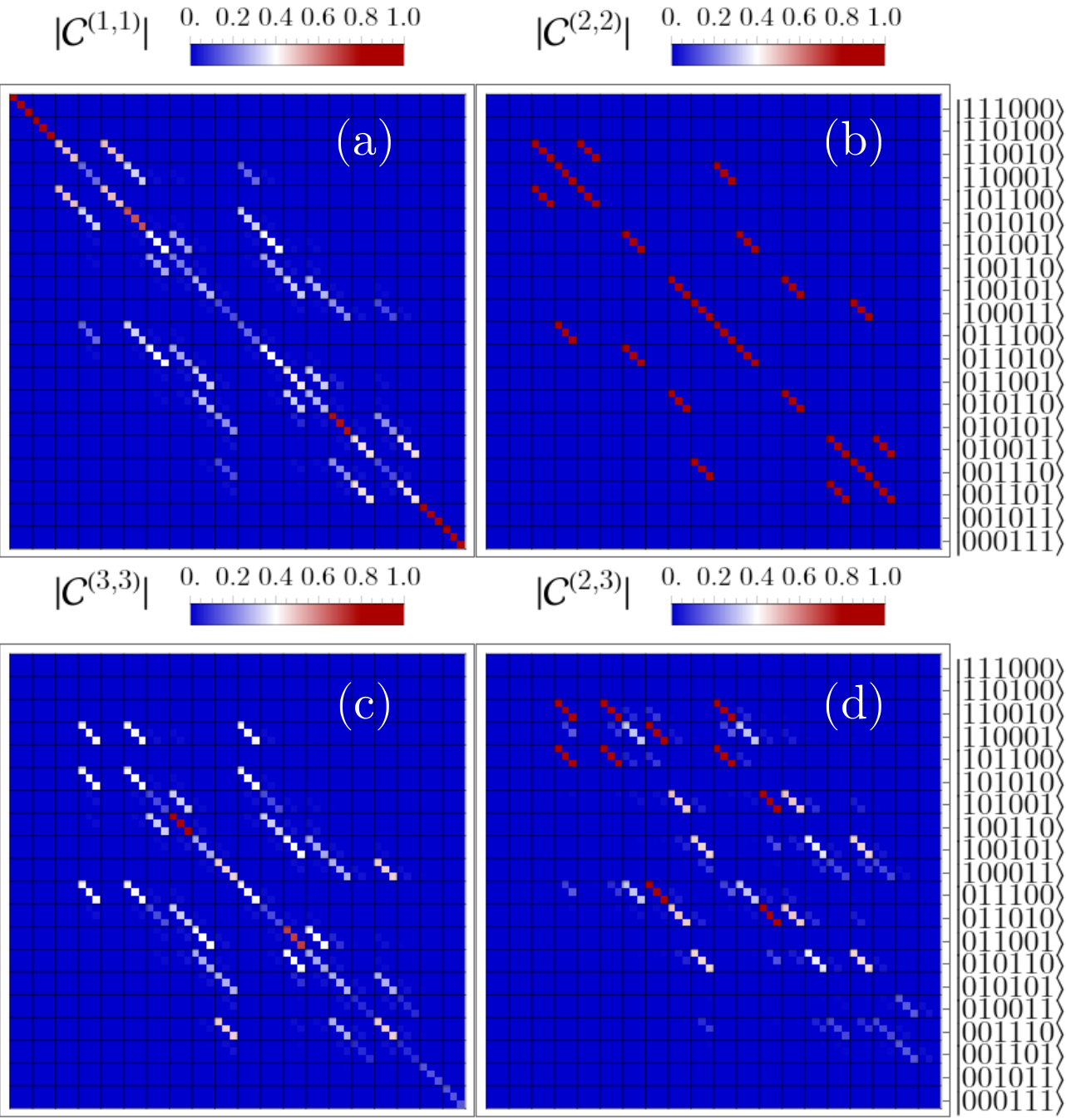}
\caption{
The absolute values of the matrix elements of the conserved quantities $\mathcal{C}_{(i,j)}$ for the fermionic case  for $N=3$ particles in $L=6$ sites. They correspond to the steady states which are constructed from the null states given in Eq.~\eqref{eq:N3_L6}.
We normalize $\mathcal{C}_{(i,j)}$ such that the largest matrix element has weight one. 
The matrix elements are computed in the atomic Fock basis, with each block separated by black lines corresponding to the Fock state written on the right (the columns of the matrix are ordered in the same way as the rows). 
Inside each block, the $3\times 3$ matrix correspond to the photonic Fock states ranging from the vacuum $\ket{0}$ to the state $\ket{2}$. Photonic Fock state $\ket{3}$ was also included in the calculations.
The results are obtained with exact diagonalization by making use of Eq.~\eqref{eq:conserved_C_final}, for the parameters $\Gamma/g=10$, $\delta/g=5$.
}
\label{fig:conserved_N3}
\end{figure}

In this section, we discuss the conserved quantities of the hybrid matter-cavity model, Eq.~\eqref{eq:Lindblad} and Eq.~\eqref{eq:operator}.
We show that the conserved quantities have a complex structure already for small system sizes and few particles, containing long-range and multi-body operators. 
We expect the complexity of the conserved quantities to increase for larger $N$ and $L$, resembling the case of closed integrable many-body quantum systems.

The conserved quantities can be constructed using the formula given in Eq.~\eqref{eq:conserved_C_final}, if one knows the steady states that span the decoherence-free subspace, see Sec.~\ref{sec:recipe_2}.
However, this is a non-trivial many-body problem as it requires the application and inversion of superoperators defined in the Hilbert space of both the photons and the matter. 
In the presence of the the strong symmetry generated by the $\eta$-pairing algebras, Sec.~\ref{sec:pairing_algebra}, we can perform the calculation within each symmetry sector.

We discuss the conserved quantities based on exact diagonalization results for $N=3$ particles in $L=6$ sites for the case of fermionic particles coupled to photons with a truncated Hilbert space. 
We begin by writing down the null states of $\mathcal{O}$ for $N=3$ and $L=6$ and the steady states that can be generated from them.
Afterwards, we discuss several conserved quantities.
We show that already for this small system and simple null states, the conserved quantities have nonlocal and multibody contributions. We argue that these contributions allow for the emergence of the complex quantum correlations discussed in Sec.~\ref{sec:results_steady_states}.
We further discuss the special case in which the steady state is unique in a sector of the $\eta$-pairing strong symmetry.

\emph{Null states}. 
For $N=3$ fermionic particles in $L=6$ sites, the operator $\mathcal{O}^\text{(f)}$, Eq.~\eqref{eq:operator}, has three null states:
\begin{align}
\label{eq:N3_L6}
\ket{\mathbf{0}_1}=& \ket{1,1,1,0,0,0}, \\
\ket{\mathbf{0}_2}=&\ket{1,1,0,0,1,0} -\ket{1,0,1,1,0,0}, \nonumber \\ 
\ket{\mathbf{0}_3}=& \ket{1,1,0,0,0,1}-\ket{1,0,1,0,1,0}+\ket{0,1,1,1,0,0}. \nonumber 
\end{align}
The fermionic null states are sketched in Fig.~\ref{fig:example_nullstates}.
Each state has a different value of the center-of-mass $X$, corresponding to a different sector of the $\mathfrak{sl}(2, \mathbb{C})$ algebra, Eq.~\eqref{eq:sl2_algebra}. 
With respect to the $\eta$-pairing algebra, Eq.~\eqref{eq:pairing_SU2_fermions}, the states $\ket{\mathbf{0}_1}$ and $\ket{\mathbf{0}_3}$ correspond to the $N_p=0$, while the state $\ket{\mathbf{0}_2}$ is in the $N_p=1$ sector.
The three states generate a decoherence-free subspace spanned by nine states, $\rho^{(i,j)}_\text{st}=\ket{0; \mathbf{0}_i}\bra{0; \mathbf{0}_j}$, with $i,j=1,2,3$.

\emph{Conserved quantities}. 
In Fig.~\ref{fig:conserved_N3} we show the associate conserved quantities, $\mathcal{C}^{(i,j)}$, corresponding to these steady states, Eq.~\eqref{eq:N3_L6}, and we observe a few striking differences between their behavior.
In particular, while $\mathcal{C}^{(1,1)}$, $\mathcal{C}^{(3,3)}$ and $\mathcal{C}^{(2,3)}$, contain contributions and coherences between a large number of the matter Fock states, $\mathcal{C}^{(2,2)}$ connects fewer states.
We first discuss the generic conserved quantities and afterwards we comment on the special case of $\mathcal{C}^{(2,2)}$.

The conserved quantities in Figs.~\ref{fig:conserved_N3}(a),(c),(d) contain off-diagonal contributions from a large number of Fock states, leading to non-local correlations. 
The largest contributions seem to stem from terms of the form $d^\dagger_{j}d^\dagger_{L+1-j} d_{L+1-j'} d_{j'}$, which can lead to the long-range pairing correlations discussed in Sec.~\ref{sec:results_corr}. 
In $\mathcal{C}^{(3,3)}$, Fig.~\ref{fig:conserved_N3}(c), we can also observe finite coherences between the states $\ket{1,0,0,1,1,0}$ and $\ket{0,1,1,0,0,1}$ in which all three particles are at different positions. 
This shows that multibody correlations, beyond correlations between pairs of particles, are also contained in the conserved quantities.
Furthermore, coherences between matter states which are accompanied by the annihilation, or creation, of photons, are present in $\mathcal{C}^{(2,3)}$, Fig.~\ref{fig:conserved_N3}(d).

The overlap of the conserved quantities with a large number of states increases the difficulty of choosing an initial state in order to target a certain steady state.
However, we can still obtain some information in this regard. We observe that $\mathcal{C}^{(1,1)}$ has overlap with all matter Fock states, while $\mathcal{C}^{(3,3)}$ has zero overlap with the three states with the lowest center-of-mass value. 
This implies that if the initial state has a very low center-of-mass value, the dissipative dynamics will mostly lead to the trivial $\ket{1,1,1,0,0,0}$. 
Thus, one should consider states with larger center-of-mass in order to arrive at nontrivial steady states (as we discuss in Sec.~\ref{sec:results_tmps}).

\emph{Special cases.}
For $N=3$ fermionic particles in $L=6$ sites, $\mathcal{C}^{(2,2)}$ is a conserved quantity with a special structure, Sec.~\ref{sec:recipe_2}. $\mathcal{C}^{(2,2)}$ corresponds to a symmetry sector containing multiple states, but a unique steady state. Thus, $\mathcal{C}^{(2,2)}=\mathbb{I}_\text{pho}\otimes \mathbb{I}_{\text{mat}, N_p=1}$, where the matter identity operator is defined inside the $N_p=1$ sector.
We expect that with increasing the system size, the number of special forms of the conserved quantities will be exponentially small compared to the generic conserved quantities discussed above.

\section{Mixed steady states of the dissipative dynamics \label{sec:results_tmps}}

In this section, we show that generic initial states of the dissipative evolution of the open hybrid system, Eq.~\eqref{eq:Lindblad} and Eq.~\eqref{eq:operator}, lead to steady states characterized by the nontrivial properties described for null states (see Sec.~\ref{sec:results_steady_states}).
In order to target a certain steady state of the dissipative evolution, one needs to know the overlap of the initial state with the associated conserved quantity.
However, in Sec.~\ref{sec:results_conserved} we have argued that the calculation of the conserved quantities is a complicated many-body problem, even after the determination of the steady states.
The structure of the conserved quantities further implies that simple initial states, e.g.~product states in Fock basis, would lead to a mixture containing contributions from a large number of the null states of $\mathcal{O}$.
Thus, the question arises if, in the absence of detailed knowledge of the conserved quantities, one can still control the dissipative evolution to obtain nontrivial physical properties.
In the following, we make inroads into this problem by investigating the full quantum dissipative time-evolution of the Liouvillian, Eq.~\eqref{eq:Lindblad}. We develop a guiding principle in choosing Fock states as initial states to obtain complex correlations in the steady state.

The numerically exact results presented in this section are obtained by a recently developed method based on time-dependent matrix product state (tMPS) techniques  \cite{HalatiKollath2020, HalatiKollath2020b, TolleHalati2025c}. This approach simulates the GSKL master equation describing the hybrid atom-cavity system, Eq.~\eqref{eq:Lindblad} and Eq.~\eqref{eq:operator}, dealing with the challenges stemming from the interplay of short- and global-range interactions, photon losses and hybrid many-body Hilbert space. The implementation used in this work makes use of the quantum trajectories approach and is based on the Trotter-Suzuki decomposition of the time evolution operator together with the dynamical deformation of the matrix product state structure \cite{HalatiKollath2020b, TolleHalati2025c}. For more details on the implementation and the convergence parameters employed see \ref{app:tmps}.

We perform tMPS simulations for both fermionic and bosonic atoms coupled to the dissipative cavity. We consider $N=8$ particles and a system size of $L=16$ for fermions and $L=9$ for bosons, following the fermions-bosons correspondence, Sec.~\ref{sec:bosons_to_fermions}, give rise to decoherence-free subspaces of the same size.

\begin{figure}[!hbtp]
\centering
\includegraphics[width=0.49\textwidth]{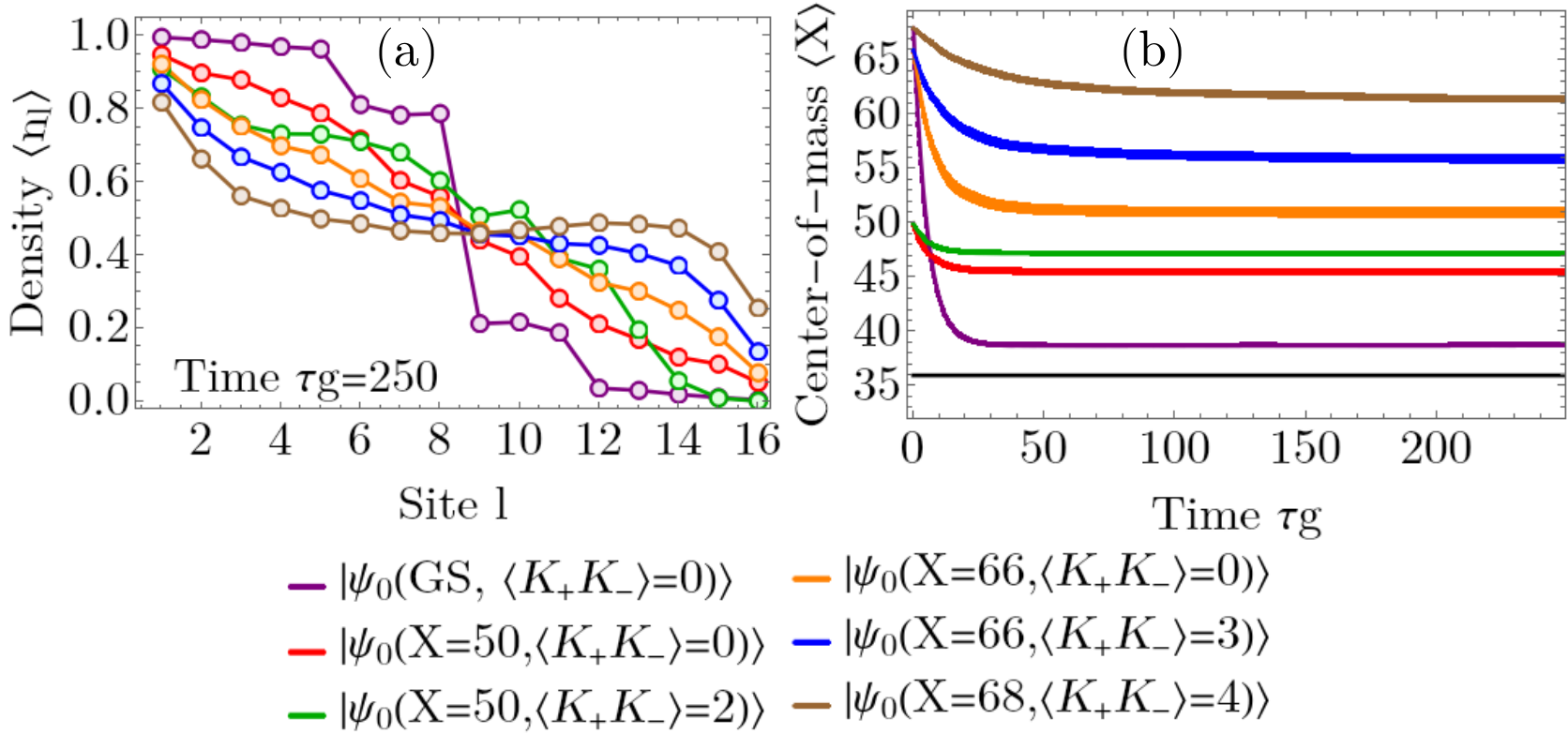}
\caption{
(a) The density profile of the long-time states at $\tau g=250$ corresponding to different initial states for fermionic atoms. (b) The time dependence of the expectation value of the center-of-mass, $\langle X \rangle$. The black line corresponds to the center-of-mass value of the fully localized state.
The initial states considered are detailed in \ref{app:initial_states}.
The numerical tMPS results are obtained by performing the dissipative time-evolution of Eq.~\eqref{eq:Lindblad}.
We consider $N=8$ fermionic particles in $L=16$ sites, with the parameters $\Gamma/g=10$, $\delta/g=5$. 
}
\label{fig:tmps_density_fermions}
\end{figure}

In order to understand how to target the nontrivial properties of the null states, Sec.~\ref{sec:results_steady_states}, we investigate several initial states: atomic ground state of the kinetic operator $(\mathcal{O}+\mathcal{O}^\dagger)$; several Fock product states characterized by different values of the center-of-mass $X$ and expectation value of $\langle K_+ K_- \rangle$. 
As the Fock states are not eigenstates of the pairing Casimir, Eq.~\eqref{eq:casimir_pairing_bosons} and Eq.~\eqref{eq:casimir_pairing_SU2}, they can contain contributions from symmetry sectors corresponding to different number of pairs. 
However, $\langle K_+ K_- \rangle$ is a conserved quantity of the dissipative dynamics due to the strong symmetry, Sec.~\ref{sec:pairing_consequences}.
The Fock initial states are explicitly given in \ref{app:initial_states}.

We begin the discussion with the fermionic case and focus on the following features:
(i) the accumulation dynamics of particle density and emergence of the symmetry-constrained Liouvillian skin effect;
(ii) the persistence of long-range kinetic correlations in the mixed states;
(iii) how $\eta$-pairing coherence manifests itself in the mixed steady states.
Afterwards, we investigate the role of the particle statistics and discuss the results obtained for bosonic atoms.
For the final time considered, $\tau g=250$, the dissipative evolution has reached the null subspace of $\mathcal{O}$, see \ref{app:numerical_results_mps}.

\begin{figure}[!hbtp]
\centering
\includegraphics[width=0.48\textwidth]{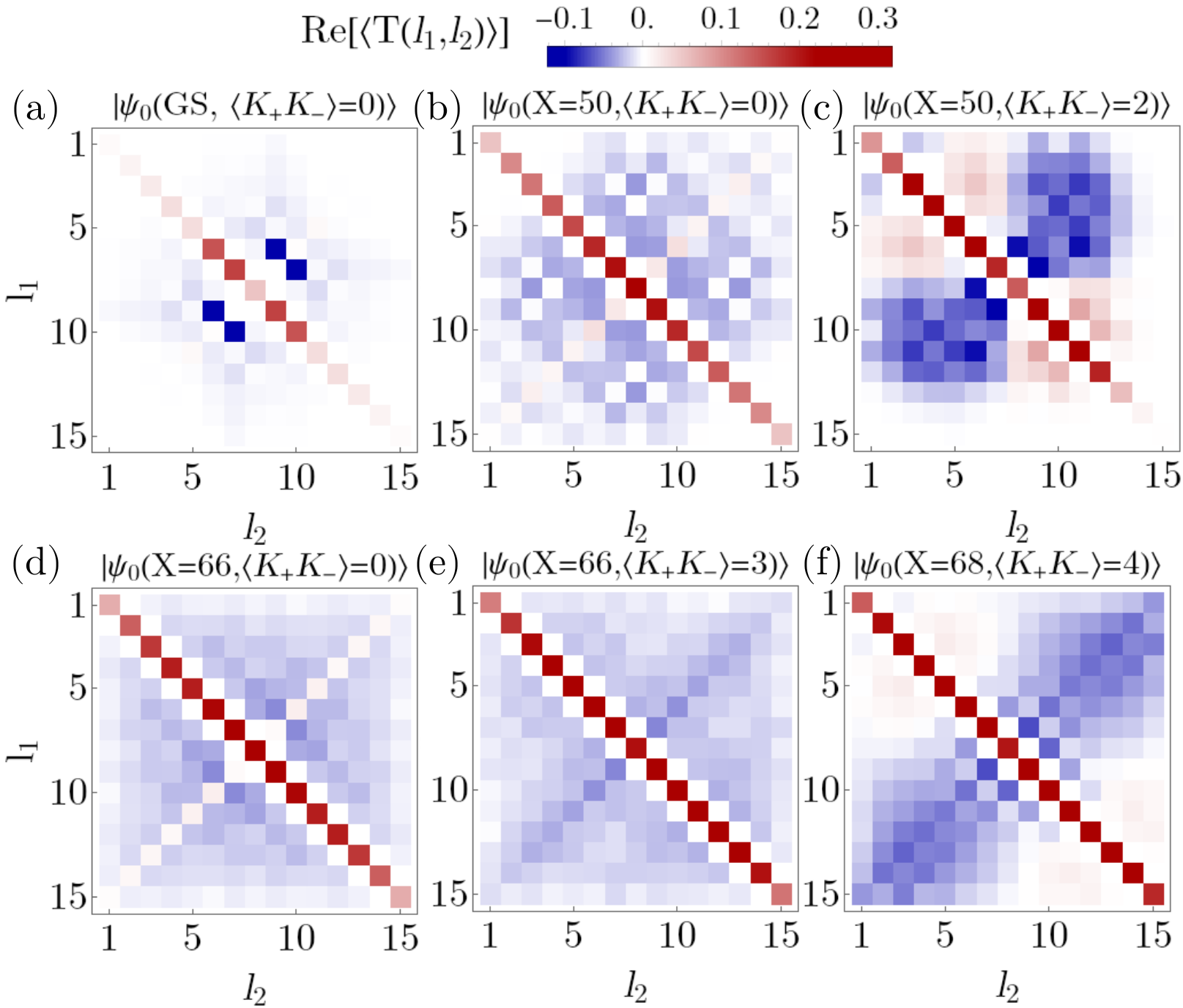}
\caption{
The expectation value of the kinetic correlations, $\Re [\langle T(l_1,l_2)\rangle]$, Eq.~\eqref{eq:kinetic_corr}, of the long-time states at $\tau g=250$ corresponding to different initial states for fermionic atoms. 
We consider $N=8$ fermionic particles in $L=16$ sites, with the parameters $\Gamma/g=10$, $\delta/g=5$. 
}
\label{fig:tmps_kinetic_fermions}
\end{figure}

\begin{figure}[!hbtp]
\centering
\includegraphics[width=0.48\textwidth]{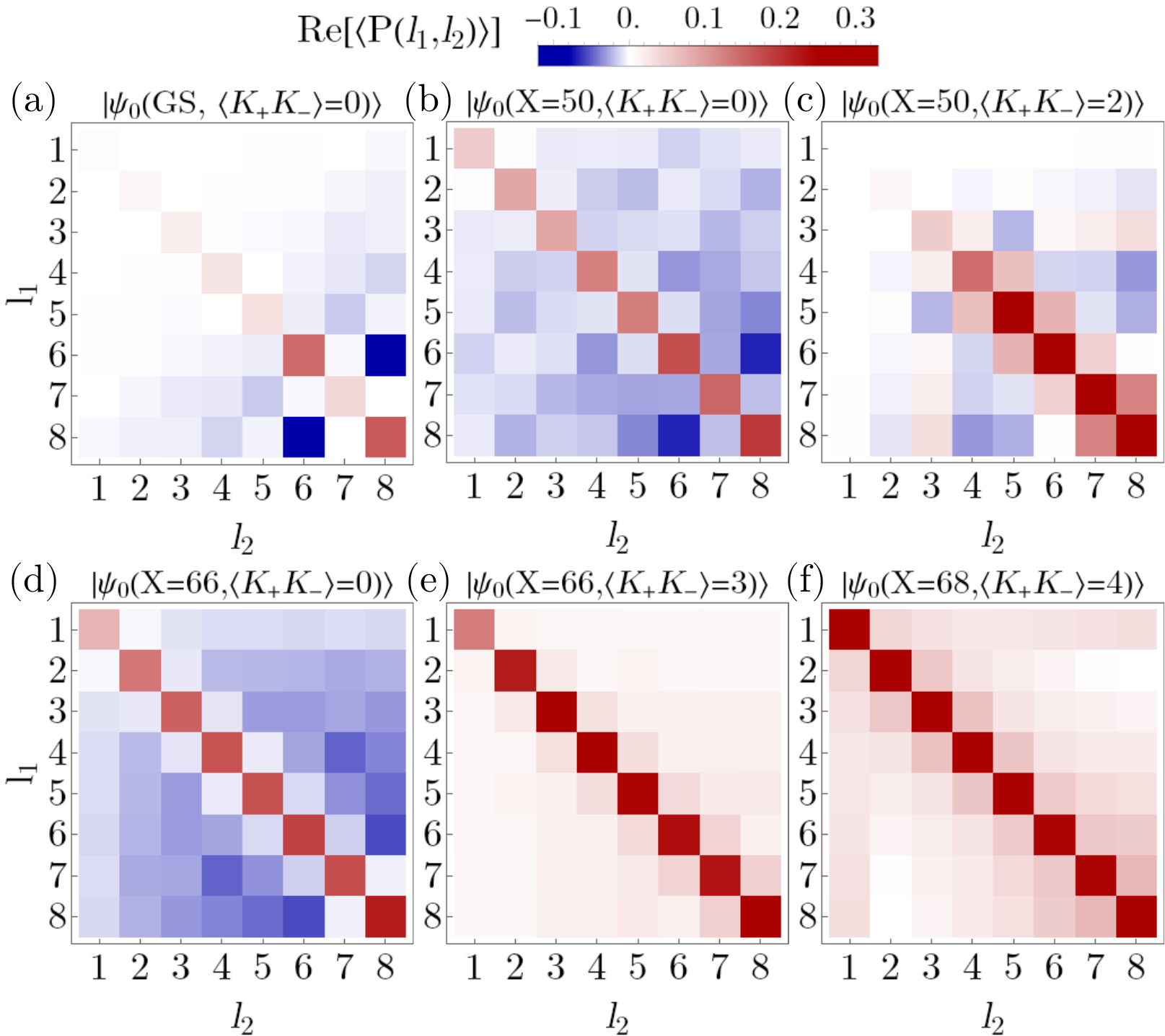}
\caption{
The expectation value of the pairing correlations, $\Re [\langle P(l_1,l_2)\rangle]$, Eq.~\eqref{eq:pairing_corr}, of the long-time states at $\tau g=250$ corresponding to different initial states for fermionic atoms. 
We consider $N=8$ fermionic particles in $L=16$ sites, with the parameters $\Gamma/g=10$, $\delta/g=5$. 
}
\label{fig:tmps_pairing_fermions}
\end{figure}

\begin{figure}[!hbtp]
\centering
\includegraphics[width=0.49\textwidth]{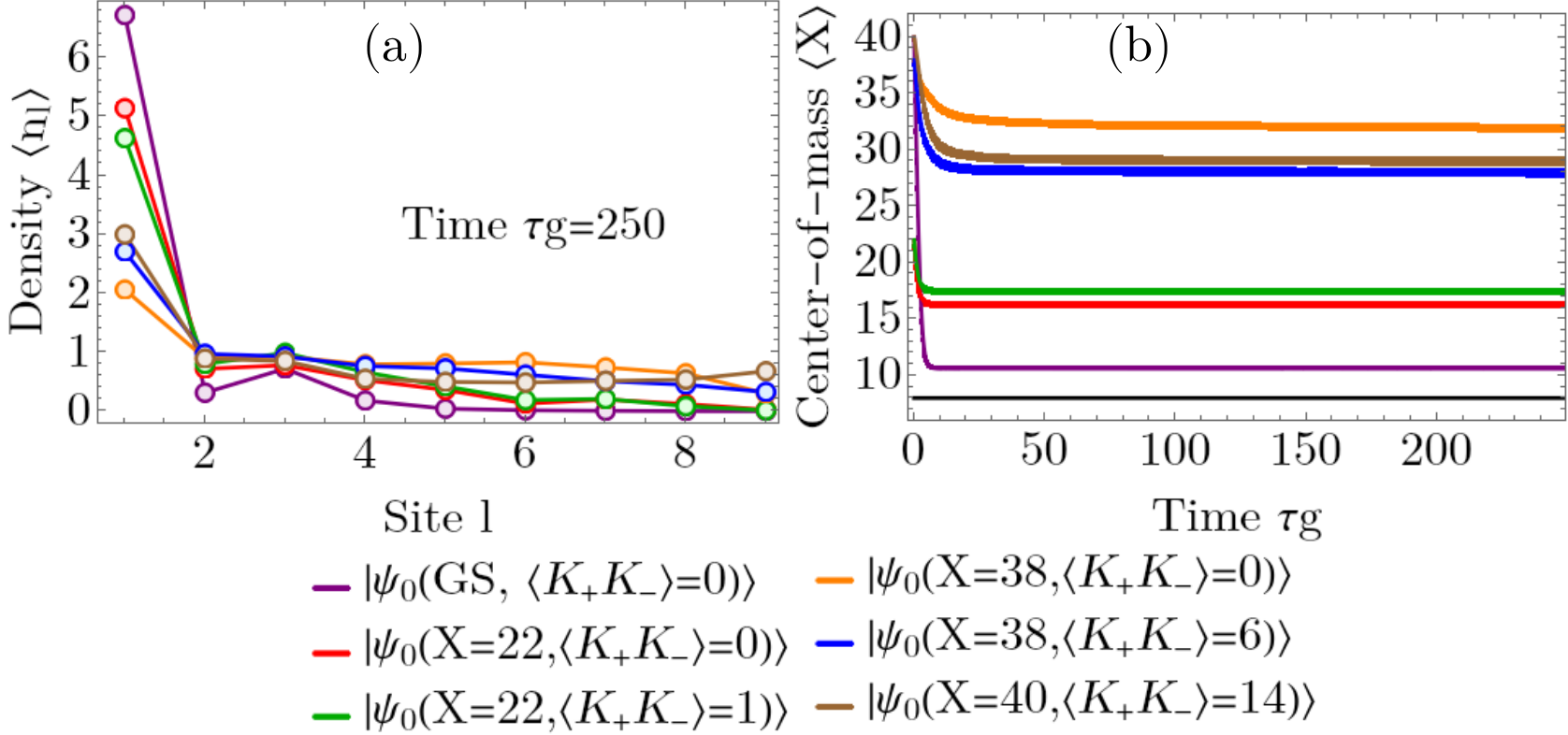}
\caption{
(a) The density profile of the long-time states at $\tau g=250$ corresponding to different initial states for bosonic atoms. (b) The time dependence of the expectation value of the center-of-mass, $\langle X \rangle$. The black line corresponds to the center-of-mass value of the fully localized state.
The initial states considered are detailed in \ref{app:initial_states}.
The numerical tMPS results are obtained by performing the dissipative time-evolution of Eq.~\eqref{eq:Lindblad}.
We consider $N=8$ bosonic particles in $L=9$ sites, with the parameters $\Gamma/g=10$, $\delta/g=5$.
}
\label{fig:tmps_density_bosons}
\end{figure}

The fermionic \emph{density profiles} reached at $\tau g=250$ from different initial states are shown in Fig.~\ref{fig:tmps_density_fermions}(a). We can observe that while a strong dependence on the initial state is present, the features of the symmetry-constrained Liouvillian skin effect (Sec.~\ref{sec:results_density}) can be identified.
Starting from the ground state, $\ket{\psi_0(\text{GS})}$, or from low center-of-mass Fock states, $X=50$, leads to a pronounced accumulation of the density at the boundary.
However, initial states with a large value of the center-of-mass and $\langle K_+ K_- \rangle >0$, lead to spread out density profile. The steady state densities resemble very well the ones of the null states (see Fig.~\ref{fig:hardcore_density}).
These effects can be understood from the dynamics of  $\langle X \rangle $ depending on the initial state, Fig.~\ref{fig:tmps_density_fermions}(b). 
The expectation value $\langle X \rangle $ shows a strictly monotonic decay in time.
For the Fock initial states, we obtain that the larger value of $X$ initially, the larger the value at long times.
Interestingly, this does not hold for the $\ket{\psi_0(\text{GS})}$ initial state, which even though it has a large $\langle X \rangle $ initially, decays close to the value of the fully localized state.
The distribution of the center-of-mass values of the single quantum trajectories in shown in \ref{app:numerical_results_mps}.

The \emph{kinetic correlations}, Eq.~\eqref{eq:kinetic_corr}, for the fermionic case at late times are shown in Fig.~\ref{fig:tmps_kinetic_fermions}. We obtain a behavior of the correlations very similar to the one of the null states, discussed in Sec.~\ref{sec:results_corr}.
The long-time states exhibit long-range kinetic correlations and, in particular, for large values of the center-of-mass the correlations only show a weak decay with the distance, see Figs.~\ref{fig:tmps_kinetic_fermions}(d),(e).
The presence of paired particles induces strong correlations on the antidiagonal of the pairing matrix, Fig.~\ref{fig:tmps_kinetic_fermions}(f).
If the steady state is characterized by a small value of $\langle X \rangle $, then the correlations will be restricted to the middle of the chain, Fig.~\ref{fig:tmps_kinetic_fermions}(a).

The \emph{$\eta$-pairing correlations}, Eq.~\eqref{eq:pairing_corr}, are plotted in Fig.~\ref{fig:tmps_pairing_fermions}. As in the results for the null states, Sec.~\ref{sec:results_corr}, the off-diagonal entries of the $\eta$-pairing matrix are finite for both vanishing and finite $\langle K_+ K_- \rangle$.
In the simulations which consider also sectors with a finite number of $\eta$-pairs, $\langle K_+ K_- \rangle>0$, the finite off-diagonal entries signal long-range $\eta$-pair coherence, Figs.~\ref{fig:tmps_pairing_fermions}(c),(e),(f).
By choosing an initial state with a larger value of $\langle K_+ K_- \rangle$ one can enhance the coherence of the $\eta$-pairs, while for the cases in which $\langle K_+ K_- \rangle=0$, the finite off-diagonal entries mark the overall coherence of the long-time state.

\begin{figure}[!hbtp]
\centering
\includegraphics[width=0.48\textwidth]{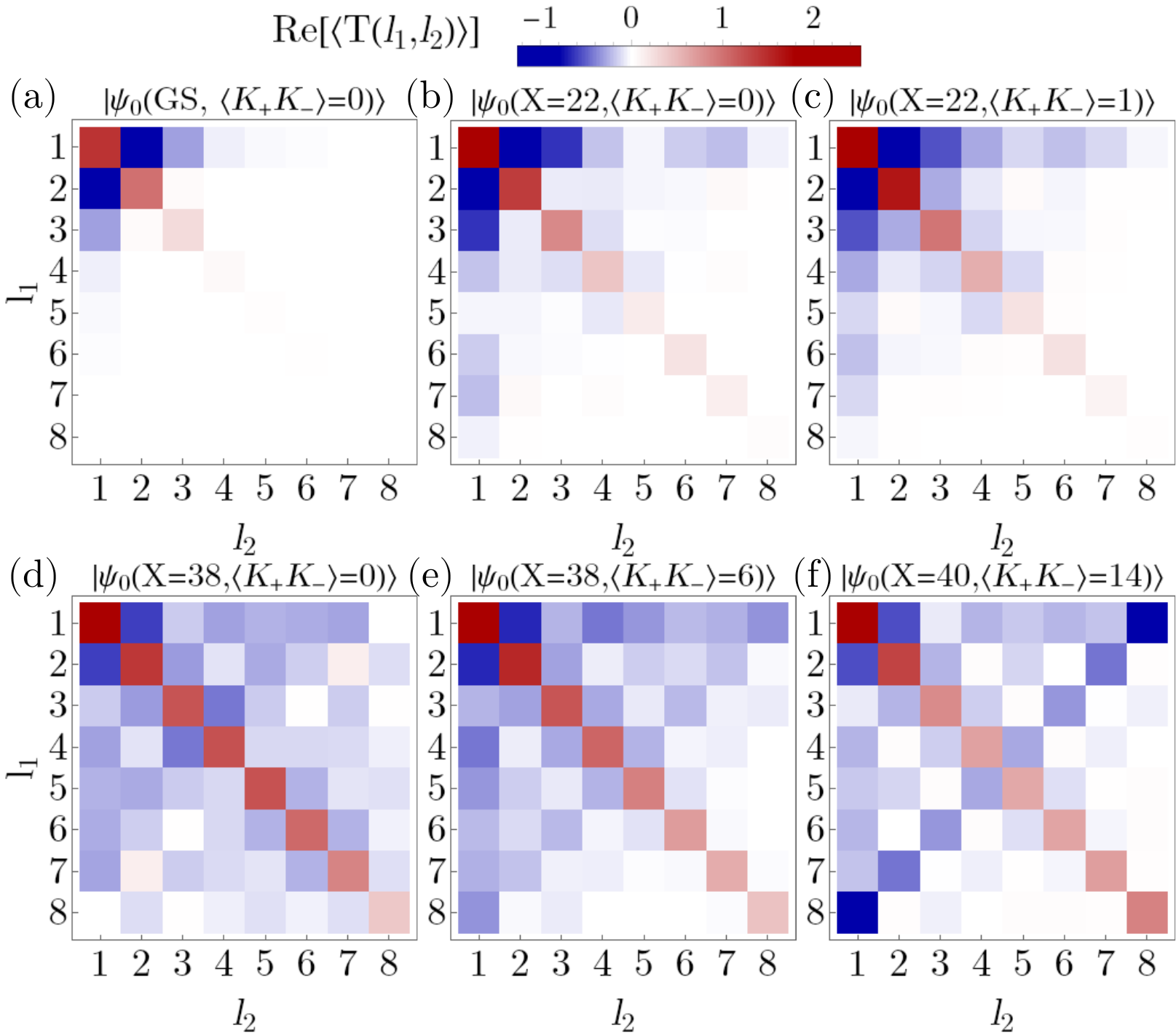}
\caption{
The expectation value of the kinetic correlations, $\Re [\langle T(l_1,l_2)\rangle]$, Eq.~\eqref{eq:kinetic_corr}, of the long-time states at $\tau g=250$ corresponding to different initial states for bosonic atoms. 
We consider $N=8$ bosonic particles in $L=9$ sites, with the parameters $\Gamma/g=10$, $\delta/g=5$. 
}
\label{fig:tmps_kinetic_bosons}
\end{figure}

\begin{figure}[!hbtp]
\centering
\includegraphics[width=0.48\textwidth]{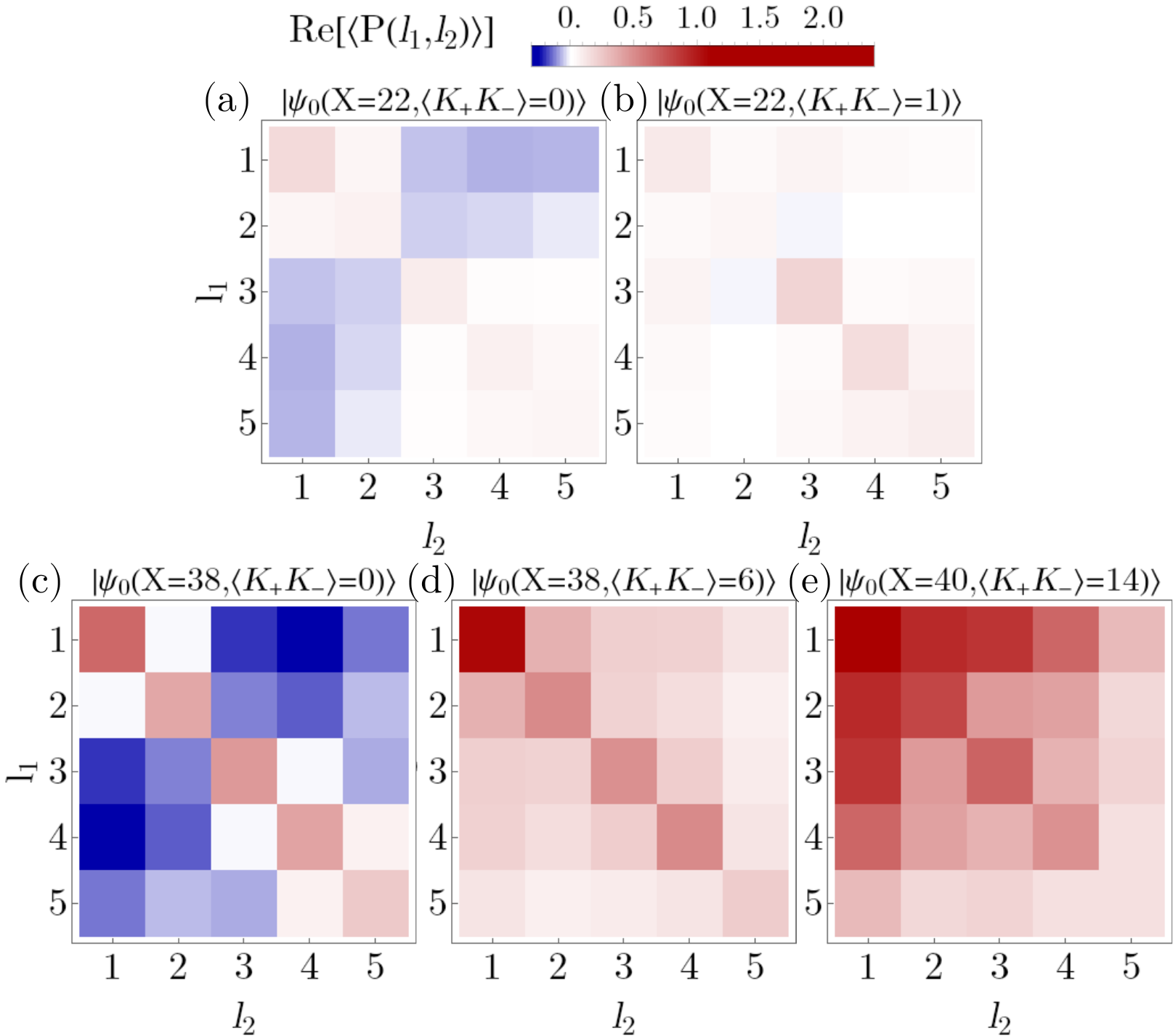}
\caption{
The expectation value of the pairing correlations, $\Re [\langle P(l_1,l_2)\rangle]$, Eq.~\eqref{eq:pairing_corr}, of the long-time states at $\tau g=250$ corresponding to different initial states for bosonic atoms. 
We consider $N=8$ bosonic particles in $L=9$ sites, with the parameters $\Gamma/g=10$, $\delta/g=5$. 
}
\label{fig:tmps_pairing_bosons}
\end{figure}

In the case of \emph{bosonic atoms}, we obtain a similar correspondence between the long-time states of the dissipative dynamics and the null states, as discussed above for fermions.
Starting from initial Fock states with a larger center-of-mass one can control the spreading of the particles away from the boundary at longer times, Fig.~\ref{fig:tmps_density_bosons}(a). 
However, in contrast with the fermionic case, having a larger number of pairs does not ensure a larger long-time value for $\langle X\rangle$, Fig.~\ref{fig:tmps_density_bosons}(b).
Figure~\ref{fig:tmps_kinetic_bosons} shows that a large value of $\langle X\rangle$ in the steady state corresponds to long-range kinetic correlations and the presence of coherence.
Similarly, $\eta$-pair coherence is present in the steady states corresponding to the initial states with $X$ large, Fig.~\ref{fig:tmps_pairing_bosons}.

From the tMPS results describe above we can derive a guiding principle in designing the initial states of the dissipative dynamics, Eq.~\eqref{eq:Lindblad}, to obtain mixed steady states with non-trivial properties. 
One should start from Fock states in which most particles are localized in the middle on the chain.
Such a state ensure a large initial $X$ and a overlap with states with finite number of $\eta$-pairs.
This allows to obtain a striking manifestation of the symmetry-constrained Liouvillian skin effect and long-rage correlations.
Furthermore, it also leads a large value of the von Neumann entropy of the quantum trajectories, \ref{app:numerical_results_mps}, however, this does not ensure mixed state entanglement.

In this section, we have focused on the long-time states of the dissipative dynamics. However, the intermediate dynamics in the presence of the complex conserved quantities offers an exciting outlook. In particular, how the structure of the decoherence-free subspace influences the generation and propagation of the long-range kinetic and pairing correlations \cite{HalatiJager2025}.

\section{Conclusions and Outlook \label{sec:conclusions}}

To summarize, we have developed a general framework for hybrid open quantum systems to construct large decoherence-free subspaces with nontrivial internal structure.
The main ingredient is the coupling of the matter operator $\mathcal{O}$ to a dissipative photonic mode. In the long-time limit the photon losses impose a global constraint onto the matter, which selects the null states of $\mathcal{O}$ as the decoherence-free subspace of the dissipative dynamics.
By employing the representation theory of Lie algebras, we have shown how to design $\mathcal{O}$, such that it posses an exponentially large null space and exhibits a prescribed algebraic structure.
This recipe encompasses previously studied models of independent spin-$1/2$ particles couple collectively to a photon field, used to generate noiseless quantum codes \cite{ZanardiRasetti1997, LidarWhaley1998}.
Furthermore, it provides a direct connection between the structure of a decoherence-free subspace and the choice of the underlying algebra.

We further have presented a fully worked example, in which the cavity field is coupled to the directed tunneling operator for bosonic and fermionic particles in one dimension.
Two commuting algebras give rise to the structure of the decoherence-free subspace.
The first $\mathfrak{sl}(2,\mathbb{C})$ algebra only characterizes the symmetry of the long-time states and yields an exponentially large number of steady states. 
The spinless $\eta$-pairing algebra instead gives rise to a strong symmetry of the full light-matter Liouvillian. 
The null states exhibit the symmetry-constrained Liouvillian skin effect, in which boundary localization competes with many-body interference effects occurring under symmetry constraints.
Furthermore, the steady states of the system are characterized by long-range kinetic and pairing correlations, and volume-law entanglement. 
Time-dependent matrix product states simulations of the open hybrid system show that the nontrivial properties emerge dynamically from generic initial states.

Several features set our framework apart from other routes proposed to generate multiple steady states.
The degeneracy of the decoherence-free subspace is not inherited from the symmetries of the Liouvillian \cite{AlbertJiang2014, BucaProsen2012}, as steady-state manifold hosts emergent symmetries that are absent from the full dissipative dynamics. 
Thus, by considering only the symmetries of the Liouvillian one would fail to account for the size and structure of the decoherence-free subspace.
Furthermore, the conserved quantities associated with the multiple steady states contain contributions from multibody and nonlocal operators. We expect that this has the potential to give rise to novel types of dissipative dynamics.

One can view the null state condition of the states forming the decoherence-free subspace as a global constraint imposed by dissipation. In contrast to Hamiltonian constrained systems in which the constraints are enforced by energy penalties, the dissipative analogue of the constrained low-energy manifold is dynamically reached.
Our results show that the coupling to the dissipative cavity actively creates complex quantum correlations.
Thus, this establishes global dissipative constraints as a design principle for open correlated quantum matter, as a non-equilibrium counterpart to the kinetic constraints considered in closed systems \cite{BernienLukin2017, TurnerPapic2018, SalaPollmann2020, KhemaniNandkishore2020, MoudgalyaRegnault2022}.

Our results open several natural outlook directions.
On one hand, within the recipe for constructing algebraic structures, by considering other Lie algebra and representations one could arrive at constraint structures beyond what we considered. For example, null spaces defined on higher-dimensional lattices can be constructed, which we expect to lead to exotic types of multi-particle pairing mechanisms.
On the other hand, one can view the decoherence-free subspace as a starting point in generating non-trivial dissipative dynamics.
By adding perturbations to the Hamiltonian, the steady state degeneracy is lifted and a metastable manifold is created with its own effective dynamics \cite{PolettiKollath2013, ZanardiCamposVenuti2014, SciollaKollath2015, MacieszczakGarrahan2016, HalatiKollath2025}.
This is further enriched by the slow relaxation associated with the Liouvillian skin effect \cite{HagaUeda2021}.
The characterization of this dynamics would offer the perspective of performing controlled operations within the long-lived manifold and manipulating complex quantum correlations \cite{ChuClerk2026}.

The required ingredients of our framework and, in particular, of our directed tunneling example, are within the reach of current experiments.
Ultracold atoms in optical cavities provide lossy photon modes coupled to collective matter operators \cite{RitschEsslinger2013, MivehvarRitsch2021} and photon-assisted tunneling terms can be obtained with Raman processes \cite{ZhengCooper2016, LaflammeZoller2017}.
Well suited to realize and probe our results are recent experiments that combine optical cavities with high-resolution microscopes \cite{OrsiBrantut2024, BologniniBrantut2025, BuhlerBrantut2026}, or with optical tweezer arrays \cite{HartungRempe2024, HoStamper-Kurn2025, SeubertDistante2025, ZhangSmith2025, RoschinskiLeonard2026}, which posses the spatially resolved control and detection of the matter degrees of freedom. 
We can envision complementary implementations in other hybrid platforms, such as superconducting circuits \cite{LEHURSchiro2016, ClerkNakamura2020, BlaisWallraff2021}.
More broadly, our framework provides an experimentally relevant route to harness nonreciprocity in a controlled manner and engineer complex quantum correlations, by using global decoherence mechanisms to shape many-body quantum states.

\section*{ACKNOWLEDGMENTS}
We thank F.~Balducci, M.~Bauer, J.-S.~Bernier, A. Clerk, T.~Giamarchi, S.~B.~J\"ager, C.~Kollath, G.~Morigi, A.~Sheikhan for fruitful discussions.
This work was supported in part by the Deutsche Forschungsgemeinschaft via the cluster of excellence ctd.qmat (EXC 2147, Project-ID 390858490) and FOR 5522 (Project-ID No.~499180199).
T.J.~is funded by the ANR-25-CE57-2088 JCJC (QuDi).
The authors used Gemini Pro 3.1 and Claude Opus 5 during the preparation of this manuscript for assistance in the analytical calculations and preparing the sketch and summary figures.
The authors assume full responsibility for the scientific content, accuracy and integrity of this paper.

\section*{DATA AVAILABILITY} 

The supporting data for this article are openly available at
Zenodo \cite{datazenodo}.

\setcounter{section}{0}
\renewcommand{\thesection}{Appendix \Alph{section}}
\renewcommand{\thesubsection}{\arabic{subsection}}

\renewcommand{\theequation}{A.\arabic{equation}}
\setcounter{equation}{0}

\section{\label{app:conserved} Conserved quantities of the hybrid quantum system}

In this Appendix, we construct a general conserved quantity for the hybrid system considered, Eq.~\eqref{eq:Lindblad}.
The derivation makes use of the assumption that an operator exists satisfying the commutation relations $[\mathcal{O}_z,\mathcal{O}]=-\alpha\mathcal{O},~[\mathcal{O}_z,\mathcal{O}^\dagger]=\alpha\mathcal{O}^\dagger$. 
We show in Sec.~\ref{sec:results_symmetries}, that this condition holds for the example we consider. 
We can split the superoperator $\mathcal{L}^\dagger$ into terms that conserve, or not, the number of excitation, $N_\text{exc}\equiv a^\dagger a +\frac{1}{\alpha} \mathcal{O}_z$, as in the following
\begin{align}
\label{eq:Lindblad_adjoint_2}
\mathcal{L}^\dagger&\equiv -\mathcal{K}+\mathcal{J},~\text{with} \\
\mathcal{J} \mathcal{C} &= \Gamma a^\dagger\mathcal{C} a,\quad \mathcal{K}=\mathcal{K}_0+\mathcal{V},\nonumber \nonumber \\
\mathcal{K}_0\mathcal{C}&=-i\left(\delta+i\frac{\Gamma}{2}\right)a^\dagger a\mathcal{C}+i\left(\delta-i\frac{\Gamma}{2}\right)\mathcal{C}a^\dagger a\nonumber \\
\mathcal{V}\mathcal{C}&=ig\left[a^\dagger \mathcal{O} +a\mathcal{O}^\dagger,\mathcal{C}\right],\nonumber
\end{align}
where the terms in $\mathcal{K}$ conserve the number of excitations $N_\text{exc}$, while $\mathcal{J}$ adds one photonic excitation. Thus, the equation that we need to solve for the conserved quantities is 
\begin{align}
\label{eq:conserverd_C}
\mathcal{K}\mathcal{C}=\mathcal{J}\mathcal{C}.
\end{align}
This structure suggests that we can expand the conserved quantities as a sum over states with a fixed number of excitations $m$, $L^{(m)}_{i,j}$,
\begin{align}
\label{eq:conserverd_C_2}
\mathcal{C}_{(i,j)}=\sum_m L^{(m)}_{i,j},
\end{align}
where $L^{(0)}_{i,j}=\rho^{(i,j)}_\text{st}=\ket{0; o_i}\bra{0; o_j}$ are the steady states of the Liouvillian $\mathcal{L}$, Eq.~\eqref{eq:rho}.
In this construction we rely on the fact that for each steady state of $\mathcal{L}$ a conserved quantity exists, which is the steady solution of $\mathcal{L}^\dagger$.
In the following, we drop the index $(i,j)$, but we keep in mind that each conserved quantity is constructed started from a certain steady state $\rho^{(i,j)}_\text{st}$.
Using the expression from Eq.~\eqref{eq:conserverd_C_2}, Eq.~\eqref{eq:conserverd_C} reduces to the following recurrence relation
\begin{align}
\label{eq:conserverd_4}
\mathcal{K}L^{(m)}_{i,j}=\mathcal{J}L^{(m-1)}_{i,j}.
\end{align}
We make the further decomposition
\begin{align}
\label{eq:conserverd_5}
L^{(m)}=L^{(m)}_0+C^{(m)},~\text{with}~L^{(m)}_0=\frac{1}{m!}\left(a^\dagger\right)^m L^{(0)} \left(a\right)^m.
\end{align}
This is justified by the fact that $L^{(m)}_0$ represents the contribution of $\mathcal{K}_0$ to the recurrence relation from Eq.~\eqref{eq:conserverd_4}
\begin{align}
\label{eq:conserverd_6}
\mathcal{K}_0L^{(m)}_0=\Gamma m L^{(m)}_0=\Gamma a^\dagger L^{(m-1)}_0 a =\mathcal{J}L^{(m-1)}_0.
\end{align}
Thus, Eq.~\eqref{eq:conserverd_4} for $m>0$ reduces to 
\begin{align}
\label{eq:conserved_7}
\mathcal{V}L^{(m)}_0+\mathcal{K}C^{(m)}=\mathcal{J} C^{(m-1)}.
\end{align}
We can compute the first term as 
\begin{align}
\label{eq:conserved_8}
\mathcal{V}L^{(m)}_0&=ig \left[a^\dagger \mathcal{O} +a\mathcal{O}^\dagger,L^{(m)}_0\right] \\ 
&=\frac{i g}{(m-1)!}\Big[\left(a^\dagger\right)^{m-1} \mathcal{O}^\dagger L^{(0)} \left(a\right)^m \nonumber\\ 
&\qquad\qquad\qquad-\left(a^\dagger\right)^{m}  L^{(0)}\mathcal{O} \left(a\right)^{m-1}\Big]. \nonumber
\end{align}
By making use of the fact that $C^{(0)}=0$, from Eq.~\eqref{eq:conserverd_5}, we can iteratively compute $C^{(m)}$, Eq.~\eqref{eq:conserved_7} and Eq.~\eqref{eq:conserved_8}, as
\begin{align}
\label{eq:conserved_9}
C^{(m)}=-\sum_{k=1}^m\left(\mathcal{K}^{-1}\mathcal{J}\right)^{m-k} \mathcal{K}^{-1} \mathcal{V}L^{(k)}_0,
\end{align}
where the inverse superoperator $\mathcal{K}^{-1}$ is well defined as it always acts on states with a finite number of excitations.
It is useful to think about the action of $\mathcal{K}^{-1}$, which we can formally expand as 
\begin{align}
\label{eq:conserved_9b}
\mathcal{K}^{-1}=\sum_{j=1}^\infty (-1)^j \left(\mathcal{K}_0^{-1}\mathcal{V}\right)^j \mathcal{K}_0^{-1},
\end{align}
where $\mathcal{K}_0^{-1}\ket{n}\bra{m}=\left[-i\delta(n-m)+\frac{\Gamma}{2}(n+m)\right]^{-1}$ for the photonic Fock state $\ket{n}\bra{m}$. 
This implies that $\mathcal{K}^{-1}$ couples all matter states that are accessible from each other by successive applications of $\mathcal{O}$ and $\mathcal{O}^\dagger$ with prefactors depending on the photonic states.

Putting everything together, we arrive at the following general form of the conserved quantities
\begin{align}
\label{eq:conserved_C_10}
&\mathcal{C}_{(i,j)}=\sum_{m\geq 0} \frac{1}{m!}\left(a^\dagger\right)^m \rho^{(i,j)}_\text{st} \left(a\right)^m\\
&\qquad-\sum_{m>0}  \sum_{k=1}^m \frac{i g}{(k-1)!}\left(\mathcal{K}^{-1}\mathcal{J}\right)^{m-k} \mathcal{K}^{-1} \nonumber\\ &\qquad\qquad\left[\left(a^\dagger\right)^{k-1} \mathcal{O}^\dagger  \rho^{(i,j)}_\text{st} \left(a\right)^k -\left(a^\dagger\right)^{k}  \rho^{(i,j)}_\text{st}\mathcal{O} \left(a\right)^{k-1}\right]. \nonumber
\end{align}
The first term can written as 
\begin{align}
\label{eq:conserved_C_11}
\sum_{m\geq 0}& \frac{1}{m!}\left(a^\dagger\right)^m \rho^{(i,j)}_\text{st} \left(a\right)^m\\
&=\sum_{m\geq 0} \frac{1}{m!}\left(a^\dagger\right)^m \ket{\alpha=0; \mathbf{0}_i}\bra{\alpha=0; \mathbf{0}_j} \left(a\right)^m \nonumber\\
&=\mathbb{I}_\text{pho}\otimes \ket{\mathbf{0}_i}\bra{\mathbf{0}_j},
\end{align}
where $\mathbb{I}_\text{pho}$ is the identity operator defined over the photonic Hilbert space. By inserting the expression from Eq.~\eqref{eq:conserved_C_11} into Eq.~\eqref{eq:conserved_C_10}, we arrive at the expression given in the main text in Eq.~\eqref{eq:conserved_C_final}.
In Sec.~\ref{sec:results_conserved} we discuss how this expression leads to multi-body and nonlocal conserved quantities for the example of $\mathcal{O}$ discussed in the main text, already for a few particles.

\setcounter{equation}{0}
\renewcommand{\theequation}{B.\arabic{equation}}
\setcounter{figure}{0}
\renewcommand{\thefigure}{B\arabic{figure}}

\section{\label{app:similarity} Checking the similarity transformation}

We show that the transformation $G$, Eq.~\eqref{eq:transformation}, represent a map between $\mathcal{O}$, Eq.~\eqref{eq:operator}, and $Q$, Eq.~\eqref{eq:sl2_Q_algebra}. 
We first present the derivation for bosons, while afterwards we derive the argument for fermionic particles.

\subsection{Transformation for bosons}

In order to show that $G^{\text{(b)}-1} \mathcal{O}^\text{(b)} G^\text{(b)} = Q^\text{(b)}$ we rewrite the transformation as
\begin{align}
&G^\text{(b)}  = \prod_{k=1}^L \lambda_k^{n_k} = \exp\left( \sum_{k=1}^L s_k n_k \right),~\text{with} \\
&~s_k = \ln \lambda_k,~\lambda_k = \sqrt{\frac{(k-1)!(L-1)!}{(L-k)!}}. \nonumber
\end{align}
By making use of the exponential form we can compute the action of the transformation $G^\text{(b)}$ onto the bosonic creation and annihilation operators
\begin{align}
&G^{\text{(b)}-1} b^\dagger_j G^\text{(b)} = e^{-s_j} b^\dagger_j = \lambda_j^{-1} b^\dagger_j, \\
&G^{\text{(b)}-1} b_{j+1} G^\text{(b)} = e^{s_{j+1}} b_{j+1} = \lambda_{j+1} b_{j+1}. \nonumber
\end{align}
This leads to
\begin{align}
G^{\text{(b)}-1} \mathcal{O}^{\text{(b)}} G^{\text{(b)}}  &=  \sum_{j=1}^{L-1} (\lambda_j^{-1} b^\dagger_j) (\lambda_{j+1} b_{j+1}) \\
&= \sum_{j=1}^{L-1} \frac{\lambda_{j+1}}{\lambda_j} b^\dagger_j b_{j+1}, \nonumber \\
&= \sum_{j=1}^{L-1} \sqrt{j(L-j)} b^\dagger_j b_{j+1}=Q^\text{(b)}.  \nonumber
\end{align}

\subsection{Transformation for fermions}

For two fermionic Fock configurations that are different just by a single tunneling event, $\ket{\mu}=c^\dagger_j c_{j+1} \ket{\lambda}$, we have $G^\text{(f)}_\lambda=\sqrt{j(L-j)}G^\text{(f)}_\mu$. Now if we compute the action of the transformed operator on a certain state Fock state
\begin{align} 
\label{eq:transformation2}
G^\text{(f)}Q^\text{(f)}G^{\text{(f)}-1}\ket{\lambda}&= G^\text{(f)}Q^\text{(f)}\frac{1}{G^\text{(f)}_\lambda}\ket{\lambda} \\
&=\frac{1}{G^\text{(f)}_\lambda} G^\text{(f)}\sum_\mu \sqrt{j(L-j)} \ket{\mu} \nonumber \\
&= \sum_\mu \sqrt{j(L-j)} \frac{G^\text{(f)}_\mu}{G^\text{(f)}_\lambda} \ket{\mu} \nonumber \\
&= \sum_\mu  \ket{\mu}\equiv\mathcal{O}^\text{(f)}\ket{\lambda}, \nonumber 
\end{align}
where the sum over $\mu$ is the sum over all configuration accessible by a tunneling event from state $\ket{\lambda}$, i.e.~all states obtained as $\ket{\mu}=c^\dagger_j c_{j+1} \ket{\lambda}$.

\renewcommand{\theequation}{C.\arabic{equation}}
\setcounter{equation}{0}
\setcounter{figure}{0}
\renewcommand{\thefigure}{C\arabic{figure}}

\section{\label{app:potential} The potential induced by the similarity transformation}

In this appendix, we aim to understand the similarity transformation between the $\mathcal{O}$-$\mathfrak{sl}(2)$ algebra, Eq.~\eqref{eq:sl2_algebra}, and the $Q$-$\mathfrak{sl}(2)$ algebra, Eq.~\eqref{eq:sl2_Q_algebra}. 
We start with the fermionic case, with the transformation defined in Eq.~\eqref{eq:transformation} as 
\begin{align} 
\label{appeq:transformation}
&G^\text{(f)}Q^\text{(f)}G^{\text{(f)}-1}=\mathcal{O}^\text{(f)},~\text{with the entries:}\\
&G^\text{(f)}_{\lambda}=\prod_{i=1}^N \prod_{j=i}^{x_i-1} \sqrt{j(L-j)}\nonumber,
\end{align}
where $\{x_1,\dots,x_N\}$ are the positions of the particles of a Fock state $\ket{\lambda}$.
In particular, we want to obtain which states are favored by this transformation, i.e.~their weight increases for the null vectors of $\mathcal{O}^\text{(f)}$, compared to the null vectors of $Q^\text{(f)}$.
In order to achieve this, we need to consider the problem of finding the maximal value of the quantity $G^\text{(f)}_{\lambda}$ for a certain distribution $\ket{\lambda}=\{x_1,\dots,x_N\}$ at a fixed center-of-mass $X=\sum x_i$.
Instead of $G^\text{(f)}_{\lambda}$ it is convenient to consider its logarithm
\begin{align} 
\label{eq:potential_1}
\ln G^\text{(f)}_{\lambda}=&\frac{1}{2}\sum_{i=1}^{N}\sum_{j=i}^{x_i-1} \ln[j(L-j)] \\
&\equiv \frac{1}{2}\sum_{i=1}^{N}\left[F^\text{(f)}(x_i)-F^\text{(f)}(i) \right],\nonumber
\end{align}
where we have introduced the function $F^\text{(f)}(x)=\sum_{j=1}^{x-1} \ln[j(L-j)]$, with $F^\text{(f)}(1)=0$.
As the last term in Eq.~\eqref{eq:potential_1} is independent of the particle distribution, the problem of maximize $G^\text{(f)}_{\lambda}$ reduces to the maximization of the sum $\sum_{i=1}^{N}F^\text{(f)}(x_i)$.
In order to get an intuition regarding the particle distributions which maximize $\sum_{i=1}^{N}F^\text{(f)}(x_i)$ we analyze the shape of the function $F^\text{(f)}(x)$.
For this we compute the first and second order discrete derivatives
\begin{align} 
\label{eq:potential_2}
\Delta F^\text{(f)}(x)&=F^\text{(f)}(x+1)-F^\text{(f)}(x)\\
&=\ln [x(L-x)]\nonumber \\
\Delta^2 F^\text{(f)}(x)&=\ln \left[\frac{(x+1)(L-x-1)}{x(L-x)}\right],\nonumber
\end{align}
thus, $F^\text{(f)}(x)$ has a convex profile for $x<\frac{L-1}{2}$ and a concave profile for $x>\frac{L-1}{2}$, leading to an S-shape potential.
This result implies that the transformation $G^\text{(f)}$ favors states with particle distributions, which have particles both localized at the boundaries and in the bulk of the chain.
The emergent potential is directly relevant for the symmetry-constrained Liouvillian skin effect (see Sec.~\ref{sec:results_density}).
Furthermore, we also obtain that $F^\text{(f)}(x)+F^\text{(f)}(L+1-x)=F^\text{(f)}(L)$, a constant. 
This implies that the distance between two particles forming a centro-symmetric pair does not change the weight of the the transformation $G$, confirming the algebraic structure discussed in Sec.~\ref{sec:results_symmetries} and Sec.~\ref{sec:pairing_algebra}.

In the bosonic case, the transformation is defined in Eq.~\eqref{eq:transformation} as 
\begin{align} 
\label{appeq:transformation_bos}
&G^\text{(b)}Q^\text{(b)}G^{\text{(b)}-1}=\mathcal{O}^\text{(b)},~\text{with the entries:}\\
&G^\text{(b)}_{\lambda}=\prod_{j=1}^L  \left[\frac{(j-1)!(L-1)!}{(L-j)!}\right]^{\frac{n_j}{2}}, \nonumber
\end{align}
where a bosonic Fock state $\ket{\lambda}$ is characterized by the local occupations $n_j$.
In this case the logarithm of $G^\text{(b)}_{\lambda}$ is given by 
\begin{align} 
\label{eq:potential_1_bos}
\ln G^\text{(b)}_{\lambda}=&\sum_{j=1}^{N}\frac{n_j}{2}\ln\left[\frac{(j-1)!(L-1)!}{(L-j)!}\right]. 
\end{align}
The $(L-1)!$ just leads to a constant, we ignore it and define the single particle potential
\begin{align} 
\label{eq:potential_2_bos}
F^\text{(b)}(j)=&\ln\left[\frac{(j-1)!}{(L-j)!}\right]. 
\end{align}
Similar to Eq.~\eqref{eq:potential_2} we compute the discrete derivative and obtain $\Delta F^\text{(b)}(x)=\ln [x(L-x)]$, which is the same expression as for the fermionic case in Eq.~\eqref{eq:potential_2}.
This means that both the fermionic and bosonic particles feel the same effective potential stemming from the similarity transformation.
Thus, the difference in the density profiles, discussed in Fig.~\ref{sec:results_density}, mainly originate in the constrains imposed by the particle statistics in constructing the null states.

\renewcommand{\theequation}{D.\arabic{equation}}
\setcounter{equation}{0}
\setcounter{figure}{0}
\renewcommand{\thefigure}{D\arabic{figure}}

\section{Basis of the translation-invariant symmetric polynomials \label{app:TISP_basis}}

We show here that every translation-invariant symmetric polynomial (TISP) is a linear combination of the products $\mathcal{S}=\prod^{N}_{r=2}q^{a_{r}}_{r}$ of Eq.~\eqref{eq:S_polynomial}, and that these products are linearly independent. The argument has three steps.

\emph{Step 1: relative coordinates.} The condition $J_{-}f=\sum_{i}\partial_{x_{i}}f=0$ states that $f$ is unchanged under a global shift $x_{i}\to x_{i}+t$ of all the variables. Choosing $t=-\bar{x}$, we get
\begin{equation}
f\left(x_{1},\dots,x_{N}\right)=f\left(y_{1},\dots,y_{N}\right),\qquad y_{i}:=x_{i}-\bar{x}.
\end{equation}
A TISP thus only depends on the coordinates relative to the center of mass, exactly as a translation-invariant many-body wavefunction would.

\emph{Step 2: power sums.} By the previous step, $f$ is a symmetric polynomial of the $y_{i}$. It is a standard result (the fundamental theorem of symmetric polynomials) that any symmetric polynomial of $N$ variables can be written as a polynomial of the first $N$ power sums of these variables. The power sums of the $y_{i}$ are precisely the centered power sums, $\sum_{i}y^{r}_{i}=q_{r}$, and the first one vanishes, $q_{1}=\sum_{i}\left(x_{i}-\bar{x}\right)=0$. Therefore $f$ is a polynomial of $q_{2},\dots,q_{N}$ only, i.e.\ a linear combination of the products $\mathcal{S}$. The same argument shows that the $q_{r}$ with $r>N$ are not new objects and can be expressed in terms of $q_{2},\dots,q_{N}$.

The case $N=2$ illustrates the mechanism: $f$ can only depend on $x_{1}-x_{2}$, and the symmetry under $x_{1}\leftrightarrow x_{2}$ only allows even powers, so that $f$ is a polynomial in $q_{2}=\frac{1}{2}\left(x_{1}-x_{2}\right)^{2}$.

\emph{Step 3: independence.} It remains to check that there is no linear relation between different products $\mathcal{S}$. The ordinary power sums $p_{1},\dots,p_{N}$ are independent variables: they can be used as coordinates in place of $x_{1},\dots,x_{N}$ and satisfy no polynomial relation. On the hyperplane $\bar{x}=0$, i.e.\ $p_{1}=0$, the centered power sums reduce to the ordinary ones, $q_{r}=p_{r}$. A relation between products of $q_{2},\dots,q_{N}$ would then be a relation between $p_{2},\dots,p_{N}$ at $p_{1}=0$, which cannot exist since fixing $p_{1}$ leaves $p_{2},\dots,p_{N}$ free. The products $\mathcal{S}$ therefore form a basis of the TISPs.

\emph{Counting.} Since $q_{r}$ has degree $r$, the basis elements of degree $d$ are labelled by the ways of writing $d$ as a sum of integers between $2$ and $N$. Let $P_{N}\left(d\right)$ be the number of partitions of $d$ into integers between $1$ and $N$. Those that contain at least one $1$ are obtained by adding a $1$ to a partition of $d-1$, so the number of TISPs of degree $d$ is
\begin{equation}
P_{N}\left(d\right)-P_{N}\left(d-1\right).
\end{equation}
For $d\leq L-1$ the box constraint of Sec.~\ref{sec:group_theory_size} plays no role, $c_{d}=P_{N}\left(d\right)$, and we recover the degeneracy $\Omega_{\mathcal{J}\left(d\right)}=c_{d}-c_{d-1}$.

\emph{Finite lattice.} A polynomial represents a state of the $L$-site lattice only if no variable appears with a power larger than $L-1$. For $d\leq L-1$ this is automatic, and every product $\mathcal{S}$ of degree $d$ is a null state. For $L-1<d\leq\frac{N\left(L-1\right)}{2}$, which requires $N\geq3$, a single product $\mathcal{S}$ generically contains $x^{d}_{i}$ and is not a state of the lattice. The null states are then the \emph{linear combinations} of products of degree $d$ in which all powers larger than $L-1$ cancel, and there are $c_{d}-c_{d-1}$ of them. For instance, for $N=4$ and $L=3$ there are two products of degree $4$, $q^{2}_{2}$ and $q_{4}$, both containing $x^{4}_{i}$ and $x^{3}_{i}x_{j}$, while $c_{4}-c_{3}=1$. The unique null state is the combination
\begin{equation}
12\,q_{4}-7\,q^{2}_{2}\;\to\;\left|2,0,2\right\rangle ^{{\rm B}}-\frac{1}{\sqrt{2}}\left|1,2,1\right\rangle ^{{\rm B}}+\frac{\sqrt{6}}{4}\left|0,4,0\right\rangle ^{{\rm B}},
\end{equation}
in which these monomials cancel.

\renewcommand{\theequation}{E.\arabic{equation}}
\setcounter{equation}{0}
\setcounter{figure}{0}
\renewcommand{\thefigure}{E\arabic{figure}}

\section{Decomposition of a single centered power sum \label{app:q_r_proof}}

We prove Eq.~\eqref{eq:q_r_decomposition}, i.e.

\begin{equation}
q_{r}=\frac{\left(-1\right)^{r}}{N^{r}}\sum_{\boldsymbol{\nu}\vdash r}\frac{r!}{\prod\nu_{a}!}\left(\sum^{\ell\left(\boldsymbol{\nu}\right)}_{a=1}\left(1-N\right)^{\nu_{a}}+N-\ell\left(\boldsymbol{\nu}\right)\right)m_{\boldsymbol{\nu}}  
\end{equation}

\subsection{Direct proof}

Using $\bar{x}=\frac{1}{N}\sum_{j}x_{j}$, we write
\begin{equation}
x_{i}-\bar{x}=-\frac{1}{N}\Bigl[\left(1-N\right)x_{i}+\sum_{j\neq i}x_{j}\Bigr].
\end{equation}
The multinomial theorem then gives
\begin{equation}
\left(x_{i}-\bar{x}\right)^{r}=\frac{\left(-1\right)^{r}}{N^{r}}\sum_{\substack{\boldsymbol{\beta}\in\mathbb{N}^{N}\\
\left|\boldsymbol{\beta}\right|=r
}
}\frac{r!}{\prod^{N}_{j=1}\beta_{j}!}\left(1-N\right)^{\beta_{i}}x^{\boldsymbol{\beta}},\label{eq:multinomial_qr}
\end{equation}
where the sum runs over all exponent vectors $\boldsymbol{\beta}=\left(\beta_{1},\dots,\beta_{N}\right)$ with $\left|\boldsymbol{\beta}\right|:=\sum_{j}\beta_{j}=r$ and $x^{\boldsymbol{\beta}}:=\prod_{j}x^{\beta_{j}}_{j}$. Summing over $i$,
\begin{equation}
q_{r}=\frac{\left(-1\right)^{r}}{N^{r}}\sum_{\left|\boldsymbol{\beta}\right|=r}\frac{r!}{\prod_{j}\beta_{j}!}\Bigl[\sum^{N}_{i=1}\left(1-N\right)^{\beta_{i}}\Bigr]x^{\boldsymbol{\beta}}.
\end{equation}
The coefficient of $x^{\boldsymbol{\beta}}$ is invariant under permutations of the entries of $\boldsymbol{\beta}$. We can therefore group the exponent vectors according to the partition $\boldsymbol{\nu}\vdash r$, $\ell\left(\boldsymbol{\nu}\right)\leq N$, obtained by sorting their entries. By definition \eqref{eq:pseudopermanent}, the sum of $x^{\boldsymbol{\beta}}$ over all \emph{distinct} rearrangements $\boldsymbol{\beta}$ of $\boldsymbol{\nu}$ is $m_{\boldsymbol{\nu}}$, while $\prod_{j}\beta_{j}!=\prod_{a}\nu_{a}!$. Finally, $\boldsymbol{\beta}$ has $\ell\left(\boldsymbol{\nu}\right)$ non-zero entries, equal to the $\nu_{a}$, and $N-\ell\left(\boldsymbol{\nu}\right)$ vanishing entries, each contributing $\left(1-N\right)^{0}=1$:
\begin{equation}
\sum^{N}_{i=1}\left(1-N\right)^{\beta_{i}}=\sum^{\ell\left(\boldsymbol{\nu}\right)}_{a=1}\left(1-N\right)^{\nu_{a}}+N-\ell\left(\boldsymbol{\nu}\right).
\end{equation}
Collecting these three facts gives Eq.~\eqref{eq:q_r_decomposition}. 

\subsection{Derivation from the power-sum expansion}

The same result follows from the route described in the main text, $q_{r}=\sum^{r}_{s=0}\left(-1\right)^{r-s}\binom{r}{s}N^{s-r}p^{r-s}_{1}p_{s}$, which only requires the matrix elements $R_{\boldsymbol{\rho\nu}}$ for the hook-shaped $\boldsymbol{\rho}=\left(s,1^{r-s}\right)$. For $s\geq1$, the coefficient of $x^{\boldsymbol{\nu}}$ in $p_{s}p^{r-s}_{1}$ is obtained by choosing the variable $x_{a}$ that $p_{s}$ contributes to (which requires $\nu_{a}\geq s$) and distributing the remaining exponents with a multinomial coefficient:
\begin{equation}
R_{\left(s,1^{r-s}\right)\boldsymbol{\nu}}=\sum_{a\,:\,\nu_{a}\geq s}\frac{\left(r-s\right)!}{\left(\nu_{a}-s\right)!\prod_{b\neq a}\nu_{b}!},\qquad s\geq1,
\end{equation}
whereas for $s=0$, $p_{0}=N$ and $R_{\left(1^{r}\right)\boldsymbol{\nu}}=r!/\prod_{a}\nu_{a}!$. Inserting this in the binomial expansion and using $\binom{r}{s}\frac{\left(r-s\right)!}{\left(\nu_{a}-s\right)!}=\frac{r!}{\nu_{a}!}\binom{\nu_{a}}{s}$, the coefficient of $m_{\boldsymbol{\nu}}$ in $q_{r}$ reads
\begin{align}
 & \frac{\left(-1\right)^{r}}{N^{r}}\frac{r!}{\prod_{b}\nu_{b}!}\Bigl[N+\sum^{\ell\left(\boldsymbol{\nu}\right)}_{a=1}\sum^{\nu_{a}}_{s=1}\binom{\nu_{a}}{s}\left(-N\right)^{s}\Bigr]\nonumber \\
 & =\frac{\left(-1\right)^{r}}{N^{r}}\frac{r!}{\prod_{b}\nu_{b}!}\Bigl[N+\sum^{\ell\left(\boldsymbol{\nu}\right)}_{a=1}\left(\left(1-N\right)^{\nu_{a}}-1\right)\Bigr],
\end{align}
which is again Eq.~\eqref{eq:q_r_decomposition}.

\subsection{Checks}

For $r=1$ the only partition is $\boldsymbol{\nu}=\left(1\right)$ and the bracket is $\left(1-N\right)+N-1=0$, consistent with $q_{1}=0$. For $N=2$, $r=2$, the partitions $\left(2\right)$ and $\left(1,1\right)$ give the brackets $1+2-1=2$ and $-1-1+0=-2$ respectively, hence
\begin{equation}
q_{2}=\frac{1}{4}\left(2\,m_{\left(2,0\right)}-4\,m_{\left(1,1\right)}\right)=\frac{1}{2}\left(m_{\left(2,0\right)}-2m_{\left(1,1\right)}\right),
\end{equation}
in agreement with Sec.~\ref{sec:example_analytic_formula}.

\renewcommand{\theequation}{F.\arabic{equation}}
\setcounter{equation}{0}
\setcounter{figure}{0}
\renewcommand{\thefigure}{F\arabic{figure}}

\begin{figure}[!hbtp]
\centering
\includegraphics[width=0.49\textwidth]{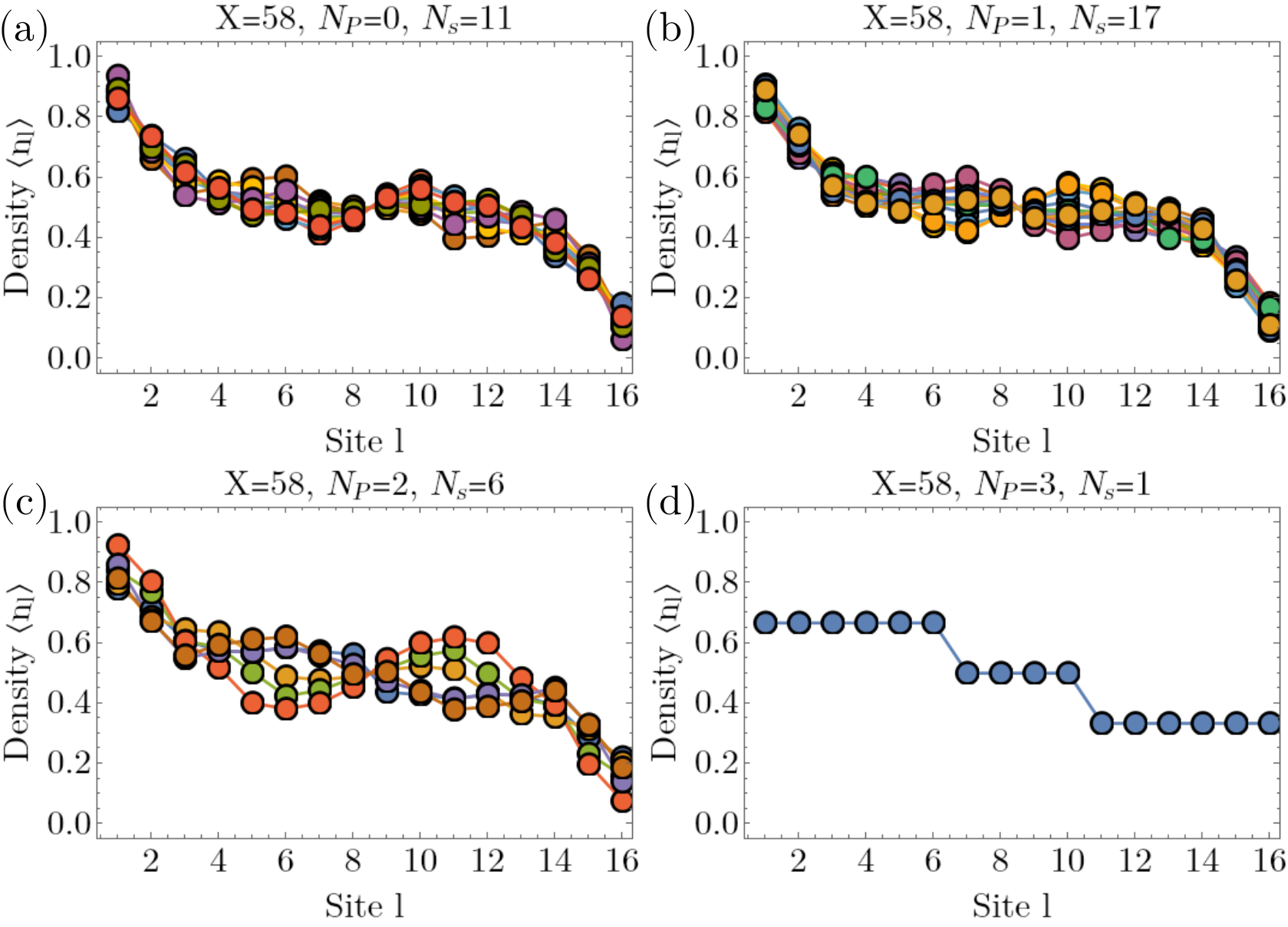}
\caption{
The density profile of the null states $\mathcal{O}$, Eq.~\eqref{eq:operator}, for fermionic particles for $N=8$ particles in $L=16$ sites, for the symmetry sectors labeled by the center-of-mass $X=58$, and the pairing quantum number $N_p$. 
As discussed around Eq.~\eqref{eq:average}, the density profiles correspond to a random basis in each $(X, N_p)$ symmetry sector, the number of states in each sector is $N_s$.
}
\label{fig:app_hardcore_density}
\end{figure}

\begin{figure}[!hbtp]
\centering
\includegraphics[width=0.49\textwidth]{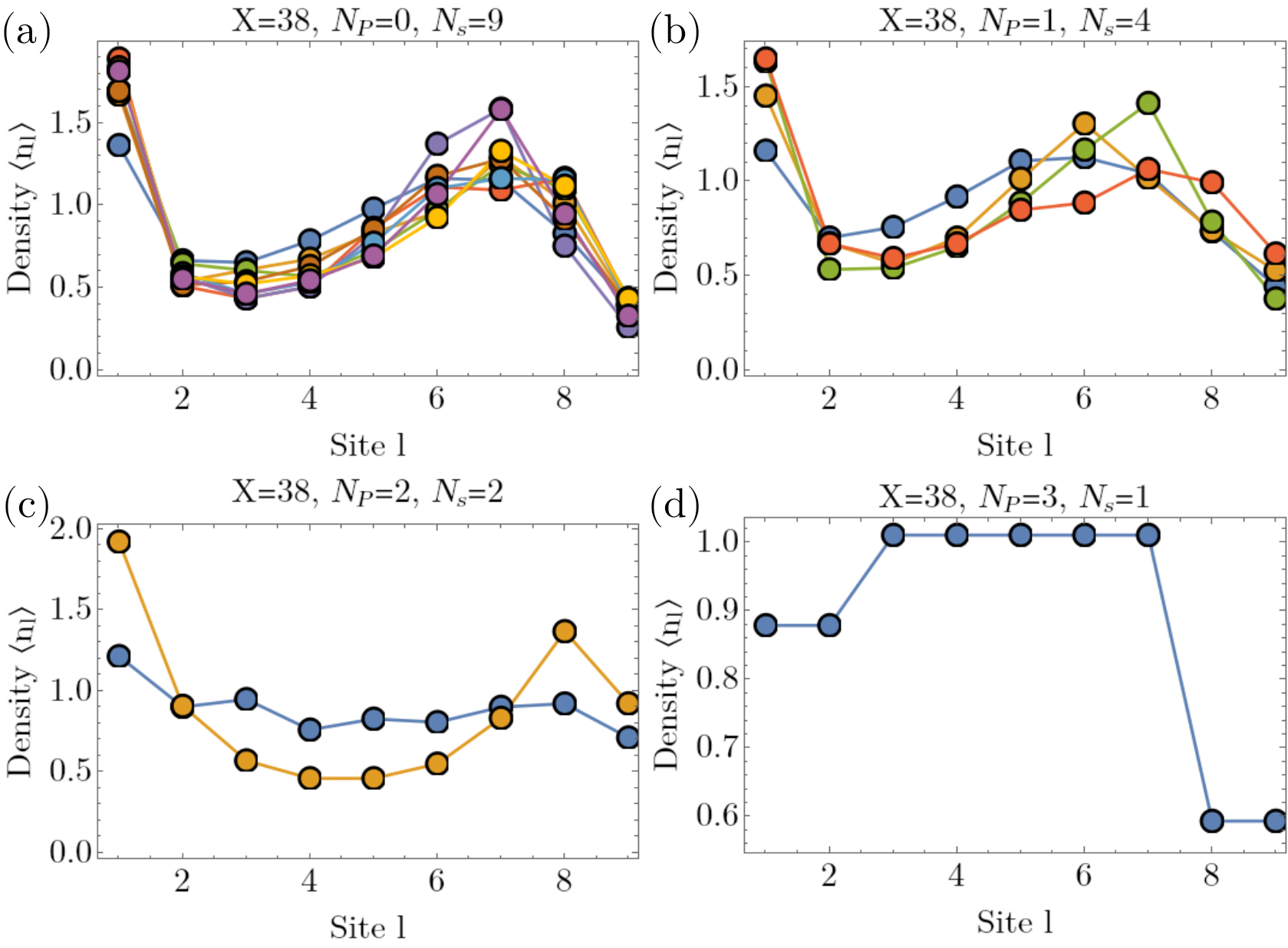}
\caption{
The density profile of the null states $\mathcal{O}$, Eq.~\eqref{eq:operator}, for bosonic particles for $N=8$ particles in $L=9$ sites, for the symmetry sectors labeled by the center-of-mass $X=38$, and the pairing quantum number $N_p$.
The results are obtained with the same procedure as for Fig.~\ref{fig:app_hardcore_density}.
}
\label{fig:app_bosons_density}
\end{figure}

\begin{figure}[!hbtp]
\centering
\includegraphics[width=0.49\textwidth]{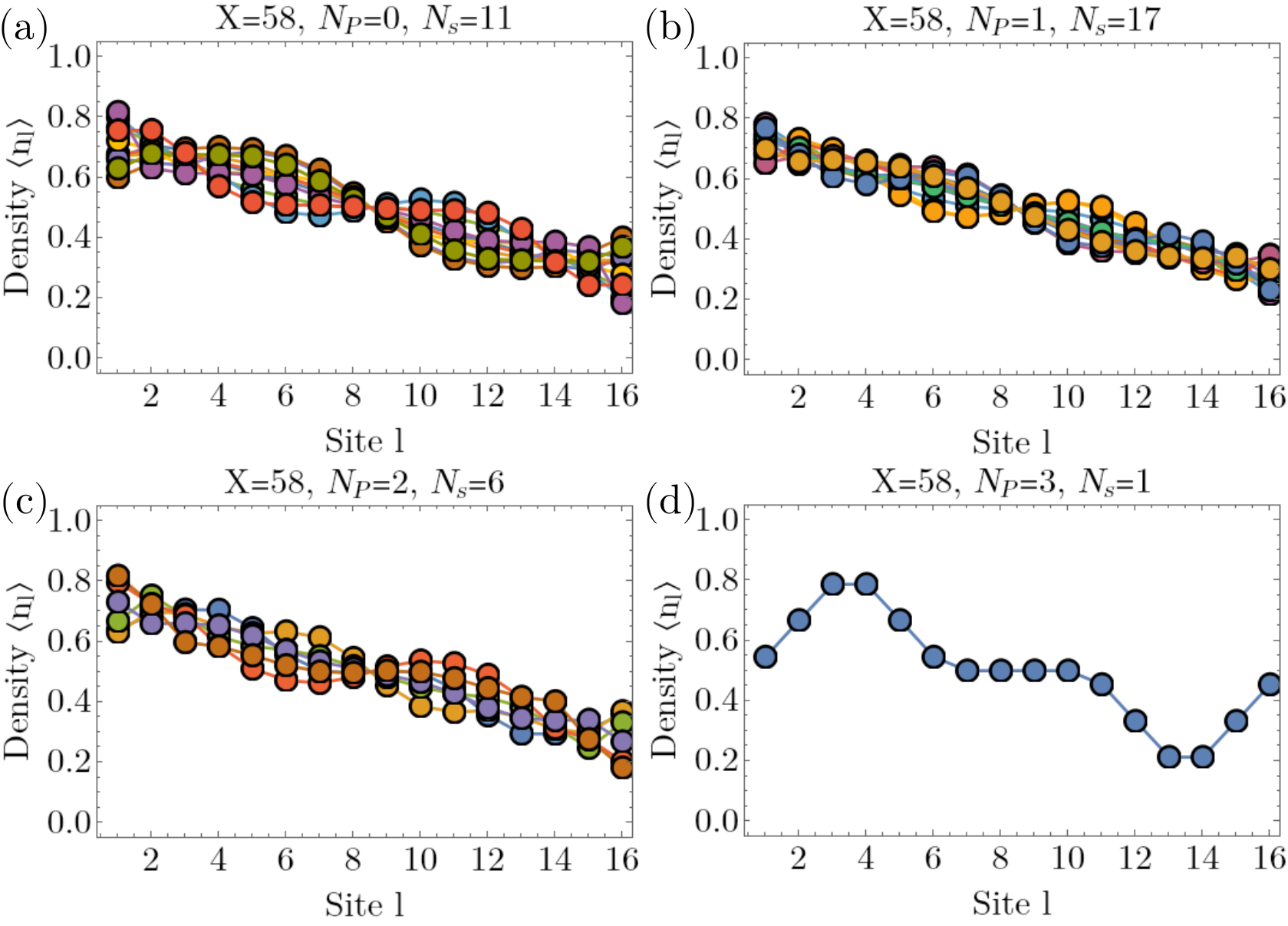}
\caption{
The density profile of the null states of the operator $Q$, Eq.~\eqref{eq:sl2_Q_algebra}, for fermionic particles for $N=8$ particles in $L=16$ sites for the symmetry sectors labeled by the center-of-mass $X=58$, and the pairing quantum number $N_p$.
The results are obtained with the same procedure as for Fig.~\ref{fig:app_hardcore_density}.
}
\label{fig:app_hardcore_Qdensity}
\end{figure}

\begin{figure}[!hbtp]
\centering
\includegraphics[width=0.49\textwidth]{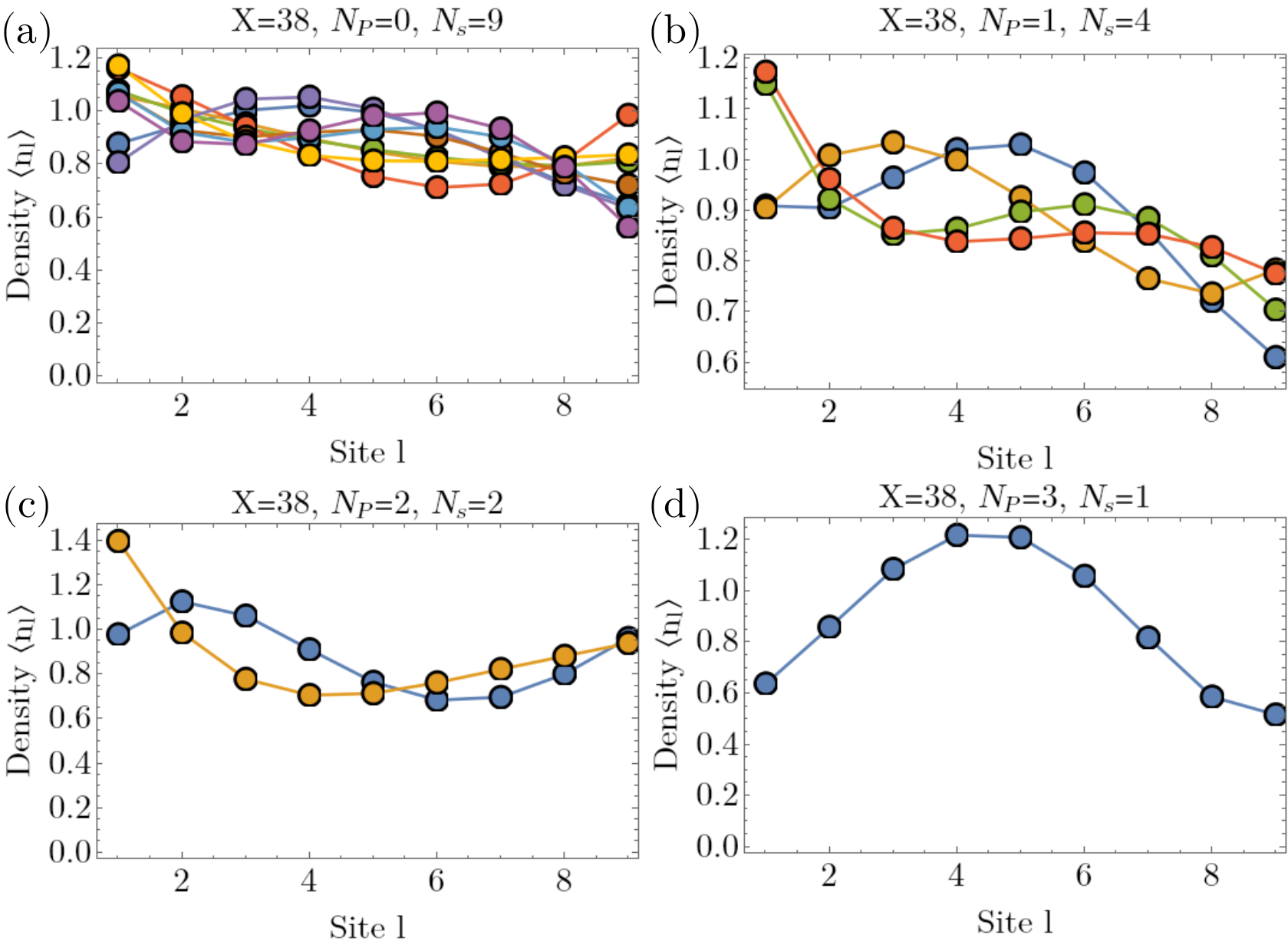}
\caption{
The density profile of the null states of the operator $\mathcal{Q}$, Eq.~\eqref{eq:sl2_Q_algebra}, for bosonic particles for $N=8$ particles in $L=9$ sites, for the symmetry sectors labeled by the center-of-mass $X=38$, and the pairing quantum number $N_p$.
The results are obtained with the same procedure as for Fig.~\ref{fig:app_hardcore_density}.
}
\label{fig:app_bosons_Qdensity}
\end{figure}

\section{\label{app:numerical_results} Additional exact diagonalization numerical results}

In this appendix, we show further results for the null states of the directed tunneling operator, Eq.~\eqref{eq:operator}, obtained with exact diagonalization.
These supplement the results discussed in Sec.~\ref{sec:results_steady_states}.
In particular, we present the density and entanglement profiles of all the null states obtained in selected symmetry sectors.
The results show that the individual null states exhibit the same non-trivial properties as the averaged states discussed in Sec.~\ref{sec:results_steady_states}, which followed the averaging procedure of Eq.~\eqref{eq:average}.
Afterwards, we further the density profiles of the null states of the $Q$ operator, Eq.~\eqref{eq:sl2_Q_algebra}, to emphasize the role of the similarity transformation, Eq.~\eqref{eq:transformation}.

\begin{figure}[!hbtp]
\centering
\includegraphics[width=0.49\textwidth]{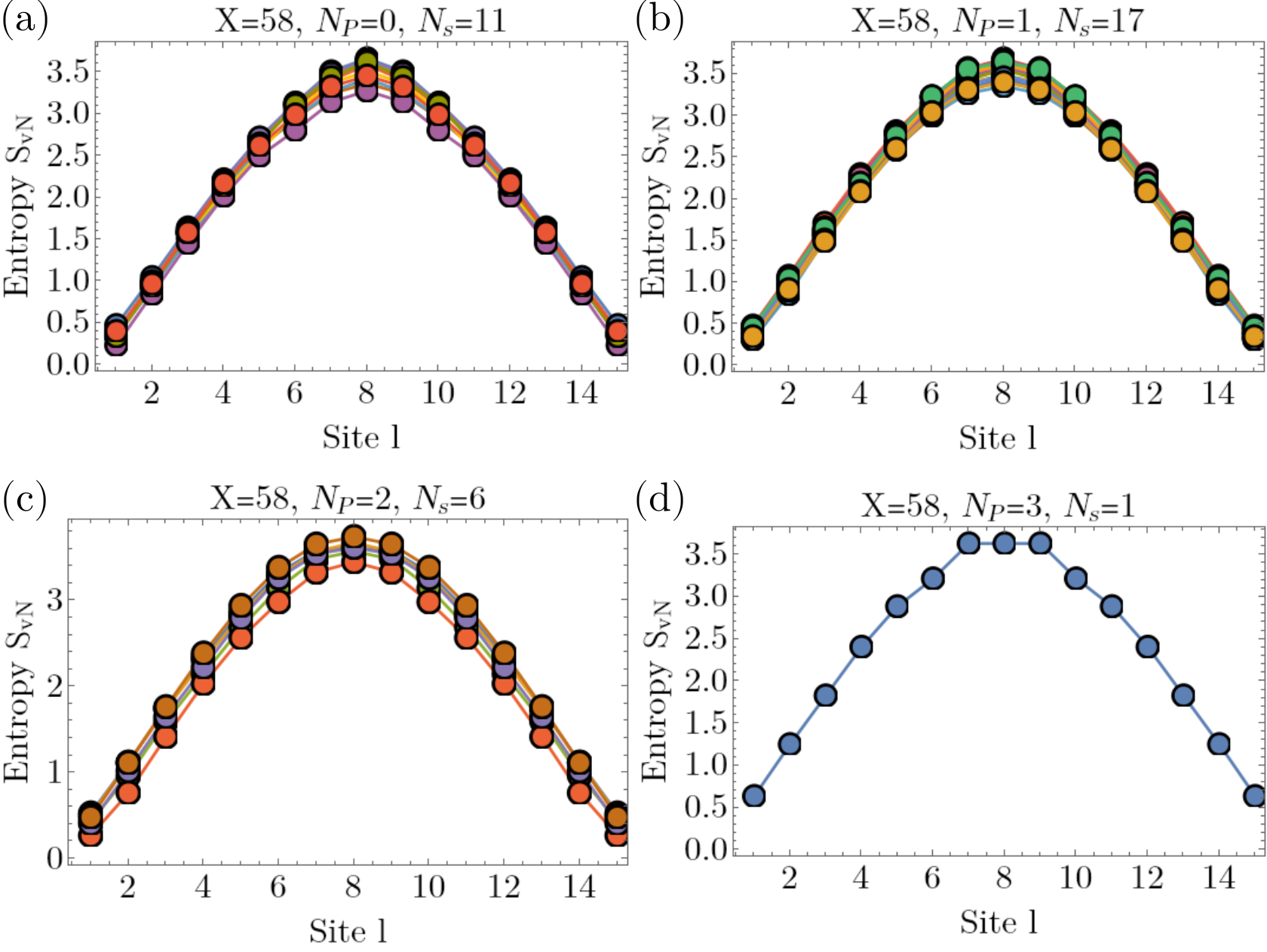}
\caption{
The entanglement entropy $S_\text{vN}$ of the null states of the operator $\mathcal{O}$, Eq.~\eqref{eq:operator}, for fermionic particles for $N=8$ particles in $L=16$ sites, for the symmetry sectors labeled by the center-of-mass $X=58$, and the pairing quantum number $N_p$.
The results are obtained with the same procedure as for Fig.~\ref{fig:app_hardcore_density}.
}
\label{fig:app_hardcore_entanglement}
\end{figure}

\begin{figure}[!hbtp]
\centering
\includegraphics[width=0.49\textwidth]{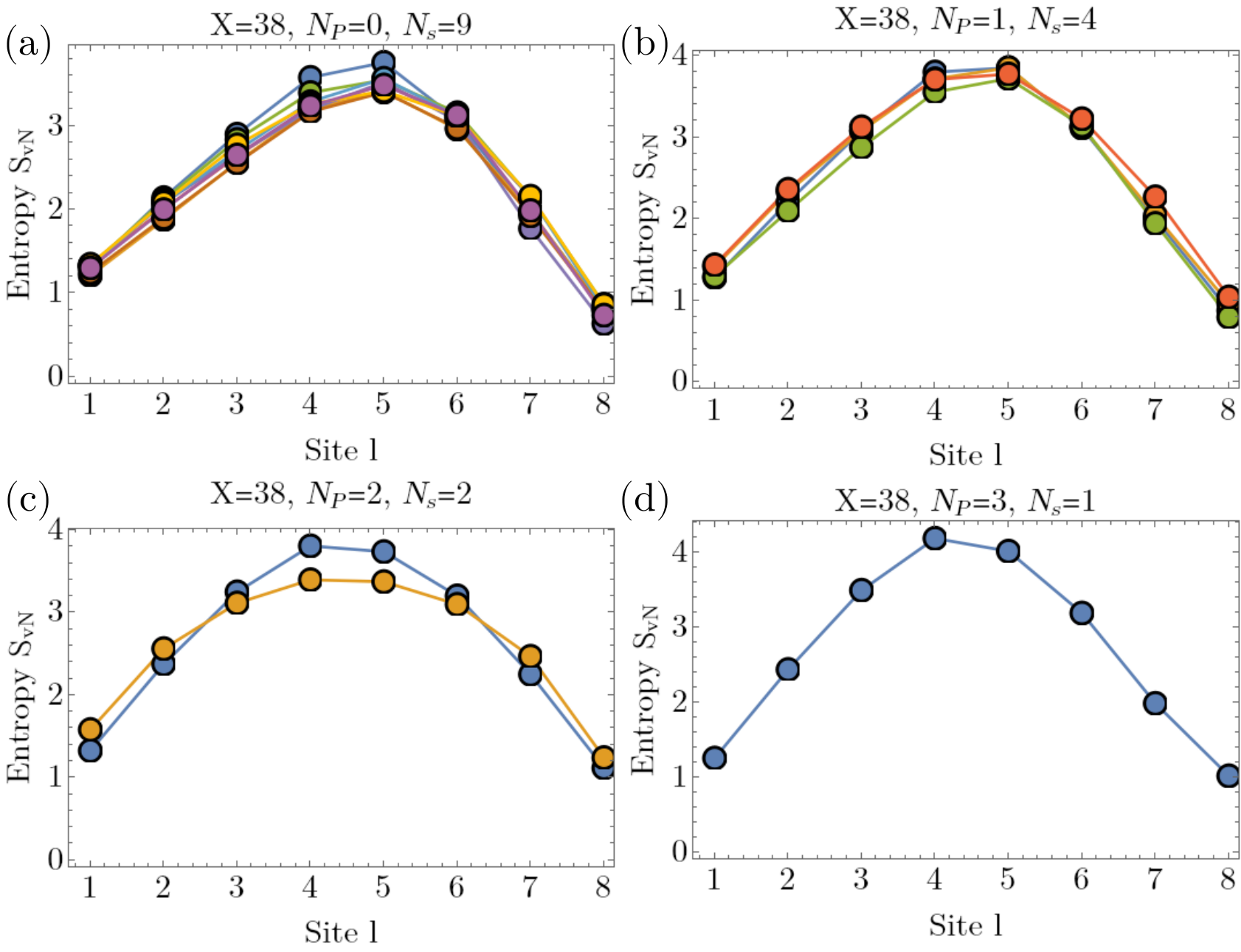}
\caption{
The entanglement entropy $S_\text{vN}$ of the null states of the operator $\mathcal{O}$, Eq.~\eqref{eq:operator}, for bosonic particles for $N=8$ particles in $L=9$ sites, for the symmetry sectors labeled by the center-of-mass $X=38$, and the pairing quantum number $N_p$.
The results are obtained with the same procedure as for Fig.~\ref{fig:app_hardcore_density}.
}
\label{fig:app_bosons_entanglement}
\end{figure}

In Fig.~\ref{fig:app_hardcore_density} we show the density profiles of the fermionic null states for $X=58$ and different number of $\eta$-pairs, $N_p$. The data corresponds to the averages shown in Fig.~\ref{fig:hardcore_density}(c). The equivalent results for the bosonic null states are shown in Fig.~\ref{fig:app_bosons_density} for $X=30$, corresponding to the average density results shown in Fig.~\ref{fig:bosons_density}(d).
The states depicted correspond to a random basis in the null subspace labeled by $X$ and $N_p$.
For both the fermionic and bosonic case we observe that the null states exhibit the same non-trivial properties discussed in Sec.~\ref{sec:results_density} which established the \emph{symmetry-constrained Liouvillian skin effect}.

In order to further emphasize the effect of the similarity transformation, Eq.~\eqref{eq:transformation}, we show the density profiles of the null states of the operator $Q$, Eq.~\eqref{eq:sl2_Q_algebra}.
In Sec.~\ref{sec:symmetry_2} we showed that $\mathcal{O}$ and $Q$ are related by a similarity transformation and one can also map their null states from one to the other employing the same transformation, Eq.~\eqref{eq:transformation3}.
Furthermore, we argued that the transformation can be viewed as an effective potential for the particles, with a direct impact on the accumulation profiles, \ref{app:potential}.
The effect of the similarity transformation is visible by comparing the equivalent null states of $\mathcal{O}$ and $Q$, Fig.~\ref{fig:app_hardcore_density} in contrast to Fig.~\ref{fig:app_hardcore_Qdensity} for fermions, and Fig.~\ref{fig:app_bosons_density} in contrast to Fig.~\ref{fig:app_bosons_Qdensity} for bosons.
While for the null states of $Q$ the density profile is closer to the average states corresponding to the same quantum numbers $X$ and $N_p$, the null states of $\mathcal{O}$ exhibit much strong the S-shape profile induced by the effective potential.

The entanglement entropy, $S_\text{vN}$, is plotted in Fig.~\ref{fig:app_hardcore_entanglement} for the fermionic null states for $X=58$ and different number of $\eta$-pairs, $N_p$. This corresponds to the averages shown in Fig.~\ref{fig:hardcore_entropy}(c). 
The entanglement entropy for the bosonic null states is shown in Fig.~\ref{fig:app_bosons_entanglement} for $X=30$, corresponding to Fig.~\ref{fig:bosons_entropy}(d).
All states show exhibit an increase of the entanglement entropy with the size of the subsystem for which $S_\text{vN}$ was computed. This shows that the states are characterized by volume-law entanglement, see Sec.~\ref{sec:results_entanglement}.
As discussed in Sec.~\ref{sec:results_entanglement}, one of the main differences between the fermionic and bosonic results is the position of the maximum of $S_\text{vN}$, which is related to the how the particles accumulate at the boundary depending on their statistics.

\renewcommand{\theequation}{G.\arabic{equation}}
\setcounter{equation}{0}
\setcounter{figure}{0}
\renewcommand{\thefigure}{G\arabic{figure}}

\section{\label{app:tmps} Details on the numerical Matrix Product States for hybrid dissipative systems}

The numerically exact quantum dissipative dynamics results presented in Sec.~\ref{sec:results_tmps} are based on matrix product state (MPS) methods \cite{Schollwoeck2011, PaeckelHubig2019}.
The dissipative dynamics is implemented via the stochastic unraveling of the master equation with quantum trajectories \cite{DalibardMolmer1992, GardinerZoller1992, Daley2014}.
In this approach, one performs the time-evolution of quantum trajectory wavefunctions and takes the Monte-Carlo average over all sampled trajectories.
The effective Hamiltonian resulting from the unraveling contains both short-range atomic terms and the global-range coupling to the cavity.
To deal with this challenge we employ a recent variant of the quasi-exact time-dependent variational matrix product state (tMPS) \cite{HalatiKollath2020, HalatiKollath2020b, TolleHalati2025c} based on both the Trotter-Suzuki decomposition of the time evolution propagator \cite{WhiteFeiguin2004, DaleyVidal2004, Schollwoeck2011} and the dynamical deformation of the MPS structure using swap gates \cite{StoudenmireWhite2010, Schollwoeck2011, WallRey2016}. 
For the model considered in this work, in which the cavity mode is coupled to a two-sites atomic operator, one needs to extend the previous implementation \cite{HalatiKollath2020, HalatiKollath2020b} in order to include three-sites coupling gates.
Additional details regarding the implementation together with convergence benchmarks can be found in Ref.~\cite{TolleHalati2025c}.
The method was implemented with the use of the ITensor Library \cite{FishmanStoudenmire2020}.

The convergence of the tMPS results shown in this work was ensured with the following convergence parameters: a maximal bond dimension of $400$ states, which determined a truncation error of at most $10^{-12}$ at the final time, and a time-step of $d\tau g/\hbar=5\times10^{-3}$. We average the results over at least $475$ quantum trajectories.
The results are shown with the error bars stemming from the standard deviation of the Monte-Carlo sampling, when the error bars are absent the stochastic errors are smaller than the size of the symbols employed.

\renewcommand{\theequation}{H.\arabic{equation}}
\setcounter{equation}{0}

\section{\label{app:numerical_results_mps} Additional \lowercase{t}MPS numerical results}

\setcounter{figure}{0}
\renewcommand{\thefigure}{H\arabic{figure}}
\begin{figure}[!hbtp]
\centering
\includegraphics[width=0.38\textwidth]{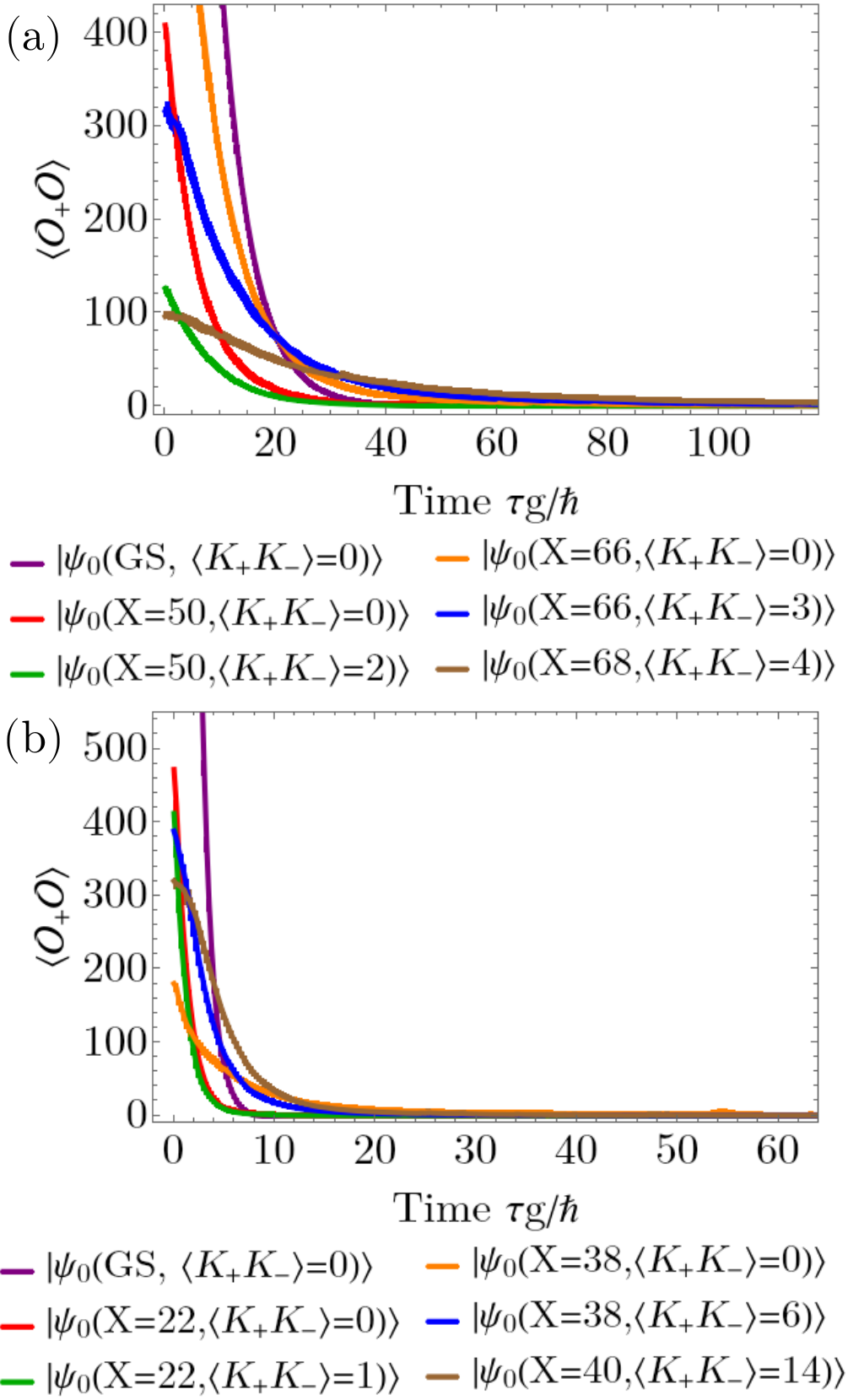}
\caption{
Dissipative time-evolution of the expectation value $\langle \mathcal{O}^\dagger\mathcal{O}\rangle$ for (a) fermions and (b) bosons.
The initial states considered are detailed in Eq.~\eqref{eq:initial_states_f} and Eq.~\eqref{eq:initial_states_b}.
The numerical tMPS results are obtained by performing the dissipative time-evolution of Eq.~\eqref{eq:Lindblad}.
We consider $N=8$ particles in (a) $L=16$ and (b) $L=9$ sites, with the parameters $\Gamma/g=10$, $\delta/g=5$.
}
\label{fig:app_opo}
\end{figure}

In this appendix, we present the initial states employed in the tMPS simulations of the dissipative time evolution and additional numerical results to complement the results discussed in Sec.~\ref{sec:results_tmps}.

\begin{figure}[!hbtp]
\centering
\includegraphics[width=0.37\textwidth]{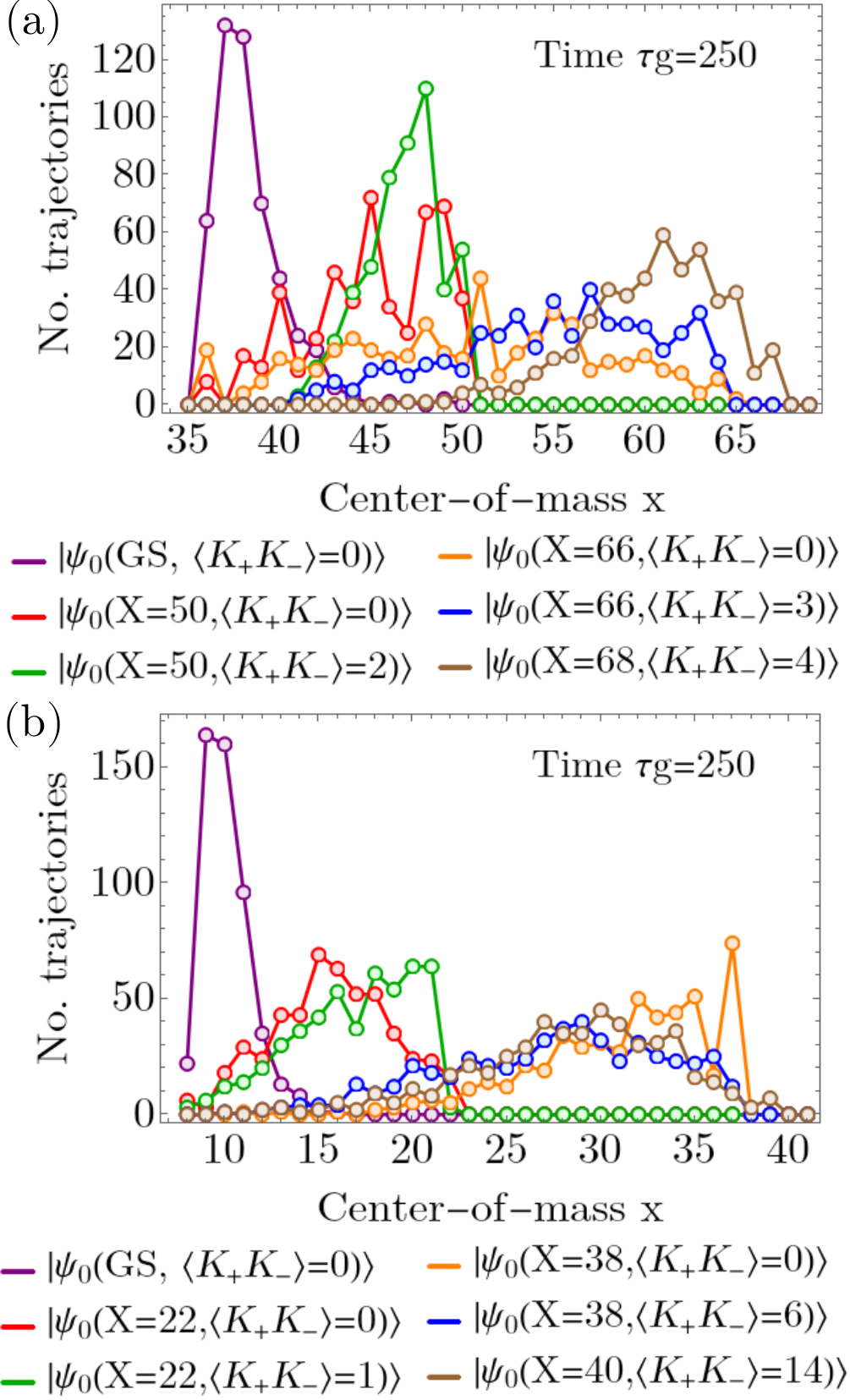}
\caption{
The distribution of the expectation value of the center-of-mass $\langle X \rangle$ among the single quantum trajectories sampled in the numerical simulation dissipative time-evolution, at $\tau g=250$ for (a) fermionic and (b) bosonic atoms. 
Each data point counts the number of trajectories for which $\langle X \rangle$ is within the interval $[x,x+1)$, with $x$ taking integer values.
The initial states considered are detailed in Eq.~\eqref{eq:initial_states_f} and Eq.~\eqref{eq:initial_states_b}.
We consider $N=8$ particles in (a) $L=16$ and (b) $L=9$ sites, with the parameters $\Gamma/g=10$, $\delta/g=5$.
}
\label{fig:app_com}
\end{figure}

\begin{figure}[!hbtp]
\centering
\includegraphics[width=0.37\textwidth]{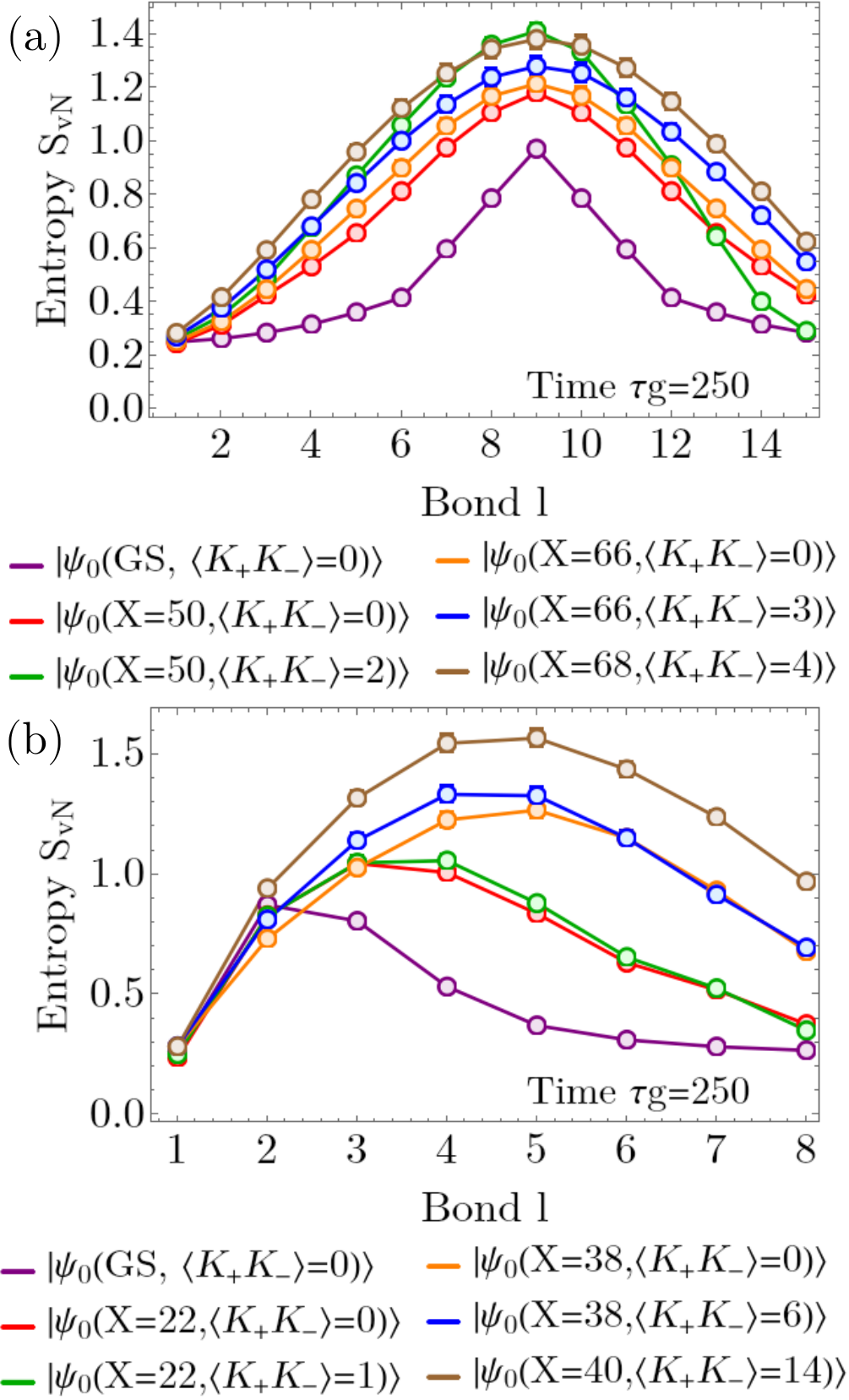}
\caption{
The MPS entanglement entropy of the quantum trajectories at the long-time $\tau g=250$, corresponding to different initial states for (a) fermionic and (b) bosonic atoms. 
The initial states considered are detailed in Eq.~\eqref{eq:initial_states_f} and Eq.~\eqref{eq:initial_states_b}.
The entanglement entropy is computed for the MPS representation which contains the cavity as the first site \cite{HalatiKollath2020, HalatiKollath2020b}.
We consider $N=8$ particles in (a) $L=16$ and (b) $L=9$ sites, with the parameters $\Gamma/g=10$, $\delta/g=5$.
}
\label{fig:app_svn}
\end{figure}

\subsection{Initial states of the tMPS simulations \label{app:initial_states}}

The fermionic Fock states employed as initial states for the case in which fermionic atoms are coupled to the dissipative cavity in Sec.~\ref{sec:results_tmps} are given by
\begin{align}
\label{eq:initial_states_f}
\ket{\psi_0(X=50, \langle K_+ K_- \rangle=0)}=& \ket{1101110101000100}, \\
\ket{\psi_0(X=50, \langle K_+ K_- \rangle=2)}=& \ket{0110111111000000}, \nonumber\\
\ket{\psi_0(X=66, \langle K_+ K_- \rangle=0)}=& \ket{0110101010101001}, \nonumber\\
\ket{\psi_0(X=66, \langle K_+ K_- \rangle=3)}=& \ket{0110001111010010}, \nonumber\\
\ket{\psi_0(X=68, \langle K_+ K_- \rangle=4)}=& \ket{0000111111110000}. \nonumber
\end{align}
We note that these Fock states are not eigenstates of the operators $K_+ K_- $ and $\Omega_K$, Eq.~\eqref{eq:casimir_pairing_bosons} and Eq.~\eqref{eq:casimir_pairing_SU2}, thus, they contain contributions from several symmetry sectors with different number of pairs. The exceptions are the states for which $\langle K_+ K_- \rangle=0$, as they are null states of $K_-$.

The bosonic Fock states employed as initial states for the case in which bosonic atoms are coupled to the dissipative cavity in Sec.~\ref{sec:results_tmps} are given by
\begin{align}
\label{eq:initial_states_b}
\ket{\psi_0(X=22, \langle K_+ K_- \rangle=0)}=& \ket{1, 3, 3, 0, 0, 1, 0, 0, 0}, \\
\ket{\psi_0(X=22, \langle K_+ K_- \rangle=1)}=& \ket{2, 2, 2, 1, 0, 1, 0, 0, 0}, \nonumber\\
\ket{\psi_0(X=38, \langle K_+ K_- \rangle=0)}=& \ket{2, 0, 2, 0, 0, 1, 0, 3, 0}, \nonumber\\
\ket{\psi_0(X=38, \langle K_+ K_- \rangle=6)}=& \ket{0, 2, 2, 0, 1, 0, 1, 2, 0}, \nonumber\\
\ket{\psi_0(X=40, \langle K_+ K_- \rangle=14)}=& \ket{0, 0, 0, 0, 8, 0, 0, 0, 0}. \nonumber
\end{align}
The discussion made above regarding the Fock states and $\eta$-pairing symmetry holds also for the bosonic case.

\subsection{\label{app:numerical_results2} Other observables}

In Sec.~\ref{sec:results_tmps} we have shown several the behavior of several observables at the time $\tau g=250$. In order to check that for this time, the dissipative dynamics is close to the decoherence-free subspace, we compute the expectation value $\langle \mathcal{O}^\dagger\mathcal{O}\rangle$, Fig.~\ref{fig:app_opo}. The expectation value $\langle \mathcal{O}^\dagger\mathcal{O}\rangle$ vanishes for the null states of $\mathcal{O}$.
In Fig.~\ref{fig:app_opo} we observe that $\langle \mathcal{O}^\dagger\mathcal{O}\rangle$ has small values for times much shorter than $\tau g=250$.
Furthermore, we can see that the different initial states exhbit different timescales in their relaxation towards the steady state.

In Fig.~\ref{fig:hardcore_density}(b) and Fig.~\ref{fig:bosons_density}(b) we have shown how the expectation value of the center-of-mass. We supplement this in Fig.~\ref{fig:app_com} where we show the distribution of the expectation values of the center-of-mass for the different quantum trajectories. We observe a large variability in the shape and width of the distributions for the different initial states considered. 
Such distributions offer valuable information for the goal of stabilizing particular steady states of the hybrid system with desired properties, e.g.~long-range correlations. One needs to consider both the average value at which  $\langle X \rangle$ saturates at long-times and also its variance and higher moments of the distribution. We leave the detailed investigation of the distributions as an outlook for further works.

In Fig.~\ref{fig:app_svn} we compute the average entropy, $S_{vN}$, of the quantum trajectories. While this is not a measure of the entanglement for the density matrix, it is still an useful quantity for understanding the complexity of each wavefunction sampled in the quantum trajectory approach.
We note that we compute $S_{vN}$ for the MPS representation which contains the cavity as the first site, including also the light-matter entanglement present in each quantum trajectory \cite{HalatiKollath2020, HalatiKollath2020b}. 
Thus, one should take care in comparing these results with the entanglement entropy of the atomic null states discussed in Sec.~\ref{sec:results_entanglement}.
However, we observe similar behavior as in Sec.~\ref{sec:results_entanglement}: entanglement entropy scaling with the size of the subsystem and larger values of $S_{vN}$ are correlated with the occupation of states with a large center-of-mass  $\langle X \rangle$.

\end{document}